%% file: main.tex
\documentclass[a4paper,11pt]{article}
\pdfoutput=1 
\usepackage{soul}
\usepackage{jcappub} 
\usepackage[T1]{fontenc} 
\usepackage[normalem]{ulem}
\usepackage{float}
\usepackage[numbers,sort&compress]{natbib}
\usepackage{tikz}
\usepackage{multirow}
\usepackage{orcidlink}

\title{Full-Shape analysis of the power spectrum and bispectrum of DESI DR1 LRG and QSO samples}
\abstract{We present the first joint analysis of the power spectrum and bispectrum using the Data Release 1 (DR1) of the Dark Energy Spectroscopic Instrument (DESI), focusing on Luminous Red Galaxies (LRGs) and quasars (QSOs) across a redshift range of $0.4\leq z\leq2.1$. By combining the two- and three-point statistics, we are able to partially break the degeneracy between the logarithmic growth rate, $f(z)$, and the amplitude of dark matter fluctuations, $\sigma_\textrm{s8}(z)$, which cannot be measured separately in analyses that only involve the power spectrum. 
In comparison with the (fiducial) Planck $\Lambda$CDM cosmology we obtain  $f/f^\textrm{fid}=\{0.888_{-0.089}^{+0.186},0.977_{-0.220}^{+0.182},1.030_{-0.085}^{+0.368}\}$,  $\sigma_{s8}/\sigma^\textrm{fid}_\textrm{s8}=\{1.224_{-0.133}^{+0.091},1.071_{-0.163}^{+0.278},1.000_{-0.223}^{+0.088}\}$ respectively for the three LRG redshift bins, corresponding to a
cumulative 10.1\% constraint on $f$, and of 8.4\% on $\sigma_\textrm{s8}$, including the systematic error budget.  Additionally, we obtain constraints for the ShapeFit compressed parameters describing the isotropic scaling parameter, $\alpha_\textrm{iso}(z)$, the Alcock-Paczyński parameter, $\alpha_\textrm{AP}(z)$, the combined growth of structure parameter $f\sigma_\textrm{s8}(z)$, and the combined shape parameter $m(z)+n(z)$. Their cumulative constraints from our joint power spectrum-bispectrum analysis are respectively $\sigma_{\alpha_\textrm{iso}}=0.9\%$ (9\% improvement with respect to our power spectrum-only analysis); $\sigma_{\alpha_\textrm{AP}}=2.3\%$ (no improvement with respect to power spectrum-only analysis, which is expected given that the bispectrum monopole has no significant anisotropic signal); $\sigma_{f\sigma_\textrm{s8}}=5.1\%$ (9\% improvement); $\sigma_{m+n}=2.3\%$ (11\% improvement). These results are fully consistent with the main DESI power spectrum analysis, demonstrating the robustness of the DESI cosmological constraints, and compatible with Planck $\Lambda$CDM cosmology.}
\input{DESI-2025-0531_author_list33}

\emailAdd{sergi.novell@icc.ub.edu, hectorgil@icc.ub.edu, liciaverde@icc.ub.edu}

\begin{document}
\maketitle
\section{Introduction}
Cosmology is in its precision era, with increasingly larger datasets providing us with huge statistical power to test our cosmological models of the Universe. The $\Lambda$-Cold Dark Matter ($\Lambda$CDM) model has proven remarkably successful in explaining observations across a broad range of epochs. However, fundamental questions persist about the nature of the dark component of the universe (dark matter and dark energy), which constitutes the majority of the Universe’s energy density. In addition, several tensions, such as the Hubble tension \cite{verde2019tensions}, persist between early and late Universe observations, thus suggesting that further investigation of the possible inconsistencies of $\Lambda$CDM is necessary.

Galaxy redshift surveys have been instrumental in probing the large-scale structure (LSS) of the Universe. Surveys such as the Sloan Digital Sky Survey (SDSS) \cite{gunn20062}, the 2-degree Field Galaxy Redshift Survey (2dFGRS) \cite{percival20012df},  the Baryon Oscillation Spectroscopic Survey (BOSS) \cite{dawson2012baryon} and extended BOSS (eBOSS) \cite{dawson2016sdss} have allowed the cosmology community to, sequentially, lay the foundations, develop the methodology, and obtain precise constraints on the baryon acoustic oscillation (BAO) feature and the growth of structure. 

The Dark Energy Spectroscopic Instrument (DESI) \cite{Snowmass2013.Levi,DESI2016a.Science,DESI2016b.Instr,DESI2022.KP1.Instr,FocalPlane.Silber.2023,Corrector.Miller.2023,Spectro.Pipeline.Guy.2023,SurveyOps.Schlafly.2023,LRG.TS.Zhou.2023,DESI2023a.KP1.SV,DESI2023b.KP1.EDR,2024arXiv240403000D,2024arXiv240403001D,adame2024desi,2024arXiv240403002D,FiberSystem.Poppett.2024} builds on this legacy, targeting over 40 million objects, including galaxies, quasars, and Ly$\alpha$ absorbers, thus mapping the cosmic web across a wide range of redshifts ($z \sim 0$ to $z \sim 4$). This dataset spans a volume several times larger than previous surveys with enough signal to go beyond two-point statistics, the primary workhorse of cosmology analyses to date. During the course of five years of observations, DESI will cover an effective volume of $V_\textrm{eff}\sim 50-60\,\textrm{Gpc}$.

The bispectrum, which is the natural next-order statistic, quantifies the correlations among triplets of points (in an analogous way as the power spectrum is obtained by correlations of pairs). Given that it is a higher-order statistic, the bispectrum encloses information about the non-Gaussian features of the matter distribution, probing non-linear structure formation, non-linear galaxy bias, and also potential signatures of primordial non-Gaussianities \cite{baldauf2011primordial,dizgah2021primordial, gualdi_matter_2021}. Hence the bispectrum complements the power spectrum by helping to break degeneracies among cosmological parameters \cite{Verde1998,Matarrese1997,Sefusatti:2006pa,fry1993biasing,Ruggeri_2018,Hahn_2020,oddo2021cosmological,bartolo2013matter,bellini2015signatures,bertacca2018relativistic,Gil_Mar_n_2011,yankelevich2019cosmological,coulton2019constraining,ruggeri2018demnuni,gagrani2017information,Gualdi:2020aniso,rizzo2023halo,ivanov2023cosmology,d2022one}, as will be discussed in what follows.

Application to real data is however challenging and computationally intensive: the bispectrum data-vector is large, with correlations between elements, its signal-to-noise is low, the covariance matrix is non-diagonal. Additionally, the modelling of the signal and the potential systematics is highly complex.  This is why, compared to the extensive theoretical efforts, application of the bispectrum to real data to constrain cosmology  has been somewhat limited to date, with few attempts by very few  research groups \cite{Verde:2001pf, Scoccimarro01, Gil-Marin:2014biasgravity, gil-marin_power_2015, HGMbispectrum15b, Gil-Marin:2016sdss, gil-marin_clustering_2017,ivanov2023cosmology, philcox2022boss,d2022boss,d2022limits,gualdi_geometrical_2019,Gualdi:2018pyw,cabass2022constraints,cabass2022constraintsmulti}.

In this work, we perform the first joint power spectrum-bispectrum analysis using the DESI Data Release 1 (DR1) \cite{desidr12024desi}, for both the Luminous Red Galaxy (LRG) and the quasar (QSO) samples. These samples span across the redshift range $0.4 \leq z \leq 2.1$, thus probing both early and intermediate epochs of structure formation. 
We model the bispectrum signal employing the GEO-FPT model, which has been calibrated and validated using a broad range of simulations \cite{novell2023geofpt}, while our power spectrum analysis is based on the renormalized perturbation theory model (RPT, \cite{gil-marinetal:2012}) and resembles the one employed in the BOSS and eBOSS analyses \cite{gil2016clustering,HGMeboss}. To accurately quantify both our statistical and systematic errors, we employ sets of realistic mock catalogues that allow us to estimate the covariance matrix and quantify observational and theoretical systematics.

We derive constraints on cosmological parameters within the template-based {\it ShapeFit} approach of ~\cite{brieden2021shapefit}. 
This approach is referred to as a type of Full-Shape analyses in the DESI papers \cite{KP5Desi}, as it exploits information on the shape of the power spectrum (and bispectrum). Other Full-Shape approaches, such as the Full Modelling (also known as direct fit) employed in other DESI papers are not considered in this work.
Within ShapeFit, the {\it shape} parameters $m$ and $n$, enclosing information about the shape of the matter power spectrum, are added to the standard parameters of interest in template-based methods: the dilation parameters along and across the line-of-sight, $\alpha_\parallel,\alpha_\bot$, the logarithmic growth rate of structure, $f$, and the amplitude of matter fluctuations, $\sigma_\textrm{s8}$. 

As detailed in \cite{brieden2021shapefit}, we adopt $\sigma_\textrm{s8}$ (rather than the more common parameter $\sigma_{8}$) for parametrising the smoothed amplitude of matter fluctuations. The essential difference is that $\sigma_8$ smooths the fluctuations on the fixed scale of $8\,h^{-1}\textrm{Mpc}$ (thus $\sigma_8$ varies  
with the change of scales via $\alpha_\parallel,\alpha_\bot$) while $\sigma_\textrm{s8}$ defines the smoothing scale as to be independent of changes in scale because it takes a fixed smoothing scale at the template cosmology, $s\equiv r_{\rm d}/r_{\rm d}^{\rm fid}$:
\begin{equation}
    \sigma_\textrm{s8}\equiv \sigma(R=s\cdot8\,h^{-1}\textrm{Mpc}),
\end{equation}
with $s$ depending on the cosmology at each step of the fitting process. This change in definition does not alter the fitting part of the analysis, only the interpretation of results. In the case where the best-fit dilation parameters $\alpha_\parallel,\alpha_\bot$ are equal to 1 and the true sound horizon scale is equal to the fiducial sound horizon scale, $\sigma_\textrm{s8}$ coincides with $\sigma_8$. 

As already noted in \cite{brieden2021shapefit}, the shape parameters $m$ and $n$ are very degenerate and we are unable to constrain them independently, thus we only constrain their combination $m+n$. Analogously, in analyses only involving two-point statistics, the parameters $f$ and $\sigma_\textrm{s8}$ have a very strong degeneracy, which is why it is common to only consider their product, $f\sigma_\textrm{s8}$.
However, the inclusion of the bispectrum breaks the degeneracy\footnote{This is because $f$ and the bias parameters $b_1,b_2$, all degenerate with $\sigma_\textrm{s8}$ in the power spectrum, are related differently in the power spectrum than in the bispectrum, as is seen in Appendix \ref{app: theory}.} and $f$ and $\sigma_{s8}$ can be inferred separately \cite{Gualdi:2018pyw,novell2023geofpt}. 

The structure of this paper is as follows. 
In Section \ref{sec: Data} we review the main aspects of the DESI DR1 data and mock catalogues. In Section~\ref{sec: set-up} we describe our adopted modelling and baseline choice for the power spectrum and bispectrum analysis and compare it with the standard approach from the DESI collaboration. In Section~\ref{sec: Systematics} we quantify the impact of the main sources of systematic errors. In Section~\ref{sec: results} we present our results, which we also compare with the official DESI full-shape power spectrum analysis of \cite{KP5Desi}. We conclude in Section~\ref{sec: conclusions}.

\section{Overview of the DESI DR1 catalogues and mocks}
\label{sec: Data}
The DESI DR1 galaxy data \cite{desidr12024desi} encompasses spectroscopic observations made by the DESI telescope in the time period between May 14, 2021 and June 14, 2022. One of the main improvements of the DESI telescope with respect to previous experiments is its structure into ten `petals', that in total comprise 5000 fibres, each guided by a robotic positioner. According to observing conditions, each fibre gets assigned a target, whose light is redirected to one of the ten spectrographs. This improvement has resulted in that the first year of observations (of total effective volume 19.5 $\textrm{Gpc}^3$) has a statistical power which surpasses that of two decades of the Sloan telescope observations at Apache Point \cite{alam2021completed}.

There are four main observed objects: Bright Galaxy Survey (BGS) \cite{bgsdesi}, Luminous Red Galaxies (LRG, divided into three redshift bins LRG1, LRG2, LRG3) \cite{lrgdesi}, Emission Line Galaxies (ELG) \cite{elgdesi} and quasars (QSO) \cite{qsodesi}. In all cases, the observations are divided into the North Galactic Cap (NGC) and the South Galactic Cap (SGC). Since both regions have been tested to be compatible, as expected, we consider the joint NGC+SGC data in all cases in this paper, as is also done in the DESI main analyses \cite{2024arXiv240403000D,KP5Desi,2024arXiv240403002D,KP7}. We use the three LRG redshift bins, which have the most signal-to-noise \cite{LRG.TS.Zhou.2023}, and the QSOs\footnote{We do not include the BGS sample due to its lower signal.}. The ELG sample features large imaging systematics \cite{KP5Desi,ruiyanginprep}, which are not easily treated within our pipeline, so we do not use these samples in this work. The details of the different samples are shown in Table \ref{tab:tracers}. The mathematical definitions used for the quoted quantities $z_\textrm{eff},V_\textrm{eff}$ in Table \ref{tab:tracers} are
\begin{equation}
\label{eq: zeffVeff}
    z_\textrm{eff}=\frac{\sum_{i>j}w_iw_j(z_i+z_j)/2}{\sum_{i>j}w_iw_j} \,\text{\cite{HGMeboss}};\quad V_\textrm{eff}=\int \left[\frac{\Bar{n}(\textbf{r})P(k)}{1+\Bar{n}(\textbf{r})P(k)}\right]^2d^3r\, \text{\cite{tegmark1997measuring}},
\end{equation}
where, respectively $w_i,w_j$ are the galaxy weights, and $\Bar{n}$ is the density of the sample.

\begin{table}[h!]
\centering
\centering
        \small
    \resizebox{\columnwidth}{!}{
\begin{tabular}{|c|c|c|c|c|c|c|}
\hline
Tracer & Redshift range & $N_\text{tracer}$ & $z_\text{eff}$ & $P_0 \, [(\text{Mpc}h^{-1})^3]$ & $V_\text{eff} \, [\text{Gpc}^3]$ & Used in this work \\ \hline\hline
BGS  & $0.1 - 0.4$ & 300,017   & 0.295 & $\sim 9.2 \times 10^3$ & 1.7 & No \\ \hline
LRG1 & $0.4 - 0.6$ & 506,905   & 0.510 & $\sim 8.9 \times 10^3$ & 2.6 & Yes \\ \hline
LRG2 & $0.6 - 0.8$ & 771,875   & 0.706 & $\sim 8.9 \times 10^3$ & 4.0 & Yes \\ \hline
LRG3 & $0.8 - 1.1$ & 859,824   & 0.920 & $\sim 8.4 \times 10^3$ & 5.0 & Yes \\ \hline
ELG1 & $0.8 - 1.1$ & 1,016,340 & 0.955 & $\sim 2.6 \times 10^3$ & 2.0 & No \\ \hline
ELG2 & $1.1 - 1.6$ & 1,415,687 & 1.317 & $\sim 2.9 \times 10^3$ & 2.7 & No \\ \hline
QSO  & $0.8 - 2.1$ & 856,652   & 1.491 & $\sim 5.0 \times 10^3$ & 1.5 & Yes \\ \hline
\end{tabular}
}
\caption{Summary of the DESI DR1 tracers main properties, including redshift ranges, tracer counts, effective redshift $z_\textrm{eff}$, power spectrum monopole amplitude at the reference scale of $k=0.14\,h\textrm{Mpc}^{-1}$, $P_0$, and effective volume $V_\textrm{eff}$. The effective redshift and volume, $z_\textrm{eff},V_\textrm{eff}$ are defined in Equation \ref{eq: zeffVeff}. In this work, we do  not consider the BGS nor the ELG samples, focusing only on the three LRG redshift bins together with the QSO redshift bin.}
\label{tab:tracers}
\end{table}

\begin{figure}[ht!]
    \centering
    \includegraphics[width=\textwidth]{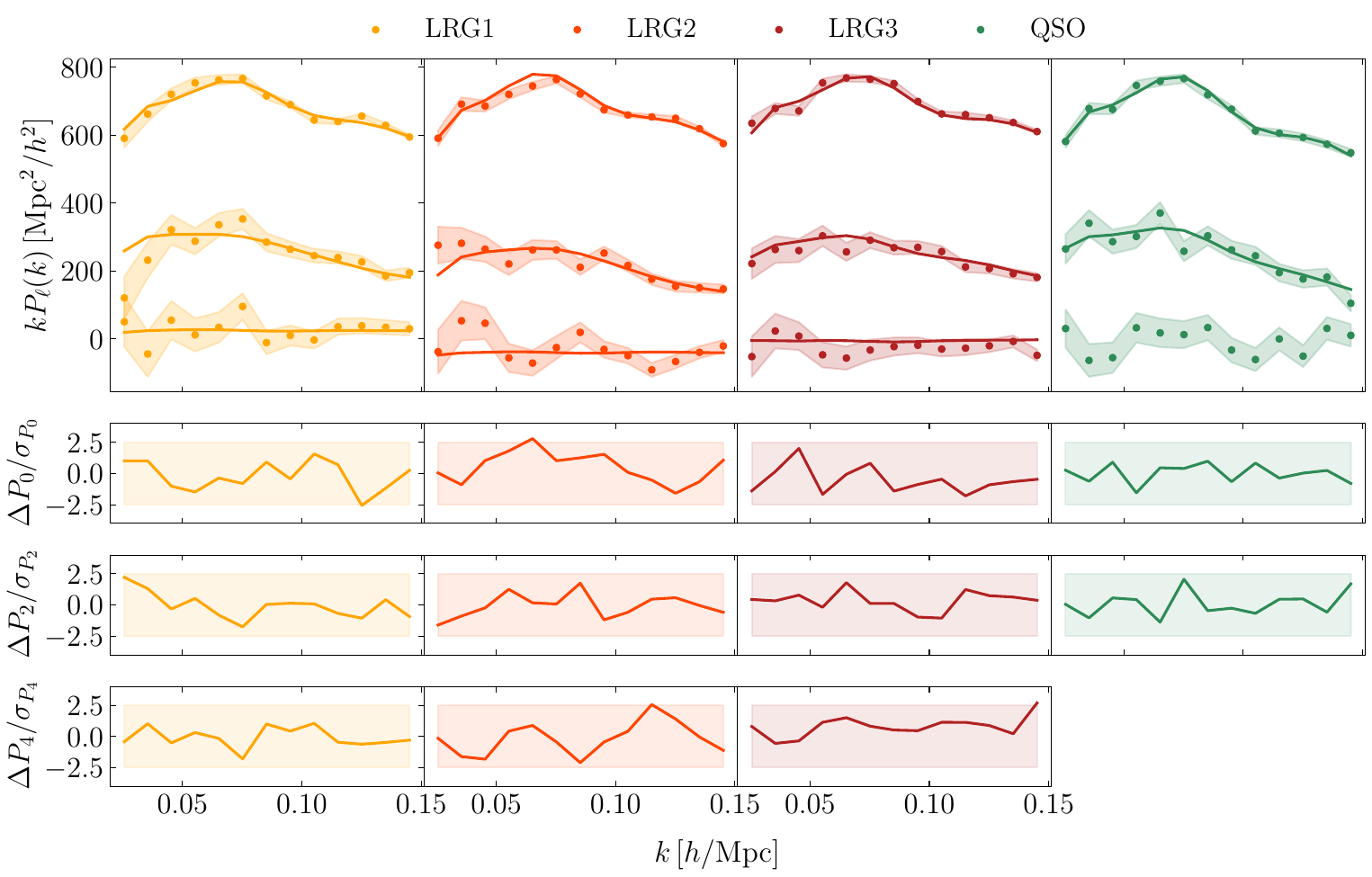}
    \includegraphics[width=\textwidth]{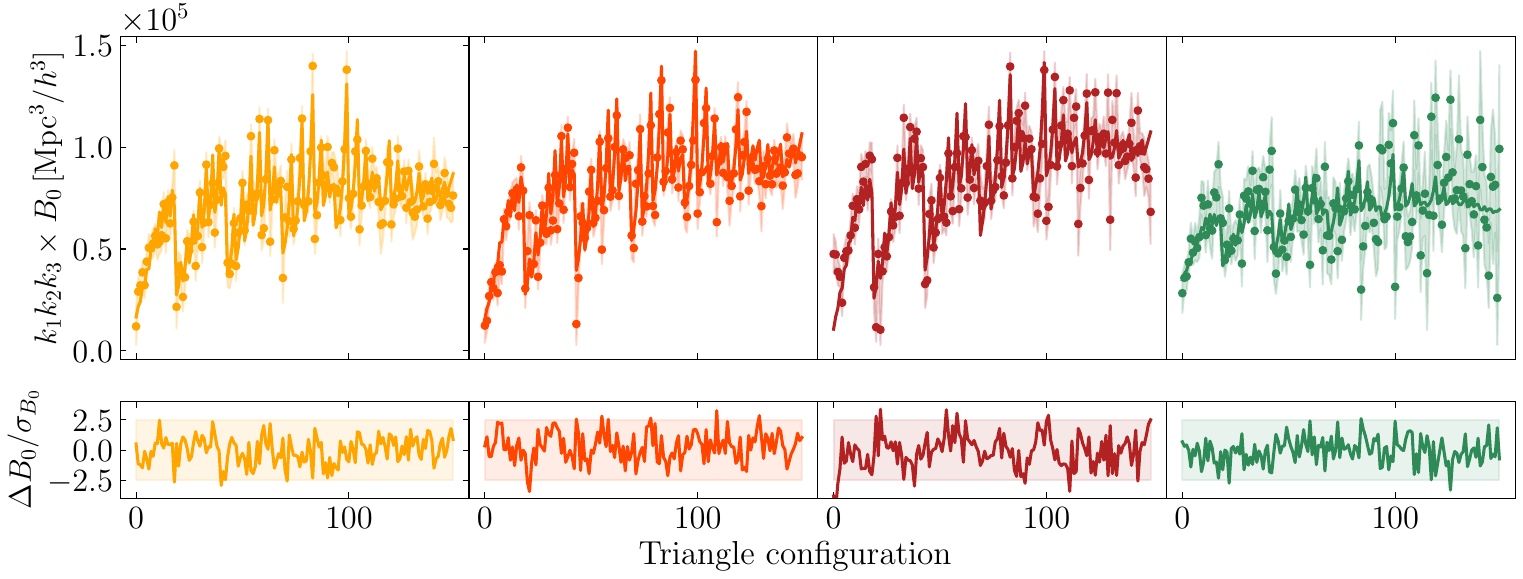}
    \caption{Power spectrum (top section) and bispectrum (bottom section) measurements and comparison with the respective model best fit. Each column corresponds to a tracer. 
    First row: In each panel, power spectrum monopole (top line), quadrupole (middle) and hexadecapole (bottom). The points and the shaded region are the measurements with their associated error, while the solid lines show the best fit of our joint $P+B$ analysis (Section \ref{sec: unblinded}). The  hexadecapole of the QSO is not used in this analysis as discussed in Appendix \ref{sec: blinded});
    Middle three rows: offset between the best fit of our $P+B$ analysis and the respective measurement, divided by the error bar of the measurement (shaded areas enclose the 2.5$\sigma$ region).  
    Bottom panels:  bispectrum monopole,  and using the same convention as above.}
    \label{fig: PBdata}
\end{figure}

In order to account for the effect of the survey geometry, a set of {\it random} catalogues are produced, consisting of a Poissonian distribution covering the same survey footprint as the data (both angular and radial), but with no structure. As it is standard practice, these random catalogues are combined with the data, following the FKP estimator approach \cite{FKP94}.  
For both data and mocks, the power spectra and bispectra measurements are obtained with the \textsc{Rustico} code\footnote{\url{https://github.com/hectorgil/Rustico}} \cite{HGMeboss}, taking into account both the observational systematics weights and the so-called FKP weights \cite{FKP94}.\footnote{The FKP weights ensure the minimum variance of the power spectrum at a certain chosen scale of reference, $0.14\,h{\rm Mpc}^{-1}$, by balancing contributions between regions with different number densities.} Figure \ref{fig: PBdata} shows the measurements of the power spectrum and bispectrum multipoles of our baseline analysis (for details and motivation see Appendix \ref{sec: blinded}), together with the ratio between the model best fit and the measurements.

Additionally, we adopt the blinding policy of the DESI key projects, according to which we only analyse the data once all analysis choices (e.g., modelling, range of scales, choice of tracers) are set and `frozen'. Until then, we only consider the mocks of the tracers of interest --presented in Section \ref{sec: Mocks} below-- and the blinded data. 
The blinding algorithm \cite{brieden2020blind} has two components, which shift the position of the BAO peak (and thus the recovered dilation parameters $\alpha_\parallel,\alpha_\bot$)  and the growth rate parameter $f$ respectively. Additionally, there is a third component of the blinding in DESI, affecting the signature of the local primordial non-Gaussianity parametrised by $f_\textrm{nl}$ \cite{chaussidon2024blinding}. In the context of our present analysis, where we kept fixed the $f_\textrm{nl}$ term to 0 in our modelling, the $f_\textrm{nl}$ blinding that is part of the blinded DR1 catalogues causes a shift in the ShapeFit $m+n$ parameter as these two parameters are very degenerate \cite{brieden2021letter}. 
Refs.~\cite{brieden2020blind,andrade2024validating} validated the blinding scheme, demonstrating how the recovered BAO and redshift-space distortions (RSD) parameters are shifted coherently with the expected blinding shift. This study was also extended for the power spectrum and bispectrum RSD analysis in \cite{novell2024blinding}, validating this catalogue blinding scheme also for bispectrum analyses.

\subsection{Mocks}
\label{sec: Mocks}
We have two main needs for using mock catalogues: 1) to assess and quantify the importance of the various systematic errors in the analysis, and 2) to estimate the covariance matrix of the data. Since there is a trade-off between accuracy and computational cost in producing simulations, two distinct sets of mocks have been generated within the DESI collaboration to address these two separate needs. 
   
\begin{figure}[ht!]
    \centering
    \includegraphics[width=\textwidth]{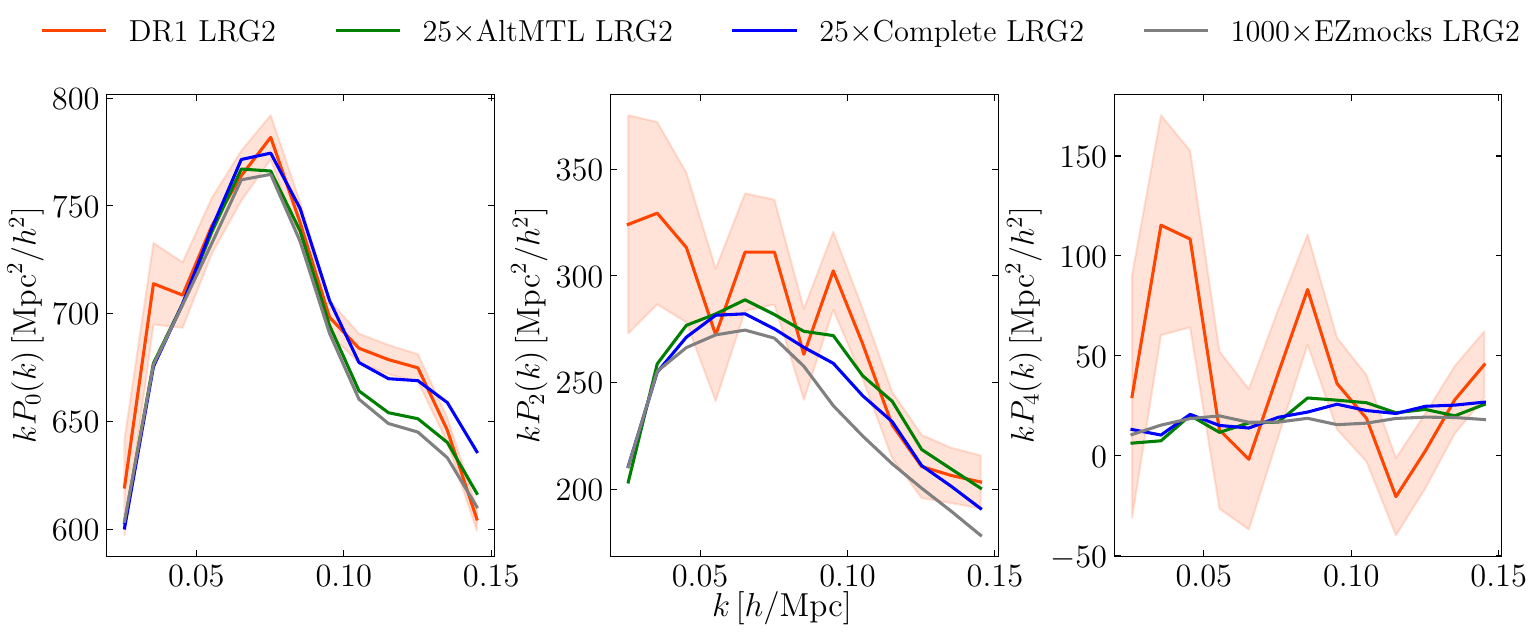}
    \includegraphics[width=\textwidth]{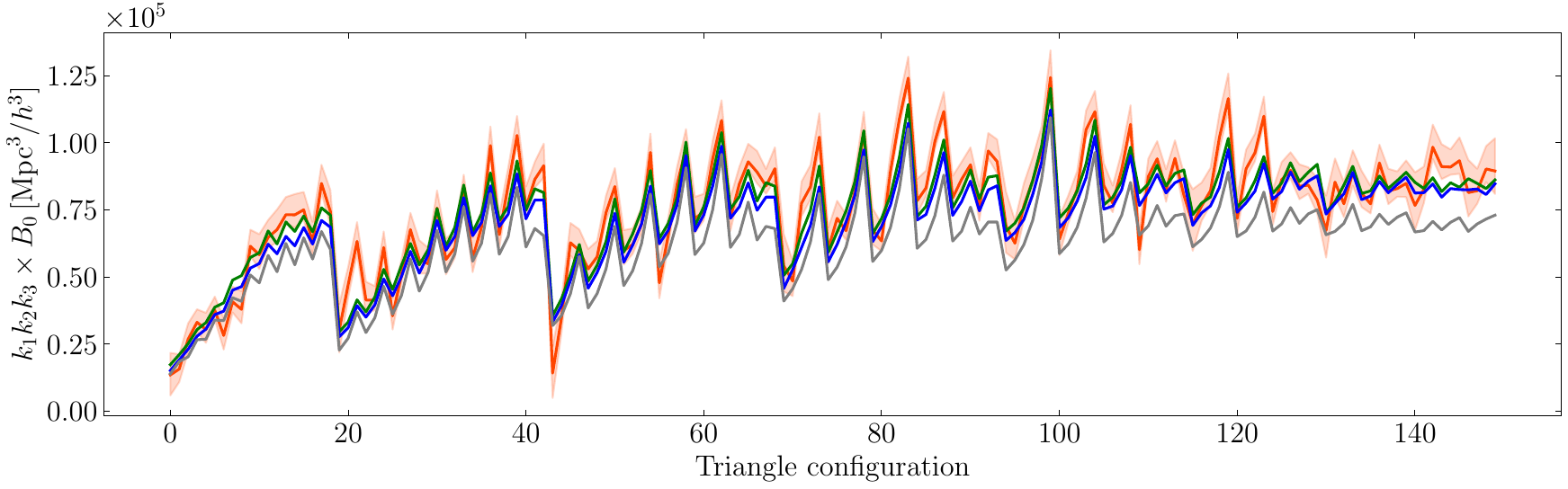}
    \caption{Power spectrum and bispectrum as measured from the LRG2 main mocks (\texttt{EZmocks} FFA, \texttt{Abacus} AltMTL and complete), together with the DESI DR1 measurement from the same sample. 
    We show the three power spectrum multipoles in the upper panel and the bispectrum monopole in the lower panel. For the mocks, we plot the mean of each data-vector across all available realisations (25 for \texttt{Abacus} and 1000 for \texttt{EZmocks}). Note that the normalisation of each of these displayed power spectra and bispectra is an arbitrary quantity that is only required (for unbiased measurements) to be consistent between each of these datasets and their corresponding window matrix. This means that across samples each normalization is in principle arbitrary and unrelated. For an insightful comparison, in this figure we (arbitrarily) choose to display {\it all} power spectra and bispectra normalization relative to the one of the AltMTL mocks.}
    \label{fig: PBmocksdata}
\end{figure}

\begin{itemize}
    \item[-] \texttt{AbacusSummit}: A suite of high-resolution N-body gravity simulations produced with the \textsc{Abacus} N-body code \cite{maksimova2021abacussummit}, each containing $6912^3$ in a physical volume of $V_\textrm{box}=8\,(\textrm{Gpc}/h)^3$.  Although there are a total of 97 cosmologies in the \texttt{AbacusSummit} suite, we only employ the 25 so-called `base' realisations in all tests. The halos are identified using the \textsc{CompasSO} halo finder, and they are populated with the appropriate tracers according to a halo occupation distribution (HOD) \cite{yuan2022abacushod,yuan2024desi}. These mocks, which we hereafter refer to as \texttt{Abacus}, represent the expected clustering of DESI tracers with high fidelity. Hence, we use them to perform tests in order to obtain the systematic error budget (see Section \ref{sec: Systematics}) and specify the analysis set-up choices  (see Section \ref{sec: set-up}).
    \item[-] \texttt{EZmocks}: These approximate mocks, which need to be calibrated to the \texttt{Abacus} mocks, directly generate the density field using the Zel'dovich approximation instead of performing the full N-body evolution. After that, the density field is populated by the corresponding tracers using an effective bias model. Since they can be obtained at a much reduced computational cost, a set of 1000 such realisations of volume $V_\textrm{box}=6^3\,(\textrm{Gpc}/h)^3$ is produced for each galaxy sample, which are used to estimate covariance matrices. As explained in detail in \cite{KP5Desi} and seen in Figure \ref{fig: PBmocksdata}, the \texttt{EZmocks} do not fully reproduce the statistics of the DESI DR1 data, which, if unaccounted for, results in an underestimate of the errors. 
    To correct for this, as prescribed by \cite{KP5Desi}, 
    we rescale the covariance matrix of each redshift sample by the factors in Table 3 of \cite{KP5Desi}. 
\end{itemize}

In both cases, there are two main versions of the mocks, usually labelled as respectively first and second generation mocks. We use the second generation mocks, whose HOD was calibrated with the full DESI Early Data Release (EDR) sample \cite{collaboration2024early}. 

The cosmology of both sets of mocks corresponds to the  Planck 2018 best fit, 
which we may denote as `base' or \texttt{c000} (following the nomenclature from \cite{maksimova2021abacussummit}) throughout this paper: the densities of cold dark matter and baryonic matter being respectively $\Omega_\textrm{cdm}h^2=0.12$ and $\Omega_bh^2=0.02237$; the amplitude of the dark matter fluctuations at $z=0$ and the slope of the primordial power spectrum take the values of $\sigma_\textrm{s8}=0.81135$ and $n_s=0.9649$, respectively; the dimensionless Hubble parameter is $h=0.6736$, and the dark energy equation of state corresponds to a cosmological constant, so $w_0=-1$ and $w_a=0$. This is also the fiducial cosmology in our analysis.

Then, the periodic boxes (both \texttt{Abacus} and \texttt{EZmocks}) are transformed into samples that account for the observed survey geometry, which we refer to as \textit{cutsky} samples. In this process the Cartesian coordinates are transformed first into sky coordinates (redshift, right ascension and declination) and then the tracers are filtered so as to obtain a density matching the DESI footprint and radial selection;\footnote{This leads to the radial integral constraint effect, described and studied in \cite{adame2024desi} for two-point statistics. As seen in \cite{Gualdi:2020aniso,novell2023geofpt}, most of the added information from the bispectrum statistic comes from linear to mildly non-linear scales, so small shifts on scales of $k\sim0.02\,h\textrm{Mpc}^{-1}$ produce a negligible effect in our analysis. This is also seen in Figure \ref{fig: QSO_posteriors_moderemoval}, where the mode correction term does not produce a significant change in the joint power spectrum-bispectrum cosmological parameters posteriors.} finally, the process of fibre assignment of the DESI instrument is simulated by running the \textsc{fibreassign} pipeline. This last step reproduces the effect of missing galaxies during the DESI observing strategy, which is caused by the limitation of placing the robotic fibre positioners on two objects which are angularly very close \cite{adame2024desi}. 

Depending on how the fibre assignment is simulated, three types of mocks are produced, which serve various purposes:
\begin{itemize}
    \item[-] Complete: Without fibre assignment incompleteness, so these can be considered  `uncontaminated' mocks. 
    \item[-] AltMTL \cite{altmtl}: The fibre assignment is implemented in the mocks using the same strategy as in the observed data. Although it is the most realistic approach, the computational cost associated only allows the AltMTL pipeline to be used with the sets of 25 \texttt{Abacus} mocks. 
    \item[-] Fast-fibreassign (FFA): This approach approximates the fibre assignment procedure, using averaged targeting probabilities from the data. Given that this procedure is significantly faster than the AltMTL pipeline, the FFA code is used to obtain the sets of 1000 \texttt{EZmocks} with fibre assignment effects.
\end{itemize}
For both AltMTL and FFA mocks, the clustering and the associated randoms are run through the corresponding fibre assignment pipeline. We plot in Figure \ref{fig: PBmocksdata} the comparison between the mean data-vectors of the \texttt{Abacus} AltMTL, Complete, and \texttt{EZmocks} FFA for the LRG2 bin, together with the corresponding DESI DR1 measurement.

As we will show in Section \ref{sec: Systematics}, we employ the complete mocks to assess the error associated with the modelling, while we can quantify the systematics originated by the fibre assignment process from the difference between the parameters recovered from the complete vs AltMTL mocks.

\section{Baseline Analysis}
\label{sec: set-up}

In this work we opt to perform a full-shape analysis of the DESI DR1 power spectrum and bispectrum data using the compression technique of ShapeFit. This approach allows us to compress all the information contained in the power spectrum and bispectrum, in a given redshift bin, into a set of parameters that do not depend on the specific choice of cosmological
 model ($\Lambda$CDM, $w$CDM, $k$CDM, etc.). In the companion paper \cite{novellmasot2025inprep} these compressed parameters are interpreted in the light of multiple models, bypassing the lengthy process of refitting again the power spectrum and bispectrum. This approach is complementary to the direct fit approach where the parameters of a given model are directly fitted to the summary statistics. We refer the reader to section 4.2 of \cite{KP5Desi} (and references therein) for a detailed description of these two methodologies. 

Our baseline choice for the analysis is the combination of the power spectrum multipoles and the bispectrum monopole of LRGs and QSO samples as follows.
For the LRG samples we consider the first three non-zero multipoles: monopole, quadrupole and hexadecapole in combination with the bispectrum monopole, $P_{024}+B_0$; whereas for the QSO sample we only consider the power spectrum monopole and quadrupole in combination with the bispectrum monopole, $P_{02}+B_0$. 
This choice is 
guided by the rationale of minimising the statistical errors while keeping the systematics (as quantified on mocks) under control, 
see Appendix \ref{sec: blinded} for more details.

\subsection{Standard compression and ShapeFit}
\label{sec: compression}
As described in \cite{brieden2021shapefit,Brieden_ptchallenge,brieden2022model}, the ShapeFit compression approach increases the information extracted from summary statistics with respect to the `standard compression' technique used previously in BOSS and eBOSS analyses \cite{gil2016clustering,HGMeboss,beutler2017clustering,satpathy2017clustering,grieb2017clustering,sanchez2017clustering,hou2021completed,bautista2021completed,neveux2020completed,de2021completed}.

The standard compression approach 
encapsulates the information content of the data into 
three main parameters: the dilation parameters along and across the line of sight, $\alpha_\parallel$ and $\alpha_\bot$, respectively; and $f\sigma_\textrm{s8}$, the product of the (smoothed) amplitude of dark matter fluctuations ($\sigma_\textrm{s8}$) and the logarithmic growth rate ($f$). 
The parameters $\alpha_\parallel, \alpha_\bot$ describe the distance dilation distortions caused by the use of a fixed fiducial cosmology when constructing the catalogue, and the isotropic shift of the BAO peak through the value of the sound horizon scale on the fixed-template fiducial cosmology:
\begin{equation}  \alpha_\parallel(z)=\frac{D_\textrm{H}(z)r_\textrm{d}^\textrm{fid}}{D_\textrm{H}^\textrm{fid}(z)r_\textrm{d}};\quad \alpha_\bot(z)=\frac{D_\textrm{M}(z)r_\textrm{d}^\textrm{fid}}{D_\textrm{M}^\textrm{fid}(z)r_\textrm{d}},
\end{equation}
where $D_\textrm{H}$ is the Hubble distance, $D_\textrm{H}(z)\equiv c/H(z)$, where $H(z)$ is the Hubble expansion history, and $D_\textrm{M}$ is the comoving angular diameter distance. 
Here $r_\textrm{d}$ denotes the sound horizon scale at radiation drag and the 
superscript `fid' stands for the fiducial cosmology chosen for creating the catalogue (through the redshift-to-distance transformation) and for the fixed-template cosmology. 
In Section \ref{sec: results} the main results 
will be presented in terms of 
an alternative re-parametrization of these dilation parameters,
\begin{equation}
    \alpha_\textrm{iso}(z)\equiv\left[\alpha_\parallel(z)\alpha_\bot(z)^2\right]^{1/3}=\frac{D_\textrm{V}(z)r_\textrm{d}^\textrm{fid}}{D_\textrm{V}^\textrm{fid}(z)r_\textrm{d}},\quad\quad \alpha_\textrm{AP}(z)\equiv\frac{\alpha_\parallel(z)}{\alpha_\bot(z)}=\frac{D_\textrm{H}(z)D_\textrm{M}^\textrm{fid}(z)}{D_\textrm{H}^\textrm{fid}(z)D_\textrm{M}(z)},
\end{equation}
where $D_\textrm{V}(z)\equiv \left[zD_\textrm{H}(z)D_\textrm{M}(z)^2\right]^{1/3}$ is sometimes referred as the isotropic BAO distance.
In this form, the effect of the absolute size of the BAO peak position in the fixed-template approach, $\propto r_d/r_d^{\rm fid}$, which only enters into the isotropic dilation scale, is separate from a purely anisotropic component, $\alpha_\textrm{AP}$, known as the Alcock-Paczyński parameter \cite{Alcock:1979mp}.\footnote{Note however, that even when considering the isotropic dilation parameter at different redshifts, the ratio between $\alpha_{\rm iso}$ at different redshifts is independent of $r_d/r_d^{\rm fid}$.}

The $f\sigma_\textrm{s8}$ parameter quantifies the growth of structure and RSD and it is the natural parameter combination constrained by the two-point statistics:
there is a strong degeneracy between  $f$ and $\sigma_\textrm{s8}$. 
Hence, when only the power spectrum multipoles are employed, we fit for $f$ with $\sigma_{s8}$ fixed at its fiducial value, $
\sigma_{s8}^{\rm fid}$, and later re-interpret the fit of $f$ as $f\sigma_{s8}$. 

Furthermore, adding a three-point statistic such as the bispectrum might allow the $f\sigma_\textrm{s8}$ degeneracy to be lifted. 
Therefore, in this work, cosmological constraints involving the bispectrum  consider $f$ and $\sigma_\textrm{s8}$ separately in all the LRG redshift bins.  The QSO sample, however, has a weak signal and thus we instead use the joint parameter $f\sigma_\textrm{s8}$ for the QSO tracers.

Additionally to the set of parameters $\{\alpha_\textrm{iso},\alpha_\textrm{AP},f,\sigma_\textrm{s8}\}$, the ShapeFit approach used in both this work and in the full-shape DESI power spectrum analysis \cite{KP5Desi,brieden2021shapefit}, includes a further parameter governing the {\it combined shape parameter}: $m+n$. The purpose of $m+n$ is to capture information on the matter-radiation equality present in the power spectrum (through $m$), but also to account for variations of the primordial spectral index (through $n$). It does so by modifying the linear power spectrum $P_\textrm{L}$ as,
\begin{equation}
\label{eq: shapefit_Plin}
    P_\textrm{L}(k)\longrightarrow P_\textrm{L}(k)\exp\left\{\frac{m}{a}\tanh\left[a\ln\left(\frac{k}{k_p}\right)\right]+n\ln\left(\frac{k}{k_p}\right) \right\},
\end{equation}
with $k_p$ and $a$ set as $k_p=\pi/r_\textrm{d}^\textrm{fid};\,a=0.6$, as justified in \cite{brieden2021shapefit}. Due to the strong degeneracy between $m$ and $n$ \cite{brieden2021shapefit,noriega2024comparing}, we opt to define the variables $m'\equiv m|_{n=0},n'\equiv n|_{m=0}$, which satisfies $m'\approx n'\approx m+n$, to which we refer as the combined shape parameter. 
For this reason we fix $m$, and only vary $n$. In the interpretation step, this $n$-at-a-fixed-$m$ parameter is equivalent to the combination $m+n$, to which we refer as the combined shape parameter. This re-interpretation of the fit can also be seen as the linear re-parametrization $\{m,\,n\}\rightarrow\{m+n,\,m-n\}$, where $m+n$ is well-constrained, whereas $m-n$ is a poorly constrained  parameter that we decide to fix to a fiducial choice $m^{\rm fid}-n^{\rm fid}\equiv 0$.\\ 
Indeed, we have checked that performing parameter inference by fitting either $n'$
or $m'$
produces indistinguishable results. 

The variation of $m+n$ in the fit modifies the $\sigma_\textrm{s8}$ parameter such that
\begin{equation}
    \sigma_\textrm{s8}=\sigma_\textrm{s8}^\textrm{fid}A^{1/2}\exp\left(\frac{m+n}{2a}\tanh\left[a\ln\left(\frac{r_\textrm{d}^\textrm{fid}}{8h^{-1}\textrm{Mpc}}\right)\right] \right),
\end{equation}
with $\sigma_\textrm{s8}^\textrm{fid}\equiv \sigma_8^\textrm{fid}$, and $A= A_\textrm{sp}/A_\textrm{sp}^\textrm{fid}$, where
\begin{equation}
    A_\textrm{sp}=\left(\frac{r_\textrm{d}^\textrm{fid}}{r_\textrm{d}}\right)^3P_\textrm{L, nw}\left(k_p\frac{r_\textrm{d}^\textrm{fid}}{r_\textrm{d}}\right),
\end{equation}
with $P_\textrm{L, nw}$ being the broadband (without BAO) linear dark matter power spectrum.

In addition to the cosmological parameters of interest, $
\{\alpha_\textrm{iso},\,\alpha_\textrm{AP},\,f,\,\sigma_\textrm{s8},\,m+n\}$, we consider the following nuisance parameters: the first and second-order bias parameters $b_1$ and $b_2$; the deviations from Poissonian shot-noise for the power spectrum and bispectrum, respectively $A_\textrm{P},A_\textrm{B}$ \cite{Gil-Marin:2014biasgravity,HGMeboss}; and the Fingers-of-God damping factors, also varied independently for the power spectrum ($\sigma_\textrm{P}$) and bispectrum ($\sigma_\textrm{B}$). The remaining non-linear bias parameters of the two-loop bias expansion are fixed according to the hypothesis of local Lagrangian bias: the tidal bias $b_{s^2}$ and third-order bias $b_{3\textrm{nl}}$ are set as functions of $b_1$, respectively $b_{s^2}=-\frac{4}{7}(b_1-1)$ and $b_{3\textrm{nl}}=32/315(b_1-1)$ \cite{baldauf2012evidence,saito2014understanding,Brieden_ptchallenge}. This choice of bias expansion, compatible with the official analysis of BOSS and eBOSS \cite{gil2016clustering,HGMeboss}, is different than the one used in the main DESI collaboration for full-shape power spectrum.  The main DESI collaboration analysis \cite{KP5Desi} includes counter-terms and stochastic parameters through the effective field theory formalism (EFTofLSS), which modify the impact of $b_{3\rm nl}$ and $b_{s^2}$ on the shape of the power spectrum.

We comment on the details of our modelling and analysis choices of the power spectrum and bispectrum in the following two subsections.

\subsection{Power spectrum modelling}
\label{sec: power spectrum}
We model the power spectrum, (both for matter, $P_{\rm NL}$, and for galaxies, $P_{g}$) at two-loop in renormalized perturbation theory (RPT)\footnote{The implementation of this code can be found in \href{https://github.com/hectorgil/PTcool}{https://github.com/hectorgil/PTcool}.}, following \cite{Crocce_2006,gil-marin_power_2015}, which closely resembles the approach taken by the eBOSS collaboration to model LRGs in Fourier space \cite{HGMeboss}. In Appendix \ref{app: theory} we provide a brief review, referencing the original sources of this formalism. This model, based on standard perturbation theory (SPT hereafter), differs
from the main modelling choices of the DESI collaboration \cite{KP5Desi,maus2024comparison,maus2024analysis,noriega2024comparing,lai2024comparison,ramirez2024full}, which are based on the EFTofLSS framework. The main effect of our choice is that the power spectrum can be used for scales not smaller than $k=0.15\,[h{\rm Mpc}^{-1}]$ in the redshifts corresponding to the LRG tracers, while the EFT approaches are used up to $k=0.20\,[h{\rm Mpc}^{-1}]$. However, it has been tested  (Figure 20 in Appendix B of \cite{KP5Desi} for the DR1 Data; and Figure 4 of \cite{maus2024analysis} for the \textsc{Abacus} N-body mocks) that the information gain of going beyond $k=0.16\,[h{\rm Mpc}^{-1}]$ is small. As we show later in Figure \ref{fig: folps}, the difference between using RPT or EFT models (FOLPS in that case) is very small even when the mock volume is of the order of $200\,[{\rm Gpc}h^{-1}]^3$. This result is in agreement with previous results such as those displayed in \cite{Brieden_ptchallenge}.

We consider in all cases two different sets of power spectrum multipoles: the monopole and quadrupole, $P_{02}=\{P_0,P_2\}$ (as in the official DESI analysis \cite{KP5Desi}), and the additional inclusion of the hexadecapole, $P_{024}=\{P_0,P_2,P_4\}$. In Appendix \ref{sec: blinded} we test the effect of adding the hexadecapole, both on the constraints and on the systematic error budget. Based on those findings, we decide (before unblinding the data) to include the power spectrum hexadecapole as our baseline model in the three LRG bins\footnote{The power spectrum hexadecapole has not been included in the official DESI power spectrum analysis due to its contribution to projection effects, and to the fact that within $\Lambda$CDM the information gain provided by $P_4$ is small. In the case of our power spectrum model, we proceed as in its previous application to BOSS and eBOSS \cite{gil2016clustering,HGMeboss} and include the hexadecapole for the LRG tracers, which does not worsen the projection effects nor the systematic error with respect to the analysis without $P_4$, as seen in Appendix \ref{sec: blinded}.}, but not in the QSO bin (see Appendix \ref{sec: blinded}).

  The effect of the survey window function is modelled as in \cite{wilson2017window,beutler2017clustering,HGMeboss}. This approach, which follows that of the BOSS and eBOSS collaborations, has some differences with the DESI official pipeline: the DESI collaboration implements a cut in the contribution of galaxy pairs separated by less than 0.05 deg (in a procedure labelled as $\theta$-cut), and a rotation of the data-vector, window matrix and covariance to better handle the resulting mode coupling \cite{KP5Desi}. The $\theta$-cut method \cite{pinon2025mitigation} mitigates fibre assignment systematic errors (which will be described in Section \ref{sec: FA}), and is increasingly necessary when small scales are included in the analysis, as in the official DESI full-shape analysis, which has $0.02<k\,[h{\rm Mpc}^{-1}]<0.20$. In our analysis we consider larger scales ($k<0.15\,[h{\rm Mpc}^{-1}]$ for the power spectrum and $k<0.12\,[h{\rm Mpc}^{-1}]$ for the bispectrum), so we opt to not use the $\theta$-cut correction. We quantify the residual effects of unmitigated fibre assignment systematics in Section \ref{sec: FA}.

\subsubsection{Comparison of the performance of power spectrum models: RPT vs EFT}
\begin{figure}
    \centering
    \includegraphics[width=0.55\textwidth]{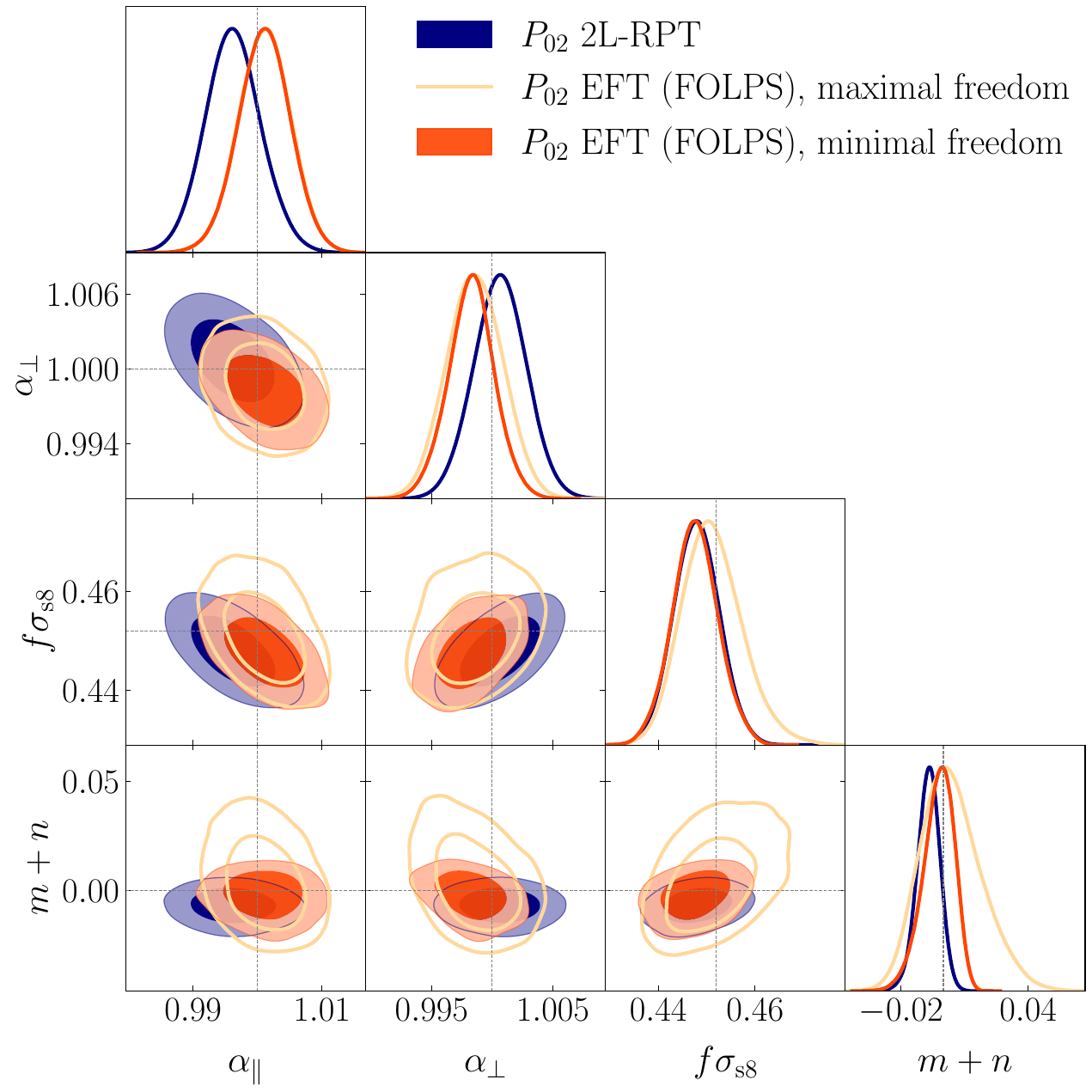}
    \includegraphics[width=0.44\textwidth]{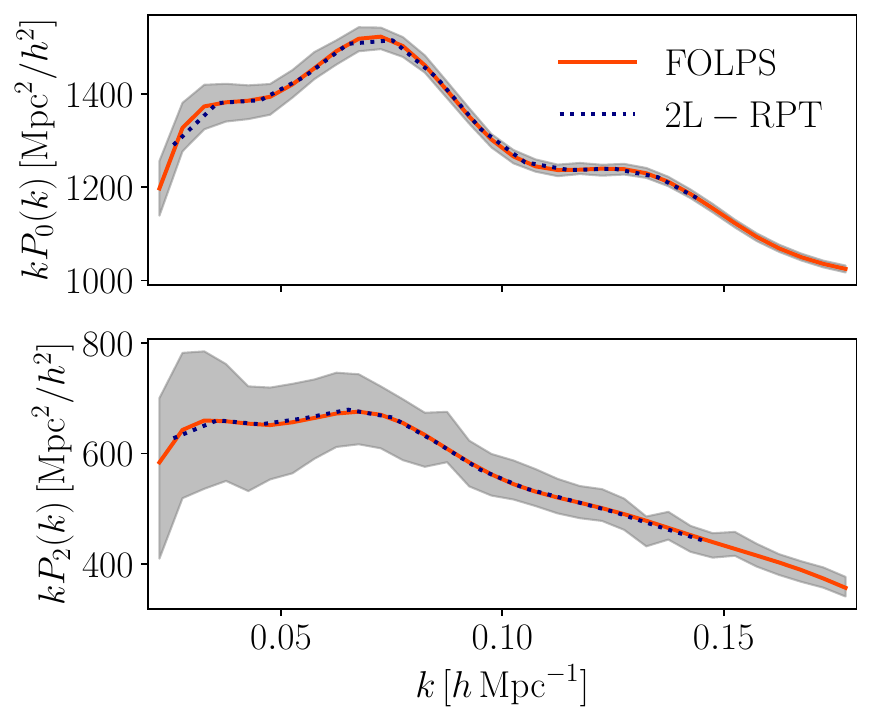}
    \caption{Left panel: posteriors of the ShapeFit parameters obtained by fitting the power spectrum monopole and quadrupole of the 25 \texttt{Abacus} mocks for the LRG sample at $z=0.8$, with a covariance corresponding to a volume of $200\,[{\rm Gpc}h^{-1}]^3$. The light orange and red contours display the performance of the EFTofLSS code \texttt{Folps} \cite{noriega2024comparing} in the range $0.02<k\,[h{\rm Mpc}^{-1}]<0.18$, for two types of galaxy bias setup: minimal freedom in filled contours and maximal freedom in empty contours, as labelled. The blue posteriors display the performance for the RPT model, which is used in this work for describing the power spectrum signal, and that has encoded a similar setup as the `minimal freedom' for \texttt{Folps}, but with $0.02<k\,[h{\rm Mpc}^{-1}]<0.15$. The horizontal and vertical black dotted lines mark the expected values for each parameter given the true cosmology. Right panel: The upper and bottom sub-panels display the power spectrum monopole and quadrupole, respectively, for the \texttt{Folps} minimal freedom and RPT best-fitting models to the measurements of the 25 \texttt{Abacus} cubic mocks (not displayed for clarity) The grey dashed region is the error associated with one \texttt{Abacus} mock, corresponding to a volume of $8\,[{\rm Gpc}h^{-1}]^3$.}
    \label{fig: folps}
\end{figure}

For completeness, we display here a direct comparison of the recovered parameters, from the power spectrum monopole and quadrupole only, for the two-loop RPT model used in this paper and \texttt{Folps}, which is an implementation of one of the EFTofLSS codes used in the DESI full-shape analysis \cite{KP5Desi,noriega2024comparing}. The left panel of Figure \ref{fig: folps} displays the posteriors of the ShapeFit parameters obtained by performing a power spectrum monopole and quadrupole fit to the mean of the 25 \texttt{Abacus} LRG cubic mocks at $z=0.8$\footnote{Only for this plot we employ a slightly different version of the \texttt{Abacus} mocks, with a different HOD parametrization than those mocks described in section~\ref{sec: Mocks}. The reason is that we aim to reproduce the results presented in the first batch of DESI full-shape comparison papers released in April 2024 \cite{maus2024analysis,maus2023comparison,lai2024comparison,noriega2024comparing,ramirez2024full} before the creation of the new set of \texttt{Abacus} mocks used in the rest of the paper.}, with the covariance rescaled to the total of the 25 mocks (volume of $200\,[{\rm Gpc}h^{-1}]^3$). In this comparison, we do not employ the hexadecapole signal to match the setup baseline choice of the main DESI full-shape analysis. 

The figure displays two setup variations of \texttt{Folps}, one with the so-called `maximal freedom' (empty contours in light orange), and `minimal freedom' (filled contours in red). As described in \cite{noriega2024comparing}, the maximal freedom leaves the bias parameters $b_{s^2}$ and $b_{3\rm nl}$ free, and the minimal freedom sets them to the Local Lagrangian predictions (or co-evolution relations, see Eq.~\ref{eq:lagrangian_local}). In both cases, the range of fitted scales is $0.02<k\,[h{\rm Mpc}^{-1}]<0.18$\footnote{This approach is slightly different from the one performed in the official DESI ShapeFit results, which is equivalent to setting the third-order bias $b_\textrm{3nl}$ to zero and leaving the tidal bias $b_{s^2}$ free.}. Our RPT model is displayed in blue, with the setup employed in this paper (except for not including the hexadecapole), with the local Lagrangian bias relations of Eq.~\ref{eq:lagrangian_local} and fitted scales of $0.02<k\,[h{\rm Mpc}^{-1}]<0.15$.  The horizontal and vertical dotted lines mark the expected values of the ShapeFit parameters. In the right panel of Figure \ref{fig: folps} we show the corresponding best-fit power spectrum multipoles (solid red and dotted blue lines) to the mean of the same 25 \texttt{Abacus} cubic mocks. The two models, which have been fitted with different binning and range of $k$, are almost indistinguishable, as expected. The figure highlights the excellent agreement between two-loop RPT and \texttt{Folps} with the minimal freedom setup, as well as their ability to recover unbiased values of the parameters. Both cases show minimal difference, well within the $1\sigma$ statistical error associated with a volume of $200\,[{\rm Gpc}h^{-1}]^3$.

We note that for the combination of parameters $m+n$ the \texttt{Folps} minimal freedom setup has a $\sim30\%$ larger errorbars than the measurements from the RPT. This difference is due to the extra nuisance parameters that \texttt{Folps} considers (2 counter-terms and 1 extra stochastic term) that allow more freedom in the shape of the power spectrum. Nevertheless, the agreement between these two models in the best-fit value of the combined shape parameter is remarkable. 
The \texttt{Folps} maximal freedom setup presents larger errorbars, especially for the $m+n$ parameter due to the strong degeneracy between the $b_{3\rm nl}$ and the shape parameters, as discussed already in \cite{noriega2024comparing}.

These results indicate that both EFTofLSS and RPT models are able to reproduce with the same level of accuracy the ShapeFit cosmological constraints when the setup of nuisance parameters is set to be the same, reinforcing the robustness of the findings to different modeling choices: both types of approaches can be indistinctly employed to retrieve the cosmological constraints from DESI data. For this paper, we choose to use the RPT model as our baseline choice for describing the power spectrum multipoles. 

\subsection{Bispectrum modelling}
Our bispectrum model has its foundations in the standard perturbation theory (SPT) formalism at tree-level, which has the form
\begin{equation}
B^{\rm SPT}(\textbf{k}_1,\textbf{k}_2)=D_\textrm{FoG}^B(\textbf{k}_1,\textbf{k}_2)\left[2Z_1^\textrm{SPT}(\textbf{k}_1)Z_1^\textrm{SPT}(\textbf{k}_2)Z_2^\textrm{SPT}(\textbf{k}_1,\textbf{k}_2)P_\textrm{L}(k_1)P_\textrm{L}(k_2) + \textrm{2perm.}\right],
\label{eq:bisp_spt}
\end{equation}
where `perm' refers to cyclic permutations of the $\textbf{k}_1,\textbf{k}_2,\textbf{k}_3$ vectors. All terms are detailed in Appendix \ref{app: theory}.

The GEO-FPT model was developed in \cite{novell2023geofpt} to 
increase the range of scales of validity of the tree-level bispectrum model, and thus extract additional non-linear information. 
In GEO-FPT, the linear matter power spectrum,  $P_\textrm{L}$,  of  Equation \ref{eq:bisp_spt} is promoted to the non-linear matter power spectrum, $P_\textrm{NL}$. Furthermore, the perturbation theory $Z_2^{\rm SPT}$ kernel is modified  as follows:
\begin{equation}\label{eq: Z2_geo}
    Z_2^{\textrm{GEO}} = Z_2^\textrm{SPT}\times\Big[f_1+f_2\frac{\cos(\theta_\textrm{med})}{\cos(\theta_\textrm{max})}+f_3\frac{\cos(\theta_\textrm{min})}{\cos(\theta_\textrm{max})}+f_4\frac{A}{A_\textrm{norm}}+f_5\frac{A^2}{A_\textrm{norm}^2}\Big]\,
\end{equation}
where the coefficients $f_1,...,f_5$ were calibrated on N-body simulations. This specific parametrisation was proposed after observing, in Figure 1 of \cite{novell2023geofpt}, that the SPT bispectrum residuals had a clear quadratic dependence on the area of the triangle, and a mild dependence on the flatness of the triangle (which can be obtained as combinations between cosines of the minimum, intermediate and maximum external angles of the triangle). Refs.~\cite{novell2023geofpt,novell2024blinding} show that this model, with the coefficients $f_1,...,f_5$ fixed to the values obtained in their initial fit, is robust against change in the underlying cosmology. Therefore, these coefficients are not free parameters in our analysis, which results in a model for the galaxy bispectrum of the form:
\begin{equation}  
\label{eq: geo_realspace}
B_\textrm{gal}^{\rm GEO}(\textbf{k}_1,\textbf{k}_2)=P_\textrm{NL}(\textbf{k}_1)P_\textrm{NL}(\textbf{k}_2)\mathcal{Z}(\textbf{k}_1,\textbf{k}_2)+\textrm{2perm.},
\end{equation}
where $\mathcal{Z}(\textbf{k}_i,\textbf{k}_j)\equiv D_\textrm{FoG}^\textrm{B}(\textbf{k}_i,\textbf{k}_j)Z_1^\textrm{SPT}(\textbf{k}_i)Z_1^\textrm{SPT}(\textbf{k}_j)Z_{2}^\textrm{GEO}(\bf{k}_i,\bf{k}_j)$ and $D_\textrm{FoG}^\textrm{B}$ is the Fingers-of-God damping factor for the bispectrum. More detail on these terms can be found in Appendix \ref{app: theory}.

The non-linear matter power spectrum $P_\textrm{NL}$ is rescaled  through the linear power spectrum scaling of 
Equation \ref{eq: shapefit_Plin}.
This accounts for the 
 dependence of the bispectrum on the combined shape parameter $m+n$. 

As for the other cosmological parameters $\alpha_\parallel,\alpha_\bot,f$, they modify the bispectrum model signal through Equations \ref{eq: z1z2}, \ref{eq: B_alpha2}, \ref{eq: B_alpha1} in Appendix \ref{app: theory} and Equation 2.11 in \cite{novell2023geofpt}, while the $\sigma_\textrm{s8}$ parameter enters as a rescaling of the different terms of the 2-loop power spectrum model, scaling as $\sigma_{\rm s8}^2\propto P_L$.

The adopted range of scales for the bispectrum in all redshift bins matches the range in which the bispectrum model has been validated: $0.02<k^B\,[h\textrm{Mpc}^{-1}]<0.12$. The binning is the same as for the power spectrum, $\Delta k=0.01\,h\textrm{Mpc}^{-1}$.

For modelling the convolution effect of the survey geometry, we make use of the approximation of \cite{Gil-Marin:2014biasgravity,Gil-Marin:2016sdss}, where the bispectrum kernels are not affected by the window.

In this approximation, only the power spectrum part of the bispectrum model Equation~\ref{eq: geo_realspace} (i.e. $P_\textrm{NL}$) is affected by the window while the other terms are left unchanged. We obtain the windowed non-linear power spectrum $P_\textrm{NL}^W$ as in \cite{wilson2017window,beutler2017clustering}, and compute the convolved bispectrum monopole $B^W_0$ as
\begin{equation}
\label{eq: win_B}
    B^W_0(k_1,k_2,k_3)=\int_{-1}^{1}d\mu_1\int_0^{2\pi}d\phi \mathcal{Z}(\textbf{k}_1,\textbf{k}_2)P_\textrm{NL}^W(\textbf{k}_1)P_\textrm{NL}^W(\textbf{k}_2)+\textrm{2perm}.
\end{equation}
In the above expression, $\mu,\theta,\phi$ are the standard choice of angles in the multipole expansion that we use \cite{Scoccimarro:1999ed}, which are detailed in Appendix \ref{app: theory}.

Equation \ref{eq: win_B} is only valid for the bispectrum monopole. A generalization to bispectrum multipoles with the effect of a window should not follow this form, and a more careful treatment is needed.\footnote{In our work we employ the bispectrum multipole expansion defined in \cite{Scoccimarro:1999ed}. In the case of the alternative base expansion presented in \cite{sugiyama2019complete}, valuable progress has been made in \cite{wang2024window,pardede2022bispectrum} towards modelling the bispectrum window convolution. However, since the multipole-expansion bases of \cite{Scoccimarro:1999ed} and \cite{sugiyama2019complete}  do not have a direct correspondence between them, the work presented in \cite{wang2024window} for the base of \cite{sugiyama2019complete} is not necessarily applicable in our framework.} Consequently, in this paper we only consider the bispectrum monopole. This has limitations, since 
\cite{novell2023geofpt} showed that, in this particular model, the inclusion of the bispectrum quadrupoles results in a gain in both precision and accuracy on most cosmological parameters. We leave the development of an appropriate window treatment for the bispectrum quadrupoles, and thus the analysis of DESI data with bispectrum monopole and quadrupoles, for future work \cite{novellmasot2025inprep}. 

\subsection{Likelihood and covariance estimation}
We sample the parameters posteriors via  Markov chain Monte Carlo (MCMC), using the \textsc{Brass}\footnote{\url{https://github.com/hectorgil/Brass}} \cite{HGMeboss,brieden2022model} code, which follows the Metropolis-Hastings algorithm \cite{metropolis1953equation,Hastings1970MonteCS}. We set the Gelman-Rubin convergence criterion to be $R-1<0.01$.\footnote{In all the MCMC runs, we first obtain a proposal parameter covariance through iterative runs where we sequentially increase the step size. The last run, with a step size of 1.9, has a burn-in of 10,000 steps for each walker (chain), and a total of 50,000 steps per walker. At that point, the code checks for convergence according to the Gelman-Rubin criterion, and in all cases it was found to be converged. With these settings, we found convergence to be obtained irrespective of the number of walkers, and in particular for the LRG redshift bins we used 4 walkers, while for the QSO we used 18 walkers.} As shown in Table \ref{tab: priors}, we specify broad, uniform priors in all parameters except for the deviations from Poissonian shot-noise, which we set to a normal prior centred at Poisson value, $\mathcal{N}(1,0.3^2)$. As motivated by \cite{novell2023approximations}, we estimate the full covariance matrix, including all off-diagonal terms. For each redshift bin and data-vector, the covariance is obtained from the 1000 independent realisations of the \texttt{EZmock} samples. Each power spectrum multipole data-vector has 13 elements, while the bispectrum monopole has 150 elements. Therefore, our largest data-vector, $P_{024}+B_0$, has a size of 189 elements, which corresponds to less than 20\% of the number of simulations used to estimate the covariance matrix, resulting in a Hartlap factor \cite{Hartlap:2006kj} of $\sim0.8$. 

As reported in \cite{adame2024desi,KP5Desi}, the Fourier-space covariance estimated from the 1000 \texttt{EZmock} has some potential inaccuracies. 
Because of this, Refs.~ \cite{KP5Desi,adame2024desi} suggest to rescale the overall amplitude of the mock-estimated covariance by the factors displayed in Table 3 of \cite{KP5Desi}. These factors were necessary to account for the mismatch between the \texttt{EZmock} and the DR1 covariances seen in configuration space. In this work, we assume the same rescaling factor that has been studied and validated for the power spectrum works as well for the bispectrum analysis. As we will discuss in Section \ref{sec:error_validation}, we find these rescaling values to be valid for our analysis as well, with a potential systematic error involved with a mis-estimation of the covariance matrix accounting for less than $\sim20\%$ of the statistical DR1 errors.

\begin{table}[h]
\centering
\renewcommand{\arraystretch}{1.4}
\begin{tabular}{|c|c|}
\hline
\textbf{Cosmological parameters} & \textbf{Prior range} \\
\hline
$\alpha_{\text{iso}}$ & $\mathcal{U}(0.7, 1.3)$ \\
$\alpha_{\text{AP}}$ & $\mathcal{U}(0.7, 1.3)$ \\
$f$ & $\mathcal{U}(0.0, 20)$ \\
$\sigma_\textrm{s8}$ & $\mathcal{U}(0.0, 10)$ \\
$m+n$ & $\mathcal{U}(-1.0, 1.0)$ \\
\hline\hline
\textbf{Nuisance parameters} & \textbf{Priors} \\
\hline
$b_1$ & $\mathcal{U}(0, 10)$ \\
$b_2$ & $\mathcal{U}(-20, 20)$ \\
$A_\textrm{P}$ & $\mathcal{N}(1, 0.3)$ \\
$A_\textrm{B}$ & $\mathcal{N}(1,0.3)$ \\
$\sigma_\textrm{P}$ & $\mathcal{U}(0, 10)$ \\
$\sigma_\textrm{B}$ & $\mathcal{U}(0, 10)$ \\

\hline
\end{tabular}
\caption{Priors for both the cosmological and nuisance parameters. We note as $\mathcal{U}(x_\textrm{min},x_\textrm{max})$ an uniform distribution between $x_\textrm{min}$ and $x_\textrm{max}$, and as $\mathcal{N}(\mu,\sigma^2)$ as a Gaussian distribution with mean $\mu$ and standard deviation $\sigma$. In all cases (except from the parameters $A_\textrm{P},A_\textrm{B}$, which have a Gaussian prior centred at 1 with $\sigma=0.3$),
the MCMC sampling is not affected by the boundaries of the priors on the remaining parameters thus being effectively improper priors.}
\label{tab: priors}
\end{table}

We consider the log-likelihood, $\log L$, to be Gaussian, and apply the Sellentin-Heavens correction \cite{Sellentin:2015waz} to account for the error in the covariance introduced by estimating it from $n=1000$ realisations, 
\begin{equation}
    \log L=-\frac{n}{2}\log\left(1+\frac{(\mathbf{D}_\textrm{meas.}-\mathbf{D}_\textrm{model})\textrm{Cov}_\textrm{mocks}^{-1}(\mathbf{D}_\textrm{meas.}-\mathbf{D}_\textrm{model})^T}{n-1} \right),
\end{equation}
where $\mathbf{D}$ is the data-vector in each case, obtained either from measurements (meas.) or from the modelling (model); 
$n$ is the number of sample realisations used for estimating the covariance matrix,
$\textrm{Cov}$.

\section{Systematic error budget}
\label{sec: Systematics}
In this section, we present the potential main sources of systematic error we have identified, and quantify their effect on parameter inference, by performing various tests and analyses of the \texttt{Abacus} mocks.

We categorise systematic errors as originating from the inaccuracies in the theoretical modelling (Section \ref{sec: theory_syst}), from the fibre assignment procedure (Section \ref{sec: FA}), from the impact of the HOD (Section \ref{sec: HOD}), and from the choice of fiducial cosmology (Section \ref{sec: fiducial}).\footnote{The systematic error associated with the covariance matrix estimation is addressed by rescaling the covariance matrices by the factors given by table 3 of \cite{KP5Desi}, which we check to be a valid approach in Section \ref{sec:error_validation}.}

As in \cite{KP5Desi}, we take into account a systematic error for a parameter once it is large enough to be correctly quantified on the mocks, which for DR1 translates into when the systematic shift is 20\% or more of the DR1 statistical error. 
Following \cite{KP5Desi} the systematics are combined quadratically into a final (statistical plus systematic) error bar.
\subsection{Theoretical systematics}
\label{sec: theory_syst}

\begin{figure}[ht!]
\centering 
\includegraphics[width = \textwidth]{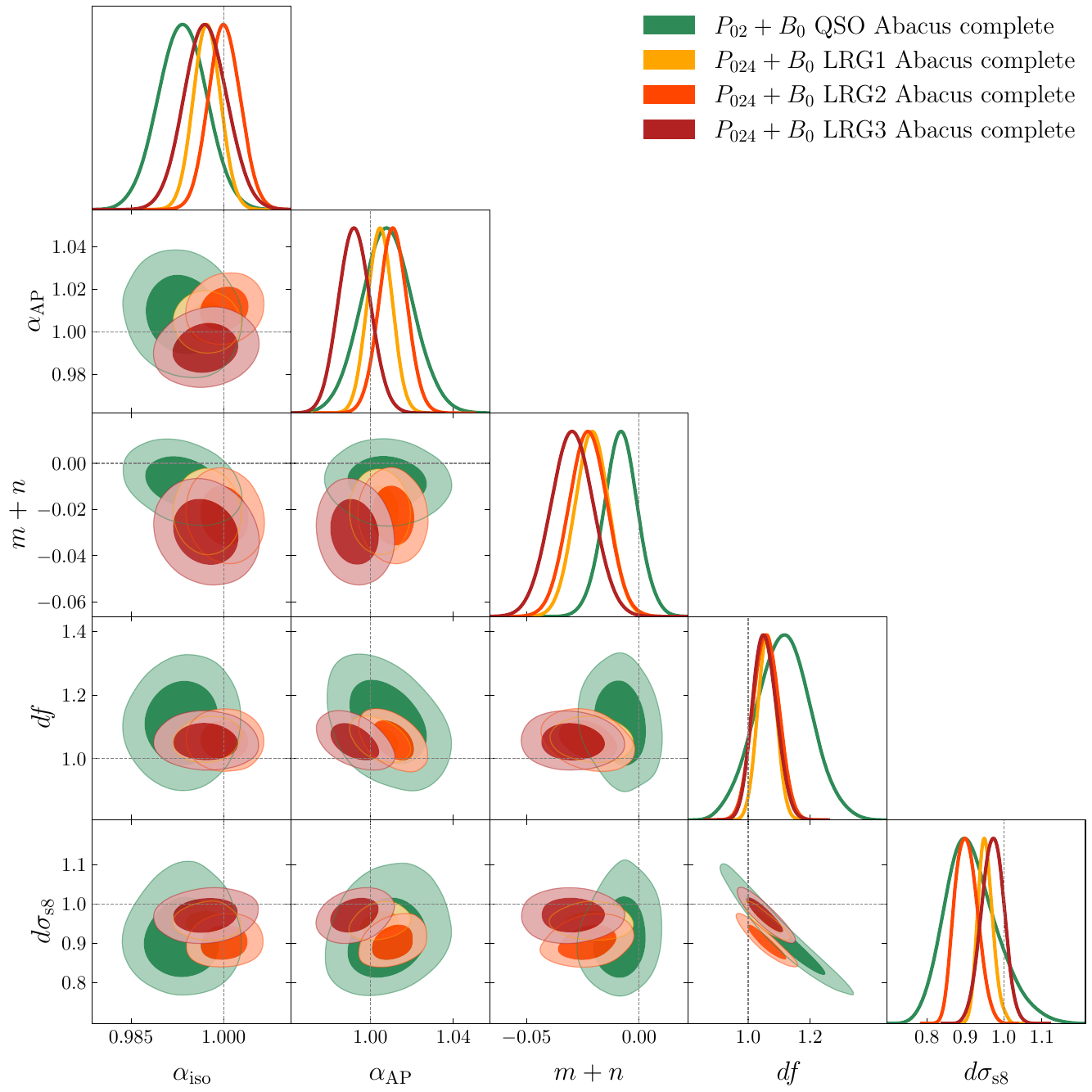}
\caption{Recovered cosmological parameters $\{df,d\sigma_\textrm{s8},\alpha_\textrm{iso},\alpha_\textrm{AP},m+n\}$ posteriors (68 and 95\% joint confidence levels C.L.) for the joint power spectrum-bispectrum data-vector $P_{024}+B_0=\{P_0,P_2,P_4,B_0\}$ from the mean of 25 \textsc{AbacusSummit} LRG complete mocks, and $P_{02}+B_0=\{P_0,P_2,B_0\}$ from the mean of 25 \textsc{AbacusSummit} QSO complete mocks, which correspond to our baseline choice of multipoles. The panels of the figure display the posteriors for the four redshift bins considered. The employed covariance has been normalised by 25 in order to match the volume of the sample. Thus, the displayed posteriors correspond to $\sim25$ times the DESI DR1 volume for each of the samples.
We quantify the theoretical systematic error as the discrepancy between the maximum a posteriori values (MAP) and the truth for each parameter and galaxy sample. }
\label{fig: P02B0_complete_samples}
\end{figure}

For each redshift bin and galaxy sample,
we consider the mean of the power spectrum and bispectrum measurements for  all the 25 corresponding `complete' \texttt{Abacus} mocks (see  Section \ref{sec: Mocks}),
corresponding to the signal of a volume of 25 times the volume of DESI DR1.
Consequently,
we rescale the covariance matrix (obtained with the corresponding set of \texttt{EZmocks}) by a factor of 25. For all the data-vectors of interest in this work, $\{P_{02},P_{024},P_{02}+B_0,P_{024}+B_0\}$, we run the pipeline presented in Section \ref{sec: set-up} on the mocks in the same way as we would do on the data.

Since they do not include any effects from observational sources of error, the `complete'  mocks can be seen as idealised simulations and we consider them as reference to assess the theoretical model systematics. 
These include both possible limitations or inaccuracies of the power spectrum and bispectrum models, as well as effects of the analysis choices such as scale cuts, and the approximation introduced by the window convolution procedure. For each sample and cosmological parameter, the effect of the theoretical model systematics is given by 
the absolute difference between the true and the recovered values.\footnote{We consider the \textit{maximum a posteriori} (MAP) from the MCMC output as the recovered parameter value instead of the central value of the posterior distribution. The MAP represents the set of cosmological and nuisance parameters that maximise the posterior distribution. However, in most cases in our analysis (especially when fitting the mean of 25 mocks), the MAP and central value are very close.}

In Figure \ref{fig: P02B0_complete_samples} we show the recovered parameter constraints from the complete \texttt{Abacus} mocks in the four redshift bins of study, compared with the true expected values (black dotted lines), for our baseline data-vector choices, $P_{024}+B_0$ for the LRG sample, and $P_{02}+B_0$ for the QSO sample. Instead of $f$ and $\sigma_\textrm{s8}$, to facilitate visualization, we display the posteriors in terms of $df\equiv f/f^\textrm{true},\,d\sigma_\textrm{s8}\equiv\sigma_\textrm{s8}/\sigma_\textrm{s8}^\textrm{true}$. 
We quantify the residuals between the MAP values of each posterior and the underlying true values for the four parameters of interest in the rows labelled as `Modelling' in Tables \ref{tab:P024B0syst}, \ref{tab:P02B0syst}, \ref{tab:P024syst}, \ref{tab:P02syst}, for all data-vector combinations (not all of them are displayed in Figure~\ref{fig: P02B0_complete_samples}).

All tracers are consistent with the true $\alpha_\textrm{iso},\alpha_\textrm{AP}$ at 2$\sigma$ (corresponding to a volume $\sim$25$\times$DR1). The LRG2 redshift bin features a systematic shift in $d\sigma_\textrm{s8}$, while all LRGs have an offset in the $m+n$ parameter with respect to its expected value. We have previously seen that systematic shifts in $df,d\sigma_\textrm{s8}$ are anticipated when not including the bispectrum quadrupoles in the analysis \cite{novell2023geofpt}. As for $m+n$, we see in the systematic tables, which we refer to above, that including $B$ causes a slightly larger systematic shift in that parameter with respect to the power spectrum-only case. Given that $m+n$ is sensitive to the freedom of the bias model, as observed in Figure \ref{fig: folps}, we leave for further work the investigation of the effect of assuming the local Lagrangian bias expansion (Section \ref{sec: compression}) in a joint power spectrum-bispectrum analysis. Another possible source for this could be the approximation that we use for the bispectrum window convolution (see Section \ref{sec: set-up}), which is part of our current work.

\subsection{Fibre assignment and imaging systematics}
\label{sec: FA}
The DESI survey has two observational effects that can impact the clustering analysis: the fibre incompleteness and the photometric angular density fluctuations.
Fibre incompleteness is caused by the inability of DESI to observe close pairs of objects within the patrol radius of a positioner. This is mitigated by performing several passes over the same area during the whole period of five years of observations. For this reason, this effect will be less important at the end of the five-year programme than for DR1.
The angular density fluctuations are already present in the imaging maps, and are mitigated by the use of imaging weights in the measurement of the power spectra and bispectra, denoted imaging systematics. A thorough description of these can be found in Refs.~\cite{ross2024construction,adame2024desi,ruiyanginprep}.

\begin{figure}[ht!]
\centering 
\includegraphics[width = \textwidth]{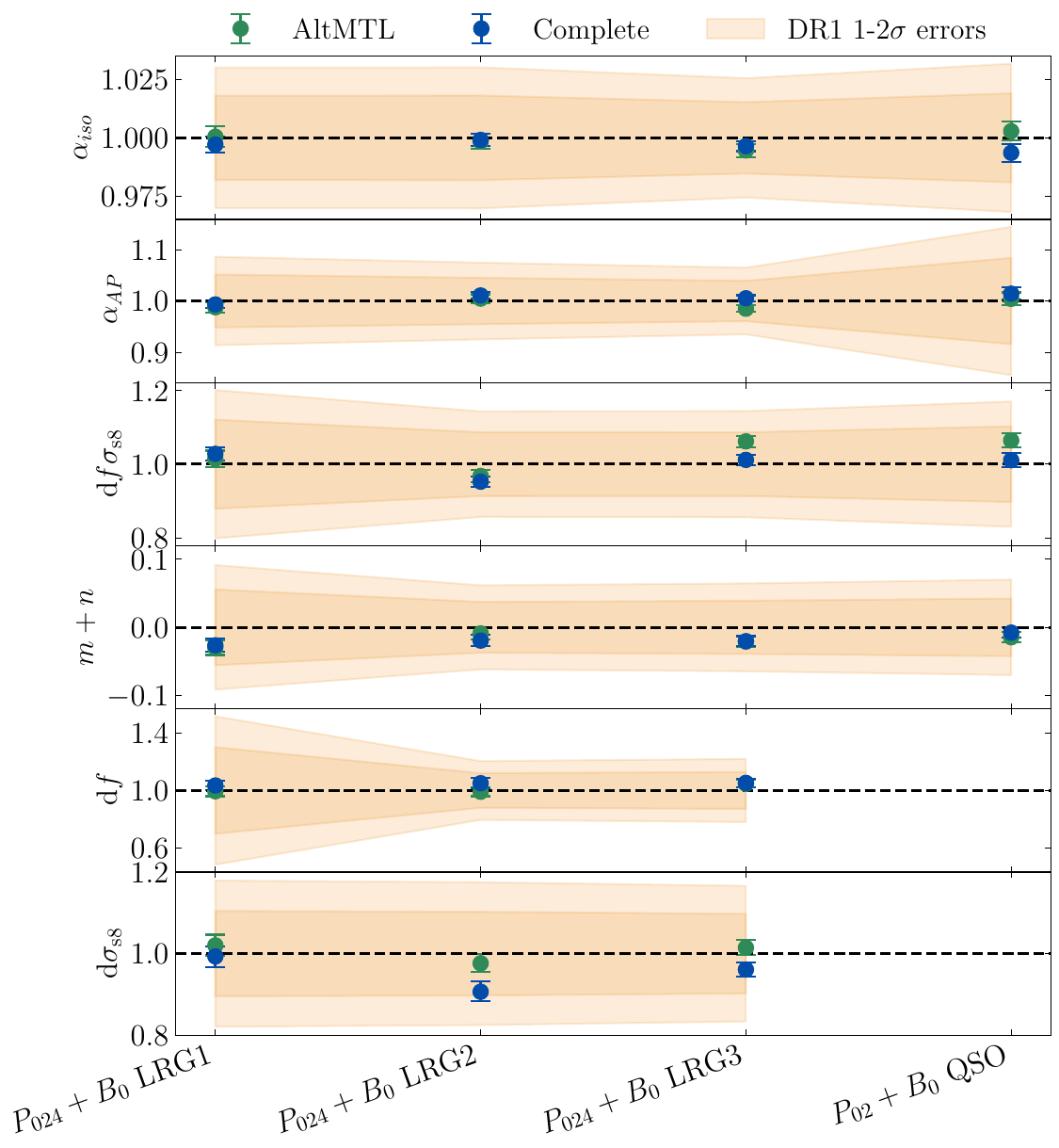}
\caption{Comparison of the MAP values for the cosmological parameters inferred from the complete and AltMTL mocks for all the tracers we consider in this work and for the choice of data-vectors of our baseline analysis. 
The dashed orange area is the $1-2\sigma$ region of the DESI DR1 analysis, that we have obtained from the blinded measurements in Appendix \ref{sec: blinded}, while the individual error bars correspond to the parameter inference with the mean of the 25 mocks (thus rescaling the associated errors by 25). For the QSO sample we only report the product $df\sigma_\textrm{s8}$ as for this sample this degeneracy is not lifted (see Section \ref{sec: set-up}).
}
\label{fig: P02B0_altmtl_complete}
\end{figure}

\begin{figure}[ht!]
\centering 
\includegraphics[width = \textwidth]{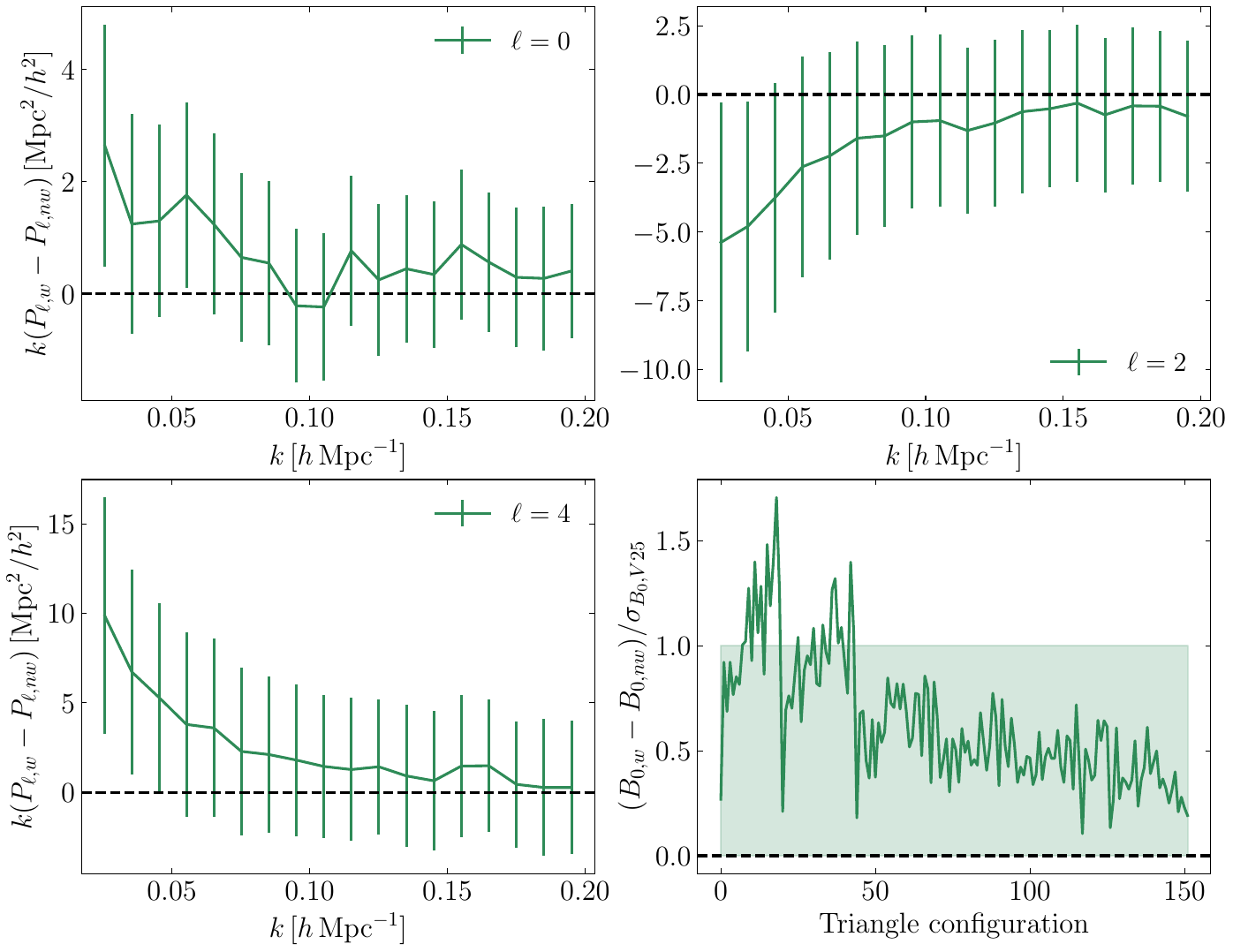}
\caption{Mode removal residuals for the QSO sample for all data-vectors considered in this work. In the top row and lower left panel we show respectively the shifts for the power spectrum monopole, quadrupole and hexadecapole, with errorbars correspondent to the volume spanned by 25 \texttt{Abacus} cubic mocks, $V=200\,(\textrm{Gpc}/h)^3$. In the lower right plot we show the mode removal correction for the bispectrum, normalized by the measurement error bars corresponding to 25 \texttt{Abacus} mocks. 
Note that the volume considered in these mocks is significantly larger than the actual DESI DR1 dataset, which makes the mode-removal effect to barely affect the ShapeFit parameters inferred from the data, as shown in Figure~\ref{fig: QSO_posteriors_moderemoval}.
}
\label{fig: QSO_RFdatavec}
\end{figure}

As for the imaging systematics, Refs.~\cite{KP5Desi,ruiyanginprep} found that in all LRG redshift bins the default linear imaging weights prescription is sufficient to capture the effect of observational conditions, whereas the QSO sample only suffers from a mild dependence on the systematic weights. The sample most affected by this effect is the ELG sample, which is the main reason not to include it in this work. 
For the QSO sample, we show in Figure \ref{fig: QSO_RFdatavec} the discrepancy (averaged over the 25 QSO AltMTL \texttt{Abacus} mocks) between the measurements with imaging weights included (e.g. $P_{\ell,{\rm weight}}$) and without them (e.g. $P_{\ell,{\rm noweight}}$), for all power spectrum multipoles and the bispectrum monopole. We observe the same trends as in Figure 9 of \cite{KP5Desi} (which displays the results for ELG tracers) and that these results are consistent with the QSO imaging shifts quantified in \cite{ruiyanginprep,KP5Desi}. 
In \cite{KP5Desi} the correction is applied by modifying the theoretical power spectrum model $P_\ell^{\rm theory}(k)$ to become an effective power spectrum $P_\ell^{\rm eff}(k)$:
\begin{align}
\label{eq: imaging_correction}
    P_\ell^{\rm eff}(k)=&P_\ell^{\rm theory}(k)+s_p\textrm{Poly}[(P^\textrm{data}_{\ell,\textrm{weight}}-P^\textrm{data}_{\ell,\textrm{noweight}})-(P^\textrm{mock}_{\ell,\textrm{weight}}-P^\textrm{mock}_{\ell,\textrm{noweight}})]+\nonumber\\
    &+\textrm{Poly}[P^\textrm{mock}_{\ell,\textrm{weight}}-P^\textrm{mock}_{\ell,\textrm{noweight}}],
\end{align}
and similarly for the bispectrum, where  `Poly' refers to a 3rd degree polynomial fit of the data points.\footnote{This polynomial fit is used instead of the data points to reduce the noise associated with the data, although its effect is minor. In the case of the bispectrum, due to its complexity, we do not use any polynomial fits to the data. }  The first correction term (varied with a nuisance parameter $s_p$) marginalises over the difference between imaging weights in mocks and in the DR1 data. The last term is a constant angular mode removal term.

Ref.~\cite{ruiyanginprep} shows that the QSO sample is only mildly affected by the mode removal term, and marginalizing over the first term has no significant effect on the recovered constraints.
The shifts presented in Figure \ref{fig: QSO_RFdatavec}, denoted as `mode removal' correction, correspond to the last term of Equation~\ref{eq: imaging_correction}. 
Therefore we do not include
the marginalisation over the non-linearities of the imaging weights (the term modulated by $s_p$ in Equation \ref{eq: imaging_correction}). We leave for future work an implementation of this marginalization, which will enable us to include the bispectrum clustering of the ELG sample.

\begin{figure}[ht!]
\centering 
\includegraphics[width = \textwidth]{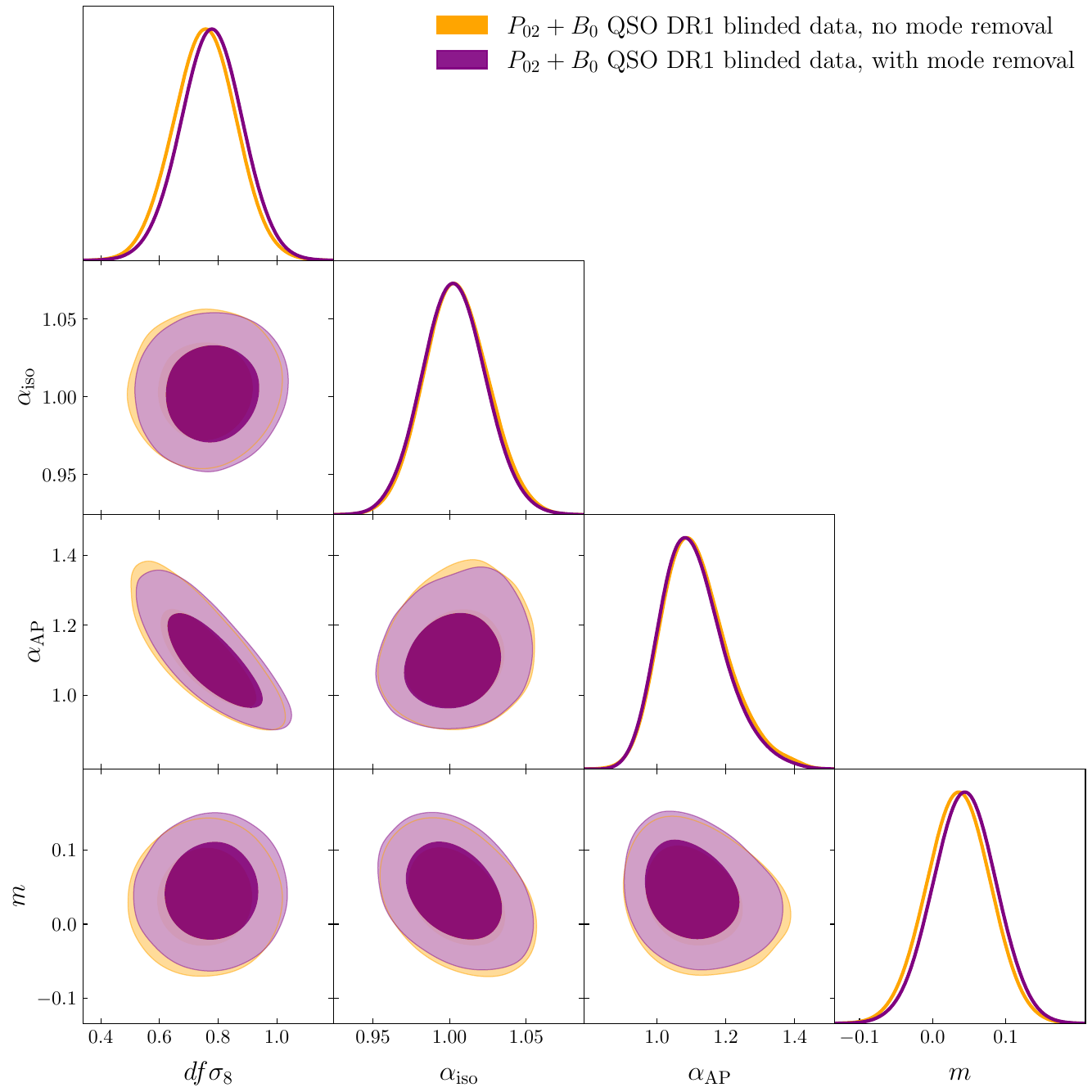}
\caption{Cosmological parameters posteriors recovered from the blinded DR1 QSO catalogue, using the combination of data-vectors $P_{02}+B_0$, with (in purple) and without (in orange) the mode removal correction defined in Equation \ref{eq: imaging_correction}. The effect of this correction is very mild, with the contours being effectively indistinguishable for DR1 statistical error bars.}
\label{fig: QSO_posteriors_moderemoval}
\end{figure}

In Figure \ref{fig: QSO_posteriors_moderemoval} we show the effect of the mode removal correction in the blinded DR1 QSO data, by comparing the recovered constraints with and without the correction computed with the last term of Equation \ref{eq: imaging_correction}. The effect of this correction is extremely small, both for the case shown in the Figure ($P_{02}+B_0$) as for the other data-vector combinations. In all cases, the shifts in the posteriors are less than  20\% of the corresponding DR1 errors, with part of the shift being absorbed by the $b_2$ and $\sigma_\textrm{B}$ parameters. In our analysis, we will model the QSO theoretical model with the mode removal correction applied, and leave the LRG tracers (which reported a smaller imaging systematic weight dependence than QSO in \cite{ruiyanginprep,KP5Desi}) uncorrected.

As for the fibre assignment systematic, we quantify it as the discrepancy between the recovered maximum a posteriori (MAP) values from the complete and AltMTL sets of 25 \texttt{Abacus} mocks. We present the shifts between the complete and AltMTL MAPs in the rows labelled as `Fibre assignment' in Tables \ref{tab:P024B0syst}, \ref{tab:P02B0syst}, \ref{tab:P024syst}, \ref{tab:P02syst}, for each of the four data-vector combinations considered in this work.

In Figure \ref{fig: P02B0_altmtl_complete}, for the baseline choice of data-vector combination as labelled in the $x$-axis, we display the MAP values for all cosmological parameters of interest and the four samples, together with their recovered error (corresponding to the volume of the 25 simulations), compared with the DESI DR1 1-2$\sigma$ error bars (obtained from the blinded data, which is shown in Appendix \ref{sec: blinded}). We include the parameter combination $f\sigma_\textrm{s8}$, for easier comparison to the QSO sample results (see Section \ref{sec: set-up}).

\subsection{HOD systematics}
\label{sec: HOD}
The systematic errors related to the HOD originate from the limitation of the bias model considered here (two-free parameters $b_1$ and $b_2$ with local-Lagrangian values for  $b_{s^2}$ and $b_{3\rm nl}$), to describe the matter-halo-galaxy connection. 

We start by considering a set of \texttt{Abacus} galaxy mocks with variations of the DESI HOD baseline, for both LRG and QSO samples. These alternative-HOD mocks were generated by the DESI collaboration by considering variations about a baseline HOD chosen for the so-called second-generation mocks, which were calibrated with the DESI Early Data Release \cite{collaboration2024early}. 
\begin{figure}[ht!]
\centering 
\includegraphics[width = \textwidth]{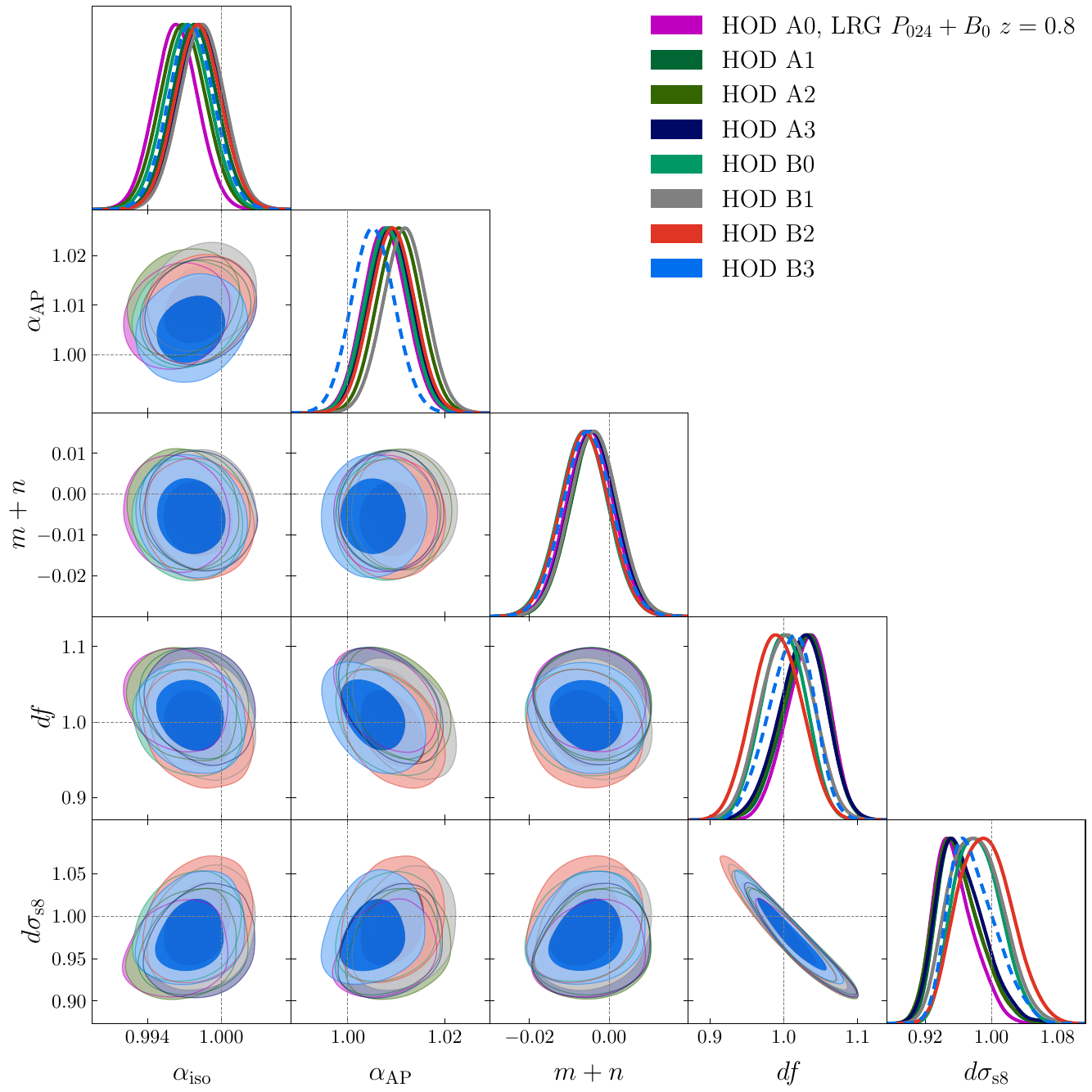}
\caption{Recovered posteriors (68 and 95 \% joint C.L.) of cosmological parameters $\{\alpha_\textrm{iso},\alpha_\textrm{AP},m+n,df,d\sigma_\textrm{s8}\}$ for the 8 alternative HOD variations (see \cite{findlay2024exploring} for a description), compared to the baseline HOD, noted as A0. The $P_{024}+B_0$ data-vector is obtained from each set of 25 \texttt{Abacus} HOD simulations, in cubic boxes describing LRG galaxies at $z=0.8$, so the covariance is normalised by 25. }
\label{fig: HODs}
\end{figure}

We perform our analysis on cubic box simulations spanning the baseline (standard) and the alternatives to the standard DESI HOD, 
which consist in variations in the modelling of velocity bias, environment-based bias and the masses of the halos hosting the central and satellite galaxies. We employ the same notation as in Ref.~\cite{KP5Desi,findlay2024exploring,mena2024hod} and refer the reader to Section 2 in \cite{findlay2024exploring} for their description. As in these works, we consider seven alternatives (A1, A2, A3, B0, B1, B2, B3) to the standard (A0) HOD  for LRG tracers, and 3 alternatives to the standard ($\textrm{QSO}_0$) HOD ($\textrm{QSO}_1$,$\textrm{QSO}_2$,$\textrm{QSO}_3$) for QSOs. We display the recovered posteriors of the LRG galaxies at $z=0.8$ in Figure \ref{fig: HODs}, where the different colours correspond to different variations of HOD.\footnote{The comparison of constraints for the QSO alternative HODs is qualitatively similar to the LRGs displayed in Figure \ref{fig: HODs}, hence we do not explicitly show the analogous plot for QSOs.}

We then estimate the contribution from the HOD systematic as the mean across HOD variations of the offset between their inferred MAP parameters
and the MAP obtained from the standard HOD simulations (labelled as A0 in Figure \ref{fig: HODs}). We see in Figure~\ref{fig: HODs} that the $\alpha_\textrm{iso},\alpha_\textrm{AP},\,m+n$ parameters do not suffer significantly from HOD-related systematics in the LRG tracers, while there is a small error budget contribution for the $f$ and $\sigma_\textrm{s8}$ parameters. The values for these systematic offsets, for both the LRG tracers and the QSO are listed in the row labelled as `HOD' in Tables \ref{tab:P024B0syst}, \ref{tab:P02B0syst}, \ref{tab:P024syst}, \ref{tab:P02syst}, for each of the four data-vector combinations considered.

\subsection{Systematics due to fiducial cosmology assumption}
\label{sec: fiducial}
The last source of systematic error that we explore is the effect of choice of fiducial cosmology. The fiducial cosmology that we use in our analysis, which is the one corresponding to the \texttt{Abacus} base simulations (\texttt{c000}, detailed in Section \ref{sec: Mocks}, first line of table \ref{tab:fidu_cosmo}) is assumed in two different steps:
\begin{enumerate}
    \item[1)] In the conversion from redshifts into distances in our galaxy and quasar catalogues.
    \item[2)] In the calculation of our power spectrum linear matter template.
\end{enumerate}

We keep fixed the power spectrum template, for which we pre-compute the perturbation theory integral terms through the \textsc{PTcool}\footnote{\url{https://github.com/hectorgil/PTcool}} code \cite{gil-marinetal:2012,Gil-Marin:2014biasgravity}, where the dependence on $\sigma_\textrm{s8}$ and $m+n$ is factorised out of the integrals. 
This approximation, which was validated in Appendix D of \cite{brieden2021shapefit}, saves significant time and resources by eliminating the need to re-compute the perturbation theory integrals in each step of the MCMC. 

Consequently, we need to quantify the effect of assuming a fiducial cosmology that could be different from the true cosmology. In this work, we only focus on the systematic error coming from the choice of fiducial cosmology in the template (i.e., in point 2). We neglect the impact of the fiducial cosmology in the catalogue creation, which has been estimated to be negligible in \cite{rafaelainprep,sanz2024bao}.

\begin{figure}[ht!]
\centering 
\includegraphics[width = \textwidth]{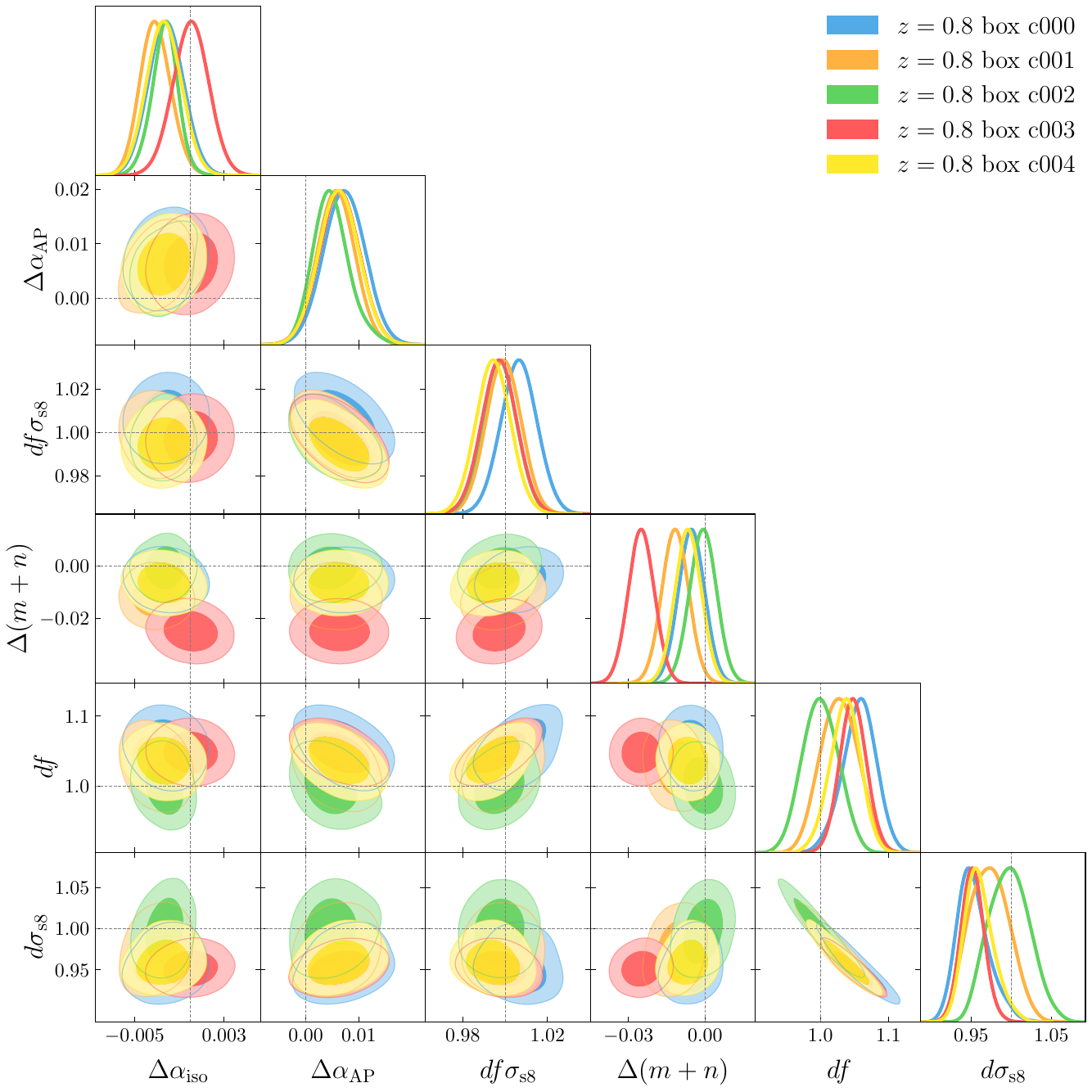}
\caption{Similar as Figure \ref{fig: HODs} 
but showing the differences between the parameter constraints (for $P_{024}+B_0$ of \texttt{Abacus} LRGs at $z$=0.8) while assuming a template based on each of the alternative fiducial cosmologies: \texttt{c001}, \texttt{c002}, \texttt{c003}, \texttt{c004}. Given that these fiducial cosmologies differ from the true one of the simulations, $\Delta$ indicates the discrepancy between the recovered and expected parameters. The posteriors have been obtained from the mean of the 25 LRG \texttt{Abacus} cubic boxes at $z=0.8$. }
\label{fig: fiducials1}
\end{figure}

Following the approach taken by the DESI collaboration in the main BAO and Full-Shape ShapeFit analyses \cite{perez2024fiducial,rafaelainprep}, we consider four alternative fiducial cosmologies, which are summarised in Table \ref{tab:fidu_cosmo}. They correspond to the first four alternative cosmologies of the \texttt{Abacus} suite of simulations, which are: \texttt{c001} (low-$\Omega_m$), \texttt{c002} (thawing dark energy), \texttt{c003} (high-$N_\textrm{eff}$), \texttt{c004} (low-$\sigma_8$).

\begin{table}[h!]
\centering
\begin{tabular}{|l|c|c|c|c|c|c|c|c|}
\hline
Name & $\omega_b$ & $\omega_\text{cdm}$ & $h$ & $10^9A_s$ & $n_s$ & $N_\text{ur}$ & $w_0$ & $w_a$ \\ \hline\hline
Planck $\Lambda$CDM -- \texttt{c000} & 0.02237 & 0.1200 & 0.6736 & 2.0830 & 0.9649 & 2.0328 & $-1$ & $0$ \\ \hline
Low-$\Omega_m$ -- \texttt{c001} & 0.02237 & 0.1134 & 0.7030 & 2.0376 & 0.9638 & 2.0328 & $-1$ & $0$ \\ \hline
Thawing DE -- \texttt{c002} & 0.02237 & 0.1200 & 0.6278 & 2.3140 & 0.9649 & 2.0328 & $-0.7$ & $-0.5$ \\ \hline
High-$N_\text{eff}$ --  \texttt{c003}& 0.02260 & 0.1291 & 0.7160 & 2.2438 & 0.9876 & 2.6868 & $-1$ & $0$ \\ \hline
Low-$\sigma_8$ -- \texttt{c004} & 0.02237 & 0.1200 & 0.6736 & 1.7949 & 0.9649 & 2.0328 & $-1$ & $0$ \\ \hline
\end{tabular}
\caption{Summary of the cosmologies considered to test the systematic error associated with the choice of fiducial cosmology (analogously as in \cite{perez2024fiducial,rafaelainprep}).  \texttt{c000} is the baseline fiducial cosmology.
}
\label{tab:fidu_cosmo}
\end{table}

The dilation parameters, $\alpha_\parallel$, 
$\alpha_\perp$, and the shape parameter $m+n$, are sensitive to mismatches between fiducial and true cosmology. This means that when the fiducial cosmology is different from the true one, we expect $\alpha_\parallel,\alpha_\bot$ to differ from 1, and $m+n$ to be non-zero, but knowing the two cosmologies we can compute their expected values.

For a given parameter $x$, we define the variable $\Delta x=x-x^\textrm{exp}$ as the shift of the obtained parameter with respect to its expected value given the true and fiducial cosmologies. For $f,\sigma_\textrm{s8}$ and $f\sigma_\textrm{s8}$ we report their obtained value, since these parameters are not expected to depend on the choice of template. Therefore, any deviation in the recovered value from the expected parameter value when assuming a different fiducial cosmology is a signal of a systematic error and must be quantified.

In Figure \ref{fig: fiducials1} we display the residuals between the recovered cosmological parameters and their expected values, when using the power spectrum templates for each of the alternative fiducial cosmologies, for LRG tracers in the average of the 25 respective cubic boxes at $z=0.8$. 
If there was no systematic error induced by the choice of fiducial cosmology, all the posterior distributions in Figure \ref{fig: fiducials1} would be coincident.
The effect of the fiducial cosmology systematics is quantified as the mean across samples of the shifts between these residuals. As seen in the row labelled as `Fiducial' in Tables \ref{tab:P024B0syst}, \ref{tab:P02B0syst}, \ref{tab:P024syst} and \ref{tab:P02syst}, the contribution of the choice of fiducial cosmology to the systematic error budget is generally minor: 
the main sources of systematic error come from the fibre assignment and modelling parts of the analysis.

\begin{table}[!ht]
\centering
    \small
    \resizebox{\columnwidth}{!}{
\begin{tabular}{|c|c|c|c|c|c|c|c|}
\hline
$[\%\sigma_\textrm{DR1}]$ & $P_{024}+B_0$ & $\sigma_{\alpha_\textrm{iso}}$ & $\sigma_{\alpha_\textrm{AP}}$& $\sigma_{f\sigma_\textrm{s8}}$  &   $\sigma_{m+n}$   & $\sigma_{f}$ & $\sigma_{\sigma_\textrm{s8}}$\\ \hline\hline
\multirow{3}{*}{Modelling} &  LRG1      &   $<20$ &     $<20$   &     $\sim20$   &  $\sim50$     &  $<20$   & $<10$    \\ 
                      &  LRG2     &    $<10$ &     $\sim20$          &     $\sim50$& $\sim50$     &      $\sim40$  & $\sim90$     \\ 
                      & LRG3      &    $\sim20$ &     $<20$          &     $<20$  & $\sim50$   &      $\sim40$  & $\sim40$  \\
                    &  QSO     &    $\sim20$ &     $<10$      &     $\sim40$ & $<20$   &   - & -      \\ \hline
\multirow{3}{*}{Observational}   & LRG1       &  $<20$  & $<20$  &     $<20$  & $<10$  &     $<20$   &      $\sim30$   \\
                      & LRG2     &   $<10$  &      $<20$         &      $<20$&  $\sim30$   & $\sim50$&      $\sim70$       \\ 
                      &  LRG3     &   $<20$  &     $\sim50$     &     $\sim60$  &  $<10$   & $<10$ &    $\sim50$   \\
                      &  QSO     &    $\sim40$ &  $\sim40$ &     $\sim90$ & $<20$         &   - & -   \\\hline
\multirow{3}{*}{HOD}   & LRG1       & $<10(16)$  & $<10(14)$     &   $<10(8)$   & $<10(8)$  &   $<10(30)$     &   $<20(39)$    \\
                      & LRG2     & $<10(16)$ & $<10(13)$         &   $<10(10)$ &  $<10(9)$    &      $<10(24)$    &   $<20(33)$     \\ 
                      &  LRG3     &  $<10(20)$ & $<10(16)$       &   $<10(10)$      &  $<10(8)$   &     $<10(19)$    &  $<20(42)$  \\
                      &  QSO     &  $<10(19)$ & $<10(22)$       &   $<20(39)$      &  $<20(39)$   &     -   & -  \\\hline
\multirow{3}{*}{Fiducial cosmology}   & LRG1  & $<10(15)$  & $<10(10)$     &   $<20(15)$   & $<20(70)$  &   $\sim30(62)$     &   $\sim30(69)$    \\
                      & LRG2     & $<10(15)$ & $<10(10)$         &   $<20(18)$ &  $<20(77)$    &      $\sim30(50)$    &   $\sim20(59)$     \\ 
                      &  LRG3     &  $<10(19)$ & $<10(12)$       &   $<20(19)$      &  $<20(75)$   &     $\sim20(39)$    &  $\sim30(75)$  \\
                      &  QSO     &  $<10(23)$ & $<10(18)$       &   $<10(19)$      &  $\sim30(71)$   &     -   & -  \\\hline\hline
\multirow{3}{*}{\textbf{Total}}&\textbf{LRG1}& $<20$&$<20$ &$\sim20$&$\sim50$   &$\sim30$&$\sim42$  \\
& \textbf{LRG2} &$<20$ & $\sim20$ & $\sim50$&  $\sim58$  & $\sim71$    &   $\sim116$ \\ 
&  \textbf{LRG3}&$\sim20$   & $\sim50$&   $\sim60$&  $\sim50$    &   $\sim45$   &  $\sim71$ \\
&  \textbf{QSO}&$\sim45$   & $\sim40$&   $\sim103$  &  $\sim30$   &  -  &  -    \\\hline

\end{tabular}
}
\caption{Contributions to the systematic error budget for each parameter and redshift bin, for the combination of summary statistics of $P_{024}+B_0$, i.e. the power spectrum monopole, quadrupole and hexadecapole, plus the bispectrum monopole. All errors are expressed here as percentage of the DR1 $1\sigma$ statistical error. The threshold to propagate a specific contribution to the systematic error budget is 20\% of the corresponding DR1 $1\sigma$ statistical error (see text for more details). For the HOD and fiducial cosmology systematics, we also show in parenthesis the maximum offsets obtained among all HOD simulations and among all fiducial cosmology choices, respectively.
The last row shows the quadratic sum of all the contributing systematic sources, which is to be added in quadrature to the statistical error bars obtained from the corresponding cosmological parameter posteriors.
\label{tab:P024B0syst}}
\end{table}

\begin{table}[!ht]
\centering
\centering
        \small
    \resizebox{\columnwidth}{!}{
\begin{tabular}{|c|c|c|c|c|c|c|c|}
\hline
$[\%\sigma_\textrm{DR1}]$ & $P_{02}+B_0$ & $\sigma_{\alpha_\textrm{iso}}$ & $\sigma_{\alpha_\textrm{AP}}$& $\sigma_{f\sigma_\textrm{s8}}$ &   $\sigma_{m+n}$ & $\sigma_{f}$ & $\sigma_{\sigma_\textrm{s8}}$   \\ \hline\hline
\multirow{3}{*}{Modelling} &  LRG1      &   $<20$ &     $<20$   &     $\sim20$  &  $\sim50$    &  $<20$   & $<20$      \\ 
                      &  LRG2     &    $<10$ &     $<10$          &     $\sim30$  & $\sim50$  &      $\sim70$  & $\sim70$      \\ 
                      & LRG3      &    $\sim20$ &     $<20$          &     $\sim30$ & $\sim50$     &      $\sim90$  & $\sim70$ \\
                    &  QSO     &    $\sim30$ &     $<20$      &     $<10$ & $<20$     &   - & -    \\ \hline
\multirow{3}{*}{Observational}   & LRG1       &  $<20$  & $<20$  &     $<10$  & $<10$  &     $<20$   &      $\sim40$   \\
                      & LRG2     &   $<20$  &      $<10$         &      $<10$ &  $\sim40$    & $\sim60$&      $\sim50$     \\ 
                      &  LRG3     &   $<20$  &     $<10$     &     $\sim30$ &  $<10$    & $\sim60$ &    $\sim90$   \\
                      &  QSO     &    $\sim40$ &  $<20$ &     $\sim50$  & $<20$        &   - & -   \\\hline
\multirow{3}{*}{HOD}   & LRG1       & $<10(16)$  & $<10(12)$     &   $<10(7)$   & $<10(11)$  &   $<10(27)$     &   $<10(36)$    \\
                      & LRG2     & $<10(15)$ & $<10(8)$         &   $<10(7)$ &  $<10(11)$    &      $<20(21)$    &   $<10(25)$     \\ 
                      &  LRG3     &  $<10(19)$ & $<10(11)$       &   $<10(8)$     &  $<10(11)$  &     $<10(21)$    &  $<20(36)$    \\
                      &  QSO     &  $<10(12)$ & $<10(7)$       &   $<20(14)$      &  $<20(26)$  &     -   & -   \\\hline
\multirow{3}{*}{Fiducial cosmology}   & LRG1  & $<10(19)$  & $<10(18)$     &   $<20(26)$   & $<20(64)$  &   $\sim30(70)$     &   $\sim30(70)$    \\
                      & LRG2     & $<10(19)$ & $<10(17$         &   $<20(30)$ &  $<20(70)$    &      $\sim30(57)$    &   $\sim20(59)$     \\ 
                      &  LRG3     &  $<10(23)$ & $<10(21)$       &   $<20(33)$      &  $<20(68)$   &     $\sim30(44)$    &  $\sim30(75)$  \\
                      &  QSO     &  $<20(36)$ & $<10(14)$       &   $<10(18)$      &  $\sim20(66)$   &     -   & -  \\\hline\hline
\multirow{3}{*}{\textbf{Total}}&\textbf{LRG1}& $<20$&$<20$ &$\sim20$&$\sim50$   &$\sim30$&$\sim50$  \\
& \textbf{LRG2} &$<20$ & $<20$ & $\sim30$  &  $\sim64$ & $\sim97$    &   $\sim88$\\ 
&  \textbf{LRG3}&$\sim20$   & $<20$&   $\sim42$ &  $\sim50$  &   $\sim112$   &  $\sim118$  \\
&  \textbf{QSO}&$\sim50$   & $<20$&   $\sim50$   &  $\sim20$   &  -  &  -   \\\hline

\end{tabular}
}
\caption{Same as Table \ref{tab:P024B0syst}, 
but for the
power spectrum monopole and quadrupole, plus the bispectrum monopole ($P_{02}+B_0$).}
\label{tab:P02B0syst}
\end{table}
\section{Results} 
\label{sec: results}

While this analysis is not required by the DESI collaboration rules to adhere to 
the DESI blinding procedure (the data has been unblinded for several months at the time of writing this paper), we opt to work with blinded data (as explained in Section \ref{sec: set-up}) until we freeze the analysis choices and obtain our systematic error budget, similarly as is done in the DESI key papers \cite{2024arXiv240403000D,KP5Desi,KP7}.
We perform all of the tests that involve data on the official blinded data of DR1 (see Section \ref{sec: Data} and references therein).
We report the results for the blinded data in Appendix \ref{sec: blinded}, and for the true, unblinded data in Section \ref{sec: unblinded}.

Considering the associated errors (as quantified on the mocks, Section~\ref{sec: Systematics})  and projection effects in blinded data (presented in Appendix \ref{sec: blinded}) we choose our bispectrum baseline analysis to consist of the data-vector $P_{024}+B_0$ for the LRG1, LRG2, LRG3 redshift bins, and $P_{02}+B_0$ for the QSOs. 
In fact, Tables \ref{tab:P024B0syst}, \ref{tab:P02B0syst}, \ref{tab:P024syst}, \ref{tab:P02syst}  and Appendix \ref{sec: blinded} show that the inclusion of the power spectrum hexadecapole for the LRGs reduces the statistical errors without introducing additional systematics, but the inclusion of the power spectrum hexadecapole for the QSO increases notably the systematic error budget without reducing the statistical errors in any appreciable way. Hereafter this baseline combination is referred to as $P+B$.

The multipole combination for the power spectrum represents another difference with the official DESI analysis, \cite{KP5Desi}, where the power spectrum hexadecapole is not included.
For comparison, we also present power spectrum only results, where the data-vectors involved are $P_{024}$ for the LRG1, LRG2, LRG3 redshift bins, and $P_{02}$ for the QSOs.

\subsection{Unblinded data}
\label{sec: unblinded}
After having frozen the pipeline, analyses choices and combinations of data-vectors, we perform the analysis on the measurements of the unblinded power spectra and bispectra 
baseline combination $P+B$.

Figure \ref{fig: unblinded_results} presents the results for the baseline $P+B$ combination on the DESI DR1 data (see Appendix \ref{sec: blinded} for detailed justification of the choice). To highlight the role of the bispectrum,  the figure also displays the power spectrum-only results ($P_{024}$ for the LRG bins, and $P_{02}$ for the QSO, labelled as $P$), in addition to the official DESI results presented in Ref.~\cite{KP5Desi}, in terms of ShapeFit parametrization labelled as `DESI SF'.
All constraints are highly compatible.

\begin{figure}[t!]
\centering 
\includegraphics[width = \textwidth]{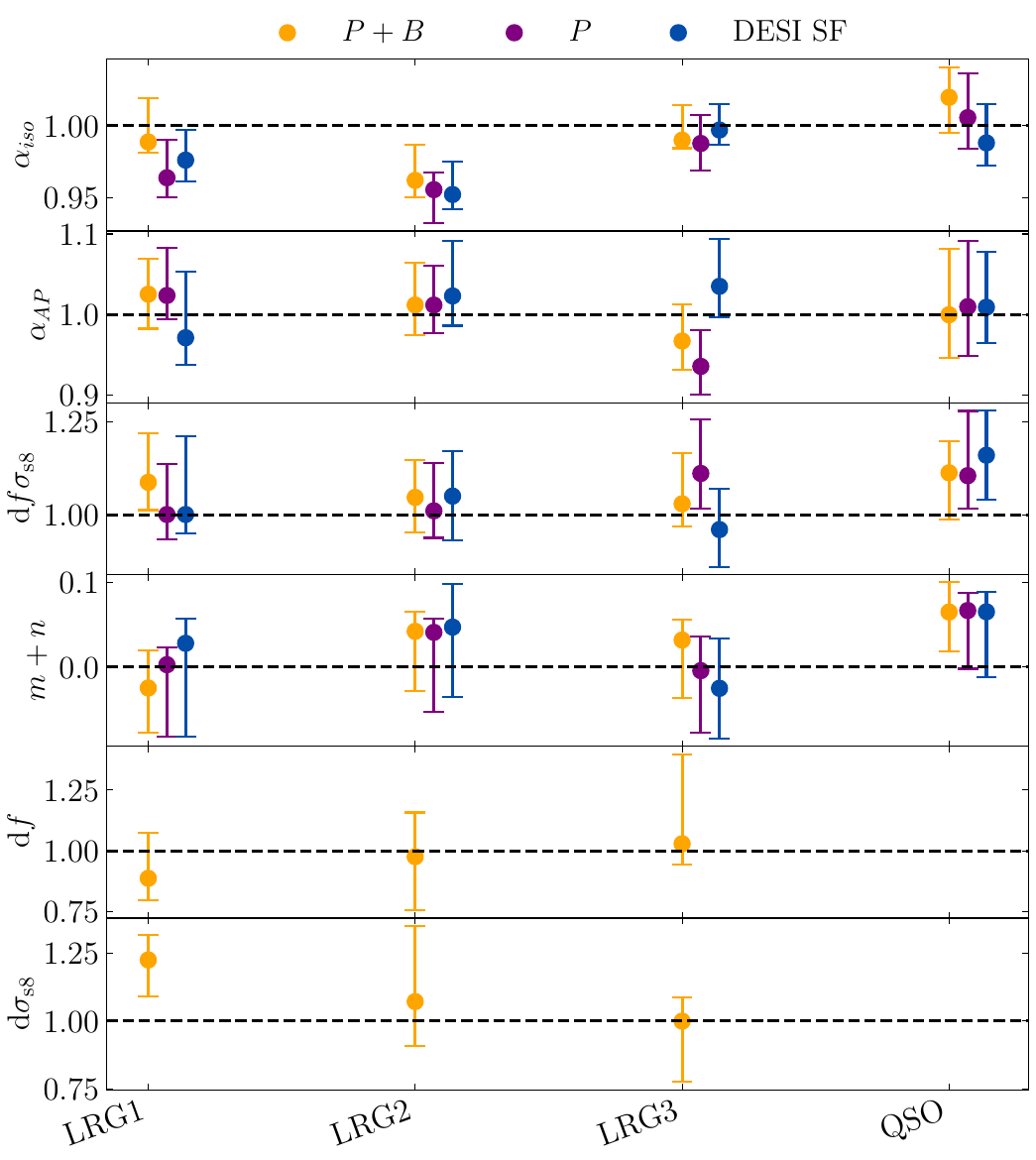}
 \vspace{-0.6cm}
\caption{
Constraints (68\% C.L.) on the parameters $\{\alpha_\textrm{iso},\alpha_\textrm{AP},df\sigma_\textrm{s8},m+n,df,d\sigma_\textrm{s8}\}$ for the unblinded DESI data in the four redshift bins (LRG1, LRG2, LRG3, QSO) as indicated in abscissa. The different colours correspond to the baseline $P+B$ (orange), $P$ only (purple) and the official DESI  SF (ShapeFit) analysis \cite{KP5Desi} results (blue). The dashed line marks the fiducial, \texttt{c000}, cosmology, and $df$, $d\sigma_{s8}$, and $df\sigma_{s8}$ are computed with respect to the fiducial model (i.e., $f/f^{\texttt{c000}}$, $\sigma_{\rm s 8}/\sigma_{\rm s 8}^{\texttt{c000}}$). 
Despite the analysis differences, the $P$ results presented here are very consistent with the official DESI ones. The addition of the bispectrum  breaks the $f\sigma_\textrm{s8}$ degeneracy, in addition it tightens the error bars especially   
on the $df\sigma_\textrm{s8},\alpha_\textrm{iso}$ and $m+n$ parameters. 
A rigorous interpretation of the ShapeFit constraints in terms of $\Lambda$CDM and extensions of it will be provided in \cite{novellmasot2025inprep}.}
\label{fig: unblinded_results}
\end{figure}
Despite the analysis differences (sample, selection, $k$-cuts, multipoles considered, different theoretical modelling), the $P$ results presented here are very consistent with the official DESI ones, highlighting the robustness of the analyses.
Compared to $P$ only,  the full $P+B$ analysis provides mild improvements on the size of the errors for the $\alpha_\textrm{iso}$ (9\% improvement), $f\sigma_\textrm{s8}$ (9\% improvement) and $m+n$ (11\% improvement) parameters. As expected, since the bispectrum monopole has no significant anisotropic signal, there is no change in the size of the constraints for the parameter $\alpha_\textrm{AP}.$
 The detailed values of the maximum a posteriori (MAP) and the total error budget (including statistical and systematic contributions) are reported in Table \ref{tab:results}. The corresponding nuisance parameters are reported in Appendix~\ref{app: nuisance}.

Notably, 
the inclusion of the bispectrum monopole breaks the $f$-$\sigma_{s8}$ degeneracy. We denote with the prefix `$d$' the ratio of $f$ or $\sigma_\textrm{s8}$ with respect to its fiducial (\texttt{c000} cosmology) value at any effective redshift. The fiducial values are: $f^\textrm{fid}=\{0.763,0.817,0.870,0.928\}$ and $\sigma^\textrm{fid}_\textrm{s8}=\{0.621,0.565,0.501,0.401\}$, corresponding to LRG1, LRG2, LRG3 and QSO. We obtain $df=\{0.888_{-0.089}^{+0.186},0.977_{-0.220}^{+0.182},1.030_{-0.085}^{+0.368}\}$,  $d\sigma_{s8}=\{1.224_{-0.133}^{+0.091},1.071_{-0.163}^{+0.278},1.000_{-0.223}^{+0.088}\}$ respectively for the LRG1, LRG2, LRG3 bins. This is the main result of this work.

\begin{figure}[ht!]
\centering 
\includegraphics[width = \textwidth]{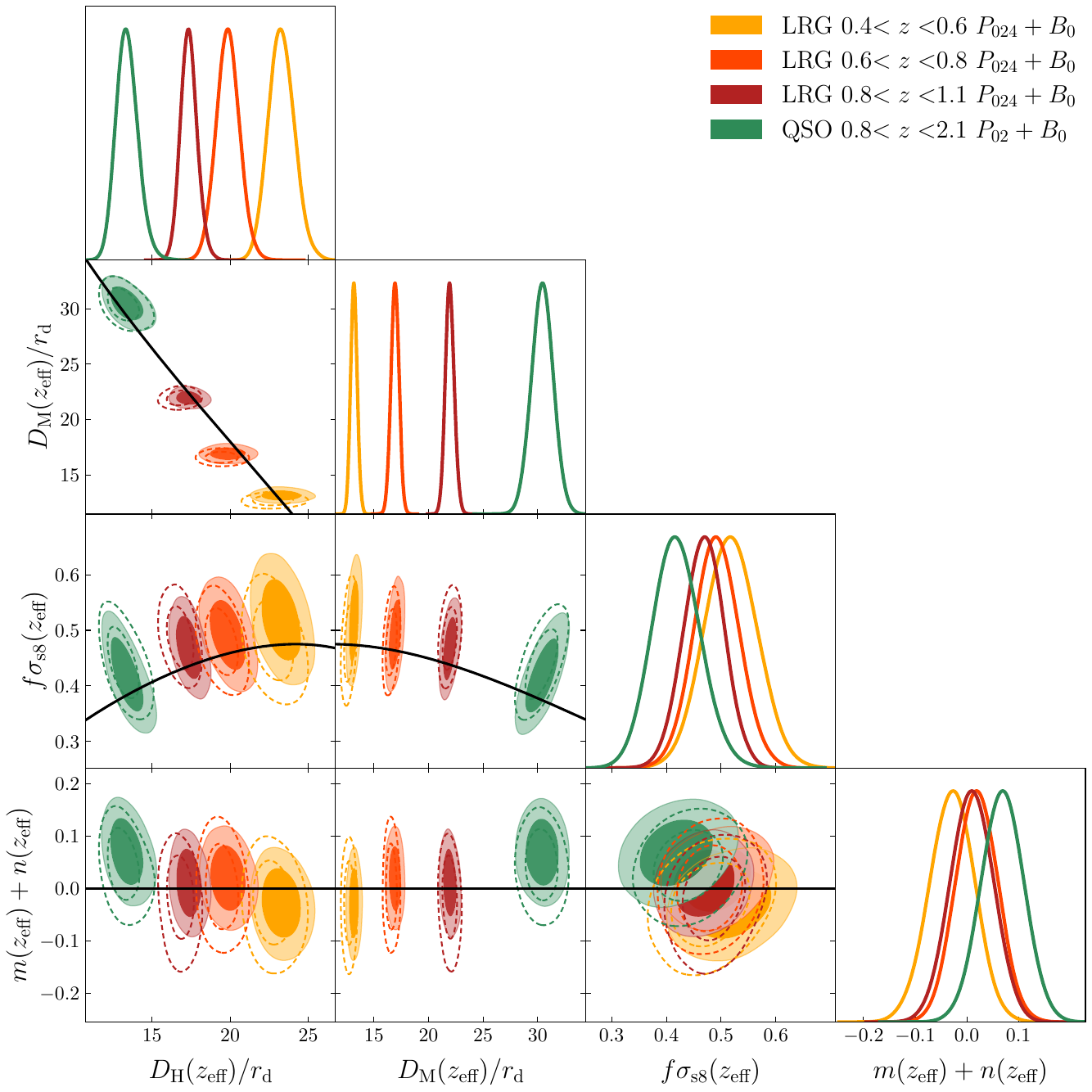}
\caption{
Posterior distributions (68 and 95\% C.L.) for the parameters $\{D_\textrm{H}(z_\textrm{eff})/r_\textrm{d},D_\textrm{M}(z_\textrm{eff})/r_\textrm{d},$ $f\sigma_\textrm{s8}(z_\textrm{eff}), m(z_\textrm{eff})+n(z_\textrm{eff})\}$ in the four redshift bins for the baseline analysis ($P+B$).  The theoretical evolution of the cosmological parameters according to Planck 2018 $\Lambda$CDM (corresponding to the fiducial \texttt{c000} cosmology) is shown as a black line.  The 2D dashed contours represent the power spectrum-only analysis highlighting how the inclusion of the bispectrum improves significantly the constraints on the $m(z)+n(z)$ combined shape parameter. The effective redshift, $z_\textrm{eff}$, is defined in Equation \ref{eq: zeffVeff}.
}
\label{fig: constraints_allbins}
\end{figure}

The combination of all the redshift bins yield a cumulative 10.1\% constraint on $f$ (with less than 10\% of the contribution to the total error bar coming from the systematic errors), and of  8.4\% on $\sigma_\textrm{s8}$ (27\% of the total error bars is the contribution from systematics). For the other compressed parameters  the cumulative errors are $\sigma_{\alpha_\textrm{iso}}=0.9\%$ (9\% improvement with respect to our power spectrum-only analysis); $\sigma_{\alpha_\textrm{AP}}=2.3\%$ (no improvement with respect to power spectrum-only analysis); $\sigma_{f\sigma_\textrm{s8}}=5.1\%$ (9\% improvement); $\sigma_{m+n}=2.3\%$ (11\% improvement). For these parameters, the contribution to the cumulative error from the systematics is always below 11\% of the total--for the $\alpha$ parameters is below 3\%. A discussion on the $\chi^2$ values can be found in sec.~\ref{sec:error_validation}.

Figure \ref{fig: constraints_allbins} provides an alternative visualization of the $P+B$ results (solid contours) as well as the $P$-only results (dashed contours),  expressed in terms of the evolution across the samples (in different colours as labelled) of the angular diameter distance $D_\textrm{M}/r_\textrm{d}$ and Hubble distance $D_\textrm{H}/r_\textrm{d}$, together with the $f\sigma_\textrm{s8}$ and $m+n$ combined shape parameter. 

It is interesting to compare 
 Figure \ref{fig: constraints_allbins} with figure 2 of Reference \cite{brieden2022model}, which displays the analogous constraints for BOSS and eBOSS when only employing the power spectrum multipoles of the LRG and QSO samples; and also including the BAO-only Ly$\alpha$ results (which we do not consider in this work).
It is remarkable that, with only one year of observations, DESI data yield 
comparable 
constraints. Subsequent DESI data releases are expected to provide significant improvements.

Within the considered redshift bins, our baseline $P+B$ analysis shows an average improvement with respect to the DESI SF results \cite{KP5Desi} of $\sim10\%$ for $\alpha_\textrm{AP}$; of $\sim20\%$ for $f\sigma_\textrm{s8}$; and of $\sim30\%$ for $m+n$. We obtain $\sim5\%$ larger constraints than DESI SF for the parameter $\alpha_\textrm{iso}$. Analogously, our $P$-only analysis results in an average improvement in the coincident redshift bins (LRG1, LRG2, LRG3, QSO) with respect to DESI SF \cite{KP5Desi} of $\sim9\%$ for $\alpha_\textrm{AP}$; of $\sim4\%$ for $f\sigma_\textrm{s8}$; of $\sim15\%$ for $m+n$; and $\sim15\%$ larger constraints than DESI SF for $\alpha_\textrm{iso}$.\footnote{The fact that the $\alpha_\textrm{iso}$ parameter is slightly less constrained in our power spectrum approach is probably due to having a lower $k_\textrm{max}$. }

The interpretation of the ShapeFit compressed parameters presented in this work in terms of constraints on cosmological parameters of a specific model, such as the primordial size of fluctuations, $A_s$, the matter density, $\Omega_m$, and the expansion parameter $H_0$; as well as the comparison with CMB data will be reported in Ref.~\cite{novellmasot2025inprep}. 

\begin{table}[!ht]
\centering
        \small
    \resizebox{\columnwidth}{!}{
\begin{tabular}{|c|c|c|c|c|c|c|c|c|}
\hline
& Sample & $\alpha_\textrm{iso}$ & $\alpha_\textrm{AP}$& $df\sigma_\textrm{s8}$ &   $m+n$ & $df$ & $d\sigma_\textrm{s8}$ & $\chi^2/{\rm dof}$   \\ \hline\hline
\multirow{4}{*}{$P+B$} &  LRG1  &   $0.989_{-0.007}^{+0.031}$ & $1.025_{-0.043}^{+0.044}$   & $1.087_{-0.075}^{+0.133}$  &  $-0.025_{-0.053}^{+0.044}$ &  $0.888_{-0.089}^{+0.186}$   & $1.224_{-0.133}^{+0.091}$  & 241/178\\ 
                      &  LRG2&    $0.962_{-0.012}^{+0.025}$ &$1.012_{-0.038}^{+0.052}$     &     $1.047_{-0.095}^{+0.099}$  & $0.042_{-0.070}^{+0.023}$  & $0.977_{-0.220}^{+0.182}$  & $1.071_{-0.163}^{+0.278}$ & 291/178 \\ 
                      & LRG3 &    $0.990_{-0.006}^{+0.025}$ &$0.967_{-0.036}^{+0.046}$     &$1.029_{-0.062}^{+0.137}$ & $0.032_{-0.068}^{+0.024}$& $1.030_{-0.085}^{+0.368}$  & $1.000_{-0.223}^{+0.088}$ & 253/178\\
                    &  QSO     &    $1.020_{-0.025}^{+0.021}$ &     $1.000_{-0.053}^{+0.082}$      &     $1.113_{-0.126}^{+0.086}$ & $0.065_{-0.046}^{+0.035}$     &   - & -  & 193/166  \\ \hline
\multirow{4}{*}{$P$}   & LRG1       &  $0.964_{-0.013}^{+0.027}$  & $1.024_{-0.030}^{+0.059}$  &     $1.001_{-0.066}^{+0.136}$  & $0.003_{-0.084}^{+0.020}$&- &-  & 33/31\\
                      & LRG2 &   $0.955_{-0.023}^{+0.012}$  &  $1.012_{-0.035}^{+0.049}$     &  $1.010_{-0.073}^{+0.128}$ &  $0.041_{-0.093}^{+0.016}$    & -&  -  & 37/31\\ 
                      &  LRG3  &   $0.988_{-0.019}^{+0.020}$  &  $0.935_{-0.035}^{+0.045}$  &  $1.111_{-0.095}^{+0.146}$ &  $-0.004_{-0.073}^{+0.040}$ & - &  - & 31/31\\
                      &  QSO     &    $1.006_{-0.022}^{+0.031}$ &  $1.010_{-0.061}^{+0.081}$ &     $1.105_{-0.088}^{+0.174}$  & $0.067_{-0.069}^{+0.021}$  &   - & -   & 20/18\\\hline

\end{tabular}
}
\caption{Results from our two main analyses, $P+B$ (including the bispectrum monopole) and $P$ (only power spectrum), for the LRG and QSO redshift bins. In each case, we show the maximum a posteriori (MAP), together with the 1$\sigma$ region (which accounts for both statistical and systematic errors) centred in the MAP value, for the ShapeFit parameters $\alpha_\textrm{iso},\alpha_\textrm{AP},f\sigma_\textrm{s8},m+n$, together with the parameters $f$ and $\sigma_\textrm{s8}$. The prefix `$d$' in a parameter indicates its ratio with respect to its fiducial value at the effective redshift of the bin. The corresponding fiducial values are: $f^\textrm{fid}=\{0.763,0.817,0.870,0.928\}$ and $\sigma^\textrm{fid}_\textrm{s8}=\{0.621,0.565,0.501,0.401\}$, for respectively LRG1, LRG2, LRG3 and QSO. We additionally report the corresponding values for $\chi^2$ over the degrees of freedom for each case (see Sec.~\ref{sec:error_validation} for discussion). 
}
\label{tab:results}
\end{table}

\subsection{Statistical error validation}
\label{sec:error_validation}
We aim to validate the statistical errors component obtained on the unblinded data with the results obtained from \texttt{Abacus} AltMTL mocks. The reasons are two-fold.
\begin{enumerate}
    \item  To check whether the DESI DR1 data catalogues have similar statistical properties to the \texttt{Abacus} mocks used to validate the modelling and to test for potential systematic errors.
\item To understand the reason for the high minimum-$\chi^2$ values found when fitting the power spectrum and bispectrum from the LRG data (reported in Table \ref{tab:results}); and to quantify the potential impact of this high minimum $\chi^2$ values on the determination of the statistical error component.
\item To validate that the rescaling of the covariance that we have applied, based on section 5.7 of Ref.~\cite{KP5Desi} is still applicable within our analysis involving the power spectrum hexadecapole and the bispectrum monopole signals.
 \end{enumerate}

\begin{figure}[t]
  \centering
    \includegraphics[width=0.49\textwidth]{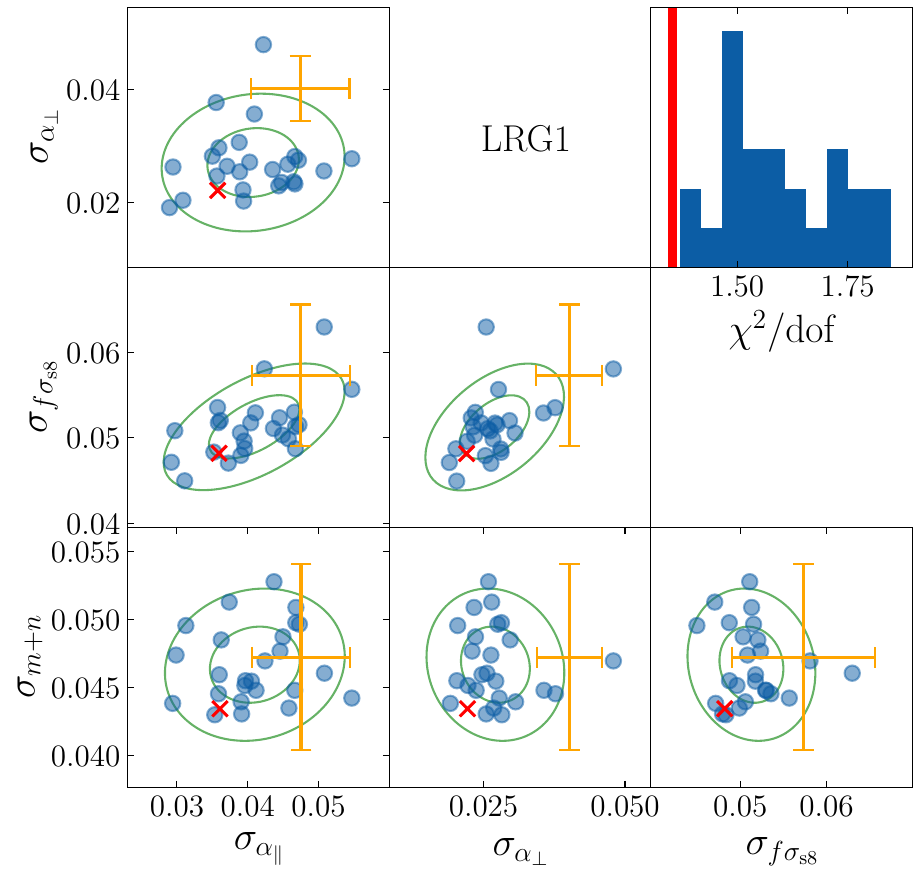}  
    \includegraphics[width=0.49\textwidth]{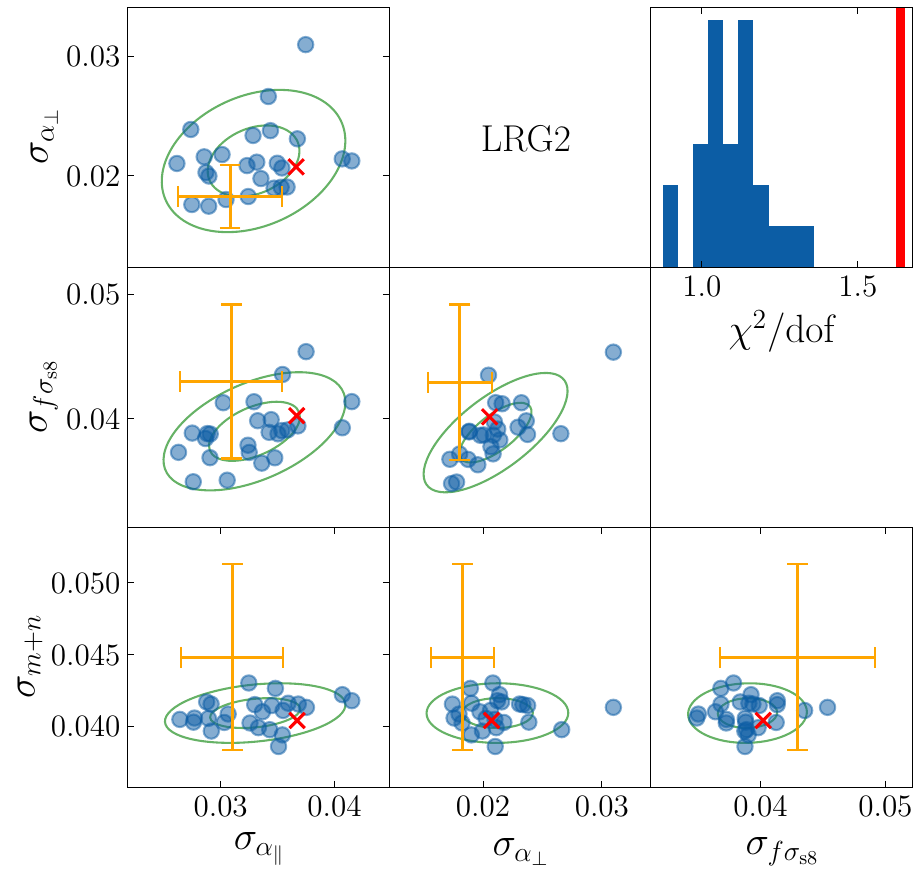} \\
    \includegraphics[width=0.49\textwidth]{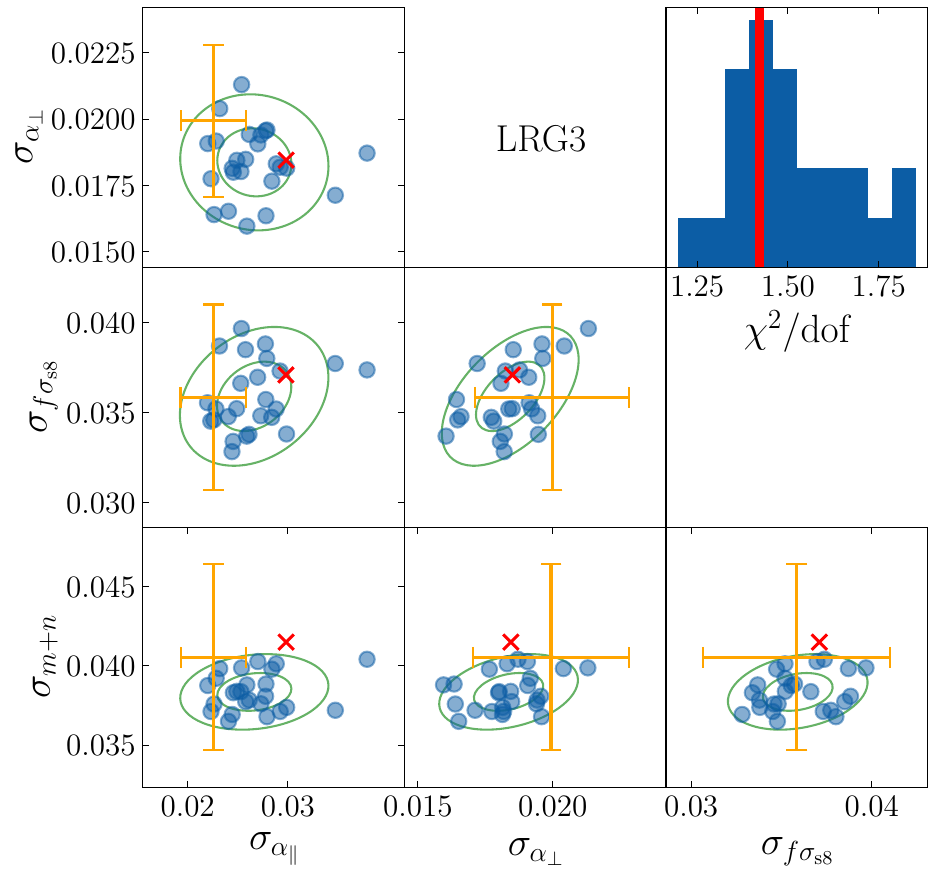}  
    \includegraphics[width=0.49\textwidth]{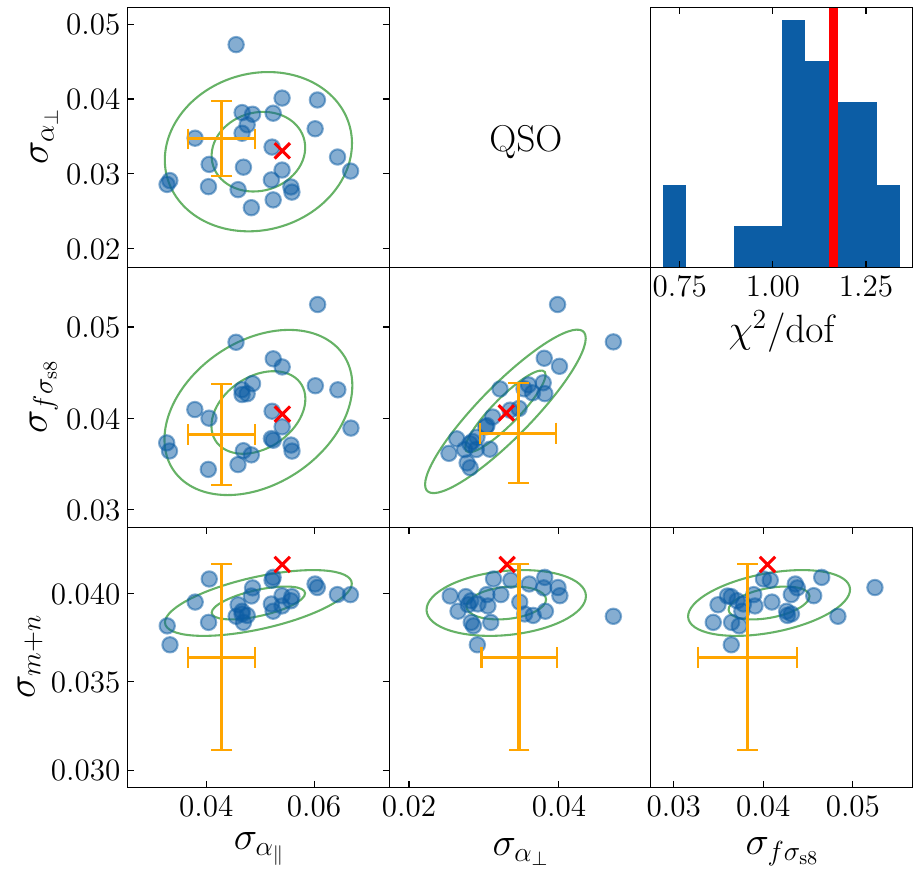} \\
  \vspace{-0.5cm}
  \includegraphics[width=0.9\textwidth]{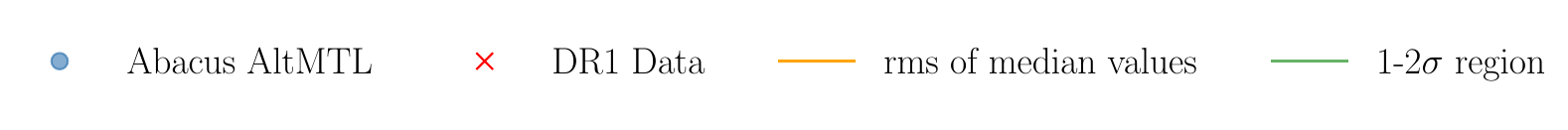}
  \vspace{-0.5cm}
  \caption{Scatter of the 1$\sigma$ statistical errors for the baseline $P+B$ analysis choice on the parameters of interest, $\{\alpha_\parallel,\,\alpha_\perp,\,f\sigma_{s8},\,m+n \}$ as well as the histogram for the $\chi^2$ distribution normalized by the number of degrees of freedom ($\chi^2/{\rm dof})$. The performance for the 25 \texttt{Abacus} AltMTL mocks is displayed in blue circles, whereas the red cross displays the actual DESI DR1 data performance. The orange points with error bars are the errors obtained as the standard deviation of the medians of the 25 individual fits (the error bars corresponding to the error on the standard deviation, estimated as $\sigma/\sqrt{2N-2}$, with $N=25$). The green curves delimit the approximated $1-2\sigma$ regions of the error distribution, estimated from the scatter of the blue data-points.   The four main panels represent the redshift bins studied in this paper, as labelled. The reported errors do not contain any systematic error budget, but the \texttt{EZmock}-derived covariance does contain the correcting factors of Table 3 of Ref.~\cite{KP5Desi}. The errors reported for the DESI data are consistent with those of the \texttt{Abacus} AltMTL mocks, and with the error derived from the medians of the 25 realisations. Only for the LRG3 and QSO redshift bins, the error on the $m+n$ parameter is on the upper side of the distribution of errors, showing a typical $\sim 2\sigma$ offset from the overall distribution. The larger-than-mocks best-fitting $\chi^2$ reported on the data for the LRG2 bin does not have a noticeable effect on the errors of the parameters of interest compared to what is found on the \texttt{Abacus} mocks. }
  \label{fig: error_validation}
\end{figure}

In order to do so, we perform the fit on the 25 individual realisations of the LRG and QSO \texttt{Abacus} AltMTL mocks, which contain the DESI DR1 data survey geometry and mimic the fibre assignment effect. We employ the same covariance used for fitting the DR1 data, derived from the 1000 FFA \texttt{EZmocks} realisations, including the correction factors of Table 3 of Ref.~\cite{KP5Desi} as explained in Section~\ref{sec: Mocks}. We do not include any extra systematic contribution, as here we are only interested in testing the {\it relative} statistical properties of the data compared to the mocks.

In Figure~\ref{fig: error_validation} we display the distribution of the $1\sigma$ (or 68\% CL) errors for the four parameters of interest, $\{\alpha_\parallel,\,\alpha_\perp,\,f\sigma_{s8},\,m+n \}$, for the four redshift bins studied in this work: top-left panel for the LRG1; top-right panel for the LRG2; bottom-left panel for the LRG3; bottom-right panel for the QSO, as labelled. For each of these panels, the blue dots display the statistical errors of the 25 individual AltMTL \texttt{Abacus} mocks. The statistical errors corresponding to the DESI DR1 data are shown as the red crosses. We compare the distribution of statistical errors with the standard deviation computed from the medians of the 25 individual posteriors, shown in orange. \footnote{In the limit of a large number of realisations, the two estimates should coincide for a Gaussian distribution: within the limits of using only 25 mocks, there is compatibility within the error associated with the estimation of the standard deviations.} In addition, the histogram of the best-fitting $\chi^2$ per number of degrees of freedom (dof) is also shown for each of the LRG and QSO samples, where the red vertical line represents the minimum $\chi^2$ found for the data (and reported in Table~\ref{tab:results}), and the blue histogram is from the 25 mocks.

Figure~\ref{fig: error_validation} shows that the errors obtained from the data are in general very consistent with the typical errors we obtain from the 25 mocks. Only for the $m+n$ parameter, for the LRG3 and QSO bins, we find that the error of the data is at the high end of the distribution of errors from the mocks. We do not consider this outlier as something statistically worrisome, as we estimate that such outliers sit in a region of  $\sim2\sigma$ from the typical size of errors. 

We also notice that for the LRG1 and the LRG3 bins, the distribution of $\chi^2/{\rm dof}$ of the \texttt{Abacus} AltMTL mocks is centred around values $\sim 1.4$ to $1.5$ while for  LRG2 and QSO samples it is centred closer to the expected value of $1.0$.
One possible explanation is
 that the model is not able to accurately describe the mocks. In some cases (see Table~\ref{tab:P024B0syst} for the LRGs and Table~\ref{tab:P02B0syst} for the QSO) we have reported modelling systematic errors of $20\%$ to $50\%$ of the total statistical error budget, that could explain the high $\chi^2$ values in some of these bins. Another possibility is that the mock-based covariance we are employing is slightly underestimating the true covariance of the mocks, even after the correction term of Table 3 of Ref.~\cite{KP5Desi}. We leave for future work a more thorough exploration of the covariance of the bispectrum, as it requires the development of more realistic mocks and more sophisticated techniques for an analytical-based covariance, which goes beyond the scope of this paper. 

Finally, we also highlight that the minimum $\chi^2$ found for the data is broadly in agreement with those obtained for the mocks, with maybe the only exception of the LRG2 bin, where the highest $\chi^2/{\rm dof}$ found on a mock is of $\sim1.4$, whereas for the data is $\sim1.6$. However, this difference in best-fitting $\chi^2$ values does not translate into a difference in the distribution of the errors of the parameters of interest, and therefore it doesn't impact the cosmology results.

We conclude that there is no evidence that the statistical errors are significantly under/over estimated, and that the rescaling factors that we have applied (following table 3 of Ref.~\cite{KP5Desi}) are appropriate for our analysis settings. 
The high values for the minimum $\chi^2$ seen in the data are also seen in the simulations and are possibly due to a combination of effects including imperfections in the model of the bispectrum, imperfections in the calibration of the covariance matrix and possibly the adoption of a Gaussian likelihood as an imperfect approximation. 
Even if we were to rescale the covariance matrix by a constant factor as to force the minimum $\chi^2$ per dof to be 1, this would not increase the errors by more than $\sim20$\%. This is below our minimal threshold to propagate systematics into the final results.

We conclude that the covariance matrix shortcomings we have seen here do not bias the final results nor the reported size of the (statistical + systematic) error-bars in a significant way.

\section{Conclusions}
\label{sec: conclusions}
We have presented the first joint analysis of the power spectrum and bispectrum signals using DESI DR1 data catalogues, focusing on the LRG and QSO samples, and thus probing the evolution of the universe through a redshift range of $0.4\leq z\leq 2.1$. The inclusion of the bispectrum allows us to break parameter degeneracies, in particular, to obtain separate constraints for the amplitude of perturbations ($\sigma_\textrm{s8}$) and growth rate ($f$) parameters, which are usually constrained only in the combination $f\sigma_\textrm{s8}$. 

We use a model and framework very similar to that of  BOSS and eBOSS analyses \cite{gil2016clustering,HGMeboss}, which differs from the official DESI pipeline \cite{KP5Desi} in several aspects:
\begin{itemize}
    \item We work with the renormalized
perturbation theory model (RPT) expansion at two-loop, while \cite{KP5Desi} uses an EFT-based model at one-loop.
    \item In the three LRG bins we use the signal from the power spectrum monopole, quadrupole and hexadecapole, while \cite{KP5Desi} uses only the power spectrum monopole and quadrupole.
    \item We consider a range of scales of $k<0.15\, h\,\textrm{Mpc}^{-1}$, while \cite{KP5Desi} considers $k<0.20\, h\,\textrm{Mpc}^{-1}$.
    \item We do not use all the systematic mitigation strategies that \cite{KP5Desi} uses, in particular the $\theta$-cut and the marginalization over the non-linearities of systematic weights (although the latter is mostly relevant for ELG galaxies which we do not include).
    \item We do not use any ELG redshift bin, given that this tracer fails to pass some of our systematic tests. We also do not use the BGS tracer.
    \item We include the bispectrum monopole.
\end{itemize}

Apart from these differences, our analysis closely matches that of the official DESI pipeline: we use the same mocks (see Section \ref{sec: Mocks}), both for validation and covariance, the same blinding procedure (Section \ref{sec: Data}), the same approach for the ShapeFit compression (Section \ref{sec: compression}), and the same treatment and characterization of systematic errors (Section \ref{sec: Systematics}).

The remarkable consistency between the $P$-only results presented here and the official DESI one highlights the exquisite fidelity of DESI data and the astounding robustness of cosmological constraints obtained from compressed-variable analyses of large-scale structure data.

The inclusion of the bispectrum monopole represents the main novelty of this work.
We use the GEO-FPT bispectrum model presented in \cite{novell2023geofpt}, for a range of scales of $0.02\leq k\, \left[h\,\textrm{Mpc}^{-1}\right]\leq0.12$. 
We limit ourselves to the bispectrum monopole ($B_0$) as we lack of a sufficiently accurate modelling of the effects of the survey window function for the bispectrum multipoles.  
The adopted bispectrum monopole window convolution is the same as that used in some BOSS and eBOSS analyses \cite{Gil-Marin:2014biasgravity,Gil-Marin:2016sdss}.

We perform a suite of systematic tests, following closely Ref.~\cite{KP5Desi}, to determine the size of the systematic error budget and to decide which summary statistics we use in each redshift bin before unblinding the data. The largest sources of systematics come from the modelling and fibre assignment process, with $\sigma_\textrm{s8}$ being the most affected parameter in terms of percentage of the DR1 statistical error bar. From the systematic error estimates and the cosmological parameter constraints performed in blinded data, we choose to use as our baseline analysis the combination $P_{024}+B_0$ for the three LRG redshift bins, and only $P_{02}+B_0$ for the QSO tracer. 

In this way (and assuming the different redshift bins are uncorrelated), we obtain cumulative constraints of 10.1\% and 8.4\% for respectively $f$ and $\sigma_\textrm{s8}$. Our constraints for the parameters $\alpha_\textrm{iso},\,\alpha_\textrm{AP},\,\,f\sigma_\textrm{s8},\,m+n$ are consistent (and competitive) with the ones obtained in the main DESI collaboration ShapeFit analysis \cite{KP5Desi}. In particular, the inclusion of the bispectrum monopole reduces the error bars of $\alpha_\textrm{iso},\,f\sigma_\textrm{s8},\,m+n$ by respectively $9\%,\,9\%\textrm{ and }11\%$ with respect to our power spectrum-only analysis. 

We envision this to be a first milestone towards the promise of tight cosmological constraints (including $f$ and $\sigma_\textrm{s8}$ when considered separately) with the combination of two-point and higher-order statistics. In particular, the ability to consider parameter constraints on $f$ and $\sigma_\textrm{s8}$ separately will allow us to constrain modifications of general relativity (where the relationship $f=\Omega_\textrm{m}^\gamma$ with $\gamma\approx6/11$ no longer holds \cite{linder2005cosmic,linder2007parameterized}). Aside from the obvious improvement in precision that will be possible with the upcoming DESI data releases, improvements will include 
strategies to mitigate the fibre assignment systematic errors and the inclusion of
the bispectrum quadrupoles in the analysis. This is a major challenge since the window function treatment is highly non-trivial, but the extra anisotropic information present may provide significant enhancement of the precision and accuracy in most cosmological parameters, as we saw in \cite{novell2023geofpt}.

In a forthcoming work \cite{novellmasot2025inprep}, we will provide the interpretation of the compressed parameter constraints presented here in terms of traditional cosmological parameters for the $\Lambda$CDM  model and its extensions.

\section*{Data Availability}
All data from the tables and figures will be available in machine-readable format at \href{https://zenodo.org/records/14944381}{10.5281/zenodo.14944381} 
upon acceptance in compliance with the DESI
data management plan.

\section*{Acknowledgements}

SNM, HGM and LV thank Pauline Zarrouk, Caroline Guandalin, Alex Krolewski and Ruiyang Zhao for the helpful discussion and suggestions.
SNM acknowledges funding from the official doctoral programme of the University of Barcelona for the development of a research project under the PREDOCS-UB grant.
HGM acknowledges support through the Leonardo programme (LEO23-1-897) of the BBVA foundation and through the programmes Ram\'on y Cajal (RYC-2021-034104) and Consolidación Investigadora (CNS2023-144605) of the Spanish Ministry of Science and Innovation. 

Funding for this work was partially provided by the Spanish MINECO under project  PID2022-141125NB-I00 MCIN/AEI,
and the ``Center of Excellence Maria de Maeztu 2020-2023'' award to the ICCUB (CEX2019-000918-M funded by MCIN/AEI/10.13039/501100011033).

This material is based upon work supported by the U.S. Department of Energy (DOE), Office of Science, Office of High-Energy Physics, under Contract No. DE–AC02–05CH11231, and by the National Energy Research Scientific Computing Center, a DOE Office of Science User Facility under the same contract. Additional support for DESI was provided by the U.S. National Science Foundation (NSF), Division of Astronomical Sciences under Contract No. AST-0950945 to the NSF’s National Optical-Infrared Astronomy Research Laboratory; the Science and Technology Facilities Council of the United Kingdom; the Gordon and Betty Moore Foundation; the Heising-Simons Foundation; the French Alternative Energies and Atomic Energy Commission (CEA); the National Council of Humanities, Science and Technology of Mexico (CONAHCYT); the Ministry of Science and Innovation of Spain (MICINN), and by the DESI Member Institutions: \url{https://www.desi.lbl.gov/collaborating-institutions}. Any opinions, findings, and conclusions or recommendations expressed in this material are those of the author(s) and do not necessarily reflect the views of the U. S. National Science Foundation, the U. S. Department of Energy, or any of the listed funding agencies.

The authors are honored to be permitted to conduct scientific research on Iolkam Du’ag (Kitt Peak), a mountain with particular significance to the Tohono O’odham Nation.

This work has made use of the following publicly available codes: \href{https://github.com/serginovell/geo-fpt}{\textsc{GEO-FPT}} \cite{novell2023geofpt}, \href{https://github.com/hectorgil/Rustico}{\textsc{Rustico}} \cite{HGMeboss}, \href{https://github.com/hectorgil/Brass}{\textsc{Brass}} \cite{HGMeboss}, \href{https://emcee.readthedocs.io/en/stable/index.html}{\textsc{Emcee}} \cite{Foreman_Mackey_2013}, \href{https://www.gnu.org/software/gsl/}{GSL} \cite{gsl}, \href{https://scipy.org/}{\textsc{SciPy}} \cite{scipy}, \href{https://numpy.org/}{\textsc{NumPy}} \cite{numpy}, \href{https://getdist.readthedocs.io/en/latest/}{\textsc{GetDist}} \cite{getdist}, \href{https://www.astropy.org}{\textsc{Astropy}} \cite{astropy}, \href{https://matplotlib.org}{\textsc{Matplotlib}} \cite{matplotlib}. We are grateful to the developers who made these codes public.

\appendix
\section{Perturbation theory modelling of the power spectrum and bispectrum}
\label{app: theory}
In this section, we review the perturbation theory theoretical formalism underlying the modelling used here for the power spectrum and bispectrum.

The redshift space galaxy power spectrum, $P_g$, is computed from the non-linear matter power spectrum, $P_{g,\delta\delta}$ (also referred to as $P_{\rm NL}$ in the main text and in Ref~\cite{GilMarin:2011form,novell2023geofpt}), the density-velocity, $P_{g,\delta\theta}$, and velocity-velocity, $P_{g,\theta\theta}$, power spectra, according to the TNS model \cite{Taruya_2010,Nishimichi_2011},
\begin{align}
\label{eq: Pnl_redshift}
     P_g(k,\mu)&=D_\textrm{FoG}^P(k,\mu,\sigma_P)\big[P_{g,\delta\delta}(k)+2f\mu^2P_{g,\delta\theta}(k)+f^2\mu^4P_{\theta\theta}(k)\nonumber\\
     &+b_1^2A^\textrm{TNS}(k,\mu,f/b_1)+b_1^4B^\textrm{TNS}(k,\mu,f/b_1)\big],
\end{align}
where $f$ denotes the  logarithmic growth rate of perturbations, and $d \ln \delta /d\ln a$ and  $P_{g,\delta\delta}, P_{g,\delta\theta}$ are computed as in \cite{beutler2014clustering} using the RPT terms of \cite{gil-marinetal:2012} for $P_{\delta\delta},\,P_{\delta\theta},\,P_{\theta\theta}$. In doing so, we are using the bias expansion $\{b_1,b_2,b_{s^2},b_\textrm{3nl}\}$ and assuming the Lagrangian local bias relations \cite{saito2014understanding,Chenetal2012,beutler2014clustering,baldauf2012evidence},
\begin{equation}
 \label{eq:lagrangian_local}   b_{s^2}= -\frac{4}{7}(b_1-1); \quad\quad b_{3\rm nl} = \frac{32}{315}(b_1-1).
\end{equation}
Additionally, the functions $A^\textrm{TNS},B^\textrm{TNS}$ are defined in \cite{Taruya_2010}, $\mu$ is the cosine of the angle of $k$ with the line of sight, and $D_\textrm{FoG}^P$ is a damping factor that accounts for the Fingers-of-God (FoG) effect of redshift space distortions (RSD) \cite{Jackson_1972}. We model the FoG damping factor for the power spectrum as,  
\begin{equation}
    D_\textrm{FoG}^P(k,\mu,\sigma_\textrm{FoG}^P)=\frac{1}{\left(1+k^2\mu^2\sigma_P^2/2\right)^2},
\end{equation}
where $\sigma_P$ is a free parameter to be constrained by the data.

The tree-level redshift space bispectrum can then be written in the following way at tree-level order:
\begin{equation}
B^{\rm SPT}(\textbf{k}_1,\textbf{k}_2,\textbf{k}_3)=D_\textrm{FoG}^B(\textbf{k}_1,\textbf{k}_2,\textbf{k}_3)\left[2Z_1^\textrm{SPT}(\textbf{k}_1)Z_1^\textrm{SPT}(\textbf{k}_2)Z_2^\textrm{SPT}(\textbf{k}_1,\textbf{k}_2)P_\textrm{L}(k_1)P_\textrm{L}(k_2) + \textrm{2perm.}\right],
\label{eq:bispredshiftsp}
\end{equation}
where the kernels $Z_1^\textrm{SPT},Z_2^\textrm{SPT}$ are computed as \cite{fry1994gravity,scoccimarro1998nonlinear,Verde1998},
\begin{align}
    Z_1^\textrm{SPT}(\textbf{k})&=b_1+f\mu^2,\nonumber\\
    Z_2^\textrm{SPT}(\textbf{k}_1,\textbf{k}_2)&=b_1F_2^\textrm{SPT}(\textbf{k}_1,\textbf{k}_2)+f\mu_{12}^2G_2^\textrm{SPT}(\textbf{k}_1,\textbf{k}_2)+\frac{b_1f}{2}\left(\mu_1^2+\mu_2^2+\mu_1\mu_2\left(\frac{k_1}{k_2}+\frac{k_2}{k_1}\right)\right)\nonumber\\
    &+f^2\mu_1\mu_2\left(\mu_1\mu_2+\frac{1}{2}\left(\mu_1^2\frac{k_1}{k_2}+\mu_2^2\frac{k_2}{k_1}\right)\right)+\frac{1}{2}\left(b_2+b_{s^2}S_2^\textrm{SPT}(\textbf{k}_1,\textbf{k}_2)\right),\label{eq: z1z2}
\end{align}
where $\mu_{ij}\equiv(k_i\mu_i+k_j\mu_j)/|\textbf{k}_i+\textbf{k}_j|$. The $G_2^\textrm{SPT}$ and $S_2^\textrm{SPT}$ kernels in Standard Perturbation Theory (SPT) are given by,
\begin{align}
    G_2^\textrm{SPT}(\textbf{k}_1,\textbf{k}_2)&=\frac{3}{7}+\frac{1}{2}\cos(\theta_{12})\left(\frac{k_1}{k_2}+\frac{k_2}{k_1}\right)+\frac{4}{7}\cos^2(\theta_{12}),\\
    S_2^\textrm{SPT}(\textbf{k}_1,\textbf{k}_2)&=\cos(\theta_{12})^2-\frac{1}{3}.
\end{align}
Additionally, for the bispectrum, we parametrise the FoG damping factor as \cite{Scoccimarro:1999ed,Verde1998},
\begin{equation}
\label{eq:fog_lorentz}
    D_\textrm{FoG}^B(\textbf{k}_1,\textbf{k}_2,\textbf{k}_3)=(1+\left[k_1^2\mu_1^2+k_2^2\mu_2^2+k_3^2\mu_3^2\right]^2\sigma_{B}^4/2)^{-2},
\end{equation}
again with $\sigma_B$ being a free parameter.

We model the deviations from Poissonian shot-noise with the parameters $A_\textrm{P},A_\textrm{B}$, which modify the Poisson prediction as in \cite{Gil-Marin:2014theory,gil-marin_clustering_2017},
\begin{align}
    P_{\textrm{noise}}&=(1-\frac{A_\textrm{P}}{\alpha_\parallel \alpha_\bot^2})P_\textrm{Poisson},\\
    \label{eq: B_alpha2}
    B_{\textrm{noise}}(k_1,k_2,k_3)&=(1-\frac{A_\textrm{B}}{\alpha_\parallel^2\alpha_\bot^4})B_\textrm{Poisson}(k_1,k_2,k_3).
\end{align}

The power spectrum and bispectrum redshift space multipoles are then obtained by integrating the expansion of the power spectrum and bispectrum dependence on the angle with respect to the line of sight in terms of Legendre polynomials $\mathcal{L}_i$, so that

\begin{align}
    P_{\ell}(k)&=\frac{2\ell+1}{2\alpha_\parallel\alpha_\bot^2}\int_{-1}^1d\mu\, P(k,\mu)\mathcal{L}_{\ell}(\mu),\label{eq: pmulti}\\
    \label{eq: B_alpha1}
    B_{\ell_i}(\textbf{k}_1,\textbf{k}_2,\textbf{k}_3)&=\frac{2\ell+1}{4\pi\alpha_\parallel^2\alpha_\bot^4}\int_{-1}^1d\mu_1\int_0^{2\pi}d\phi \,B(\textbf{k}_1,\textbf{k}_2,\textbf{k}_3)\mathcal{L}_{\ell}(\mu_i),
\end{align}
Here $\phi$ is defined as the angle fulfilling $\mu_2=\mu_1\cos\theta_{12}-\sqrt{(1-\mu_1^2)(1-\cos\theta_{12}^2)}\cos\phi$, and $\ell_i$ refers to the multipole of order $\ell$ ($\ell=0,2$ corresponding respectively to the monopole and quadrupole). For the bispectrum quadrupoles ($\ell=2$), $\ell_i$ is the quadrupole corresponding to integrating over the Legendre polynomial applied to the cosine of the $i$-th angle. In this work, we only use the bispectrum monopole, due to the current limitations of the modelling of the window function. The power spectrum multipole expansion of \ref{eq: pmulti} was proposed in \cite{hamilton1992measuring,cole1994fourier}, while the bispectrum expansion and choice of variables was first used in \cite{Scoccimarro:1999ed}. 

\section{Blinded data}
\label{sec: blinded}
\begin{figure}[bht]
\centering 
\includegraphics[width = \textwidth]{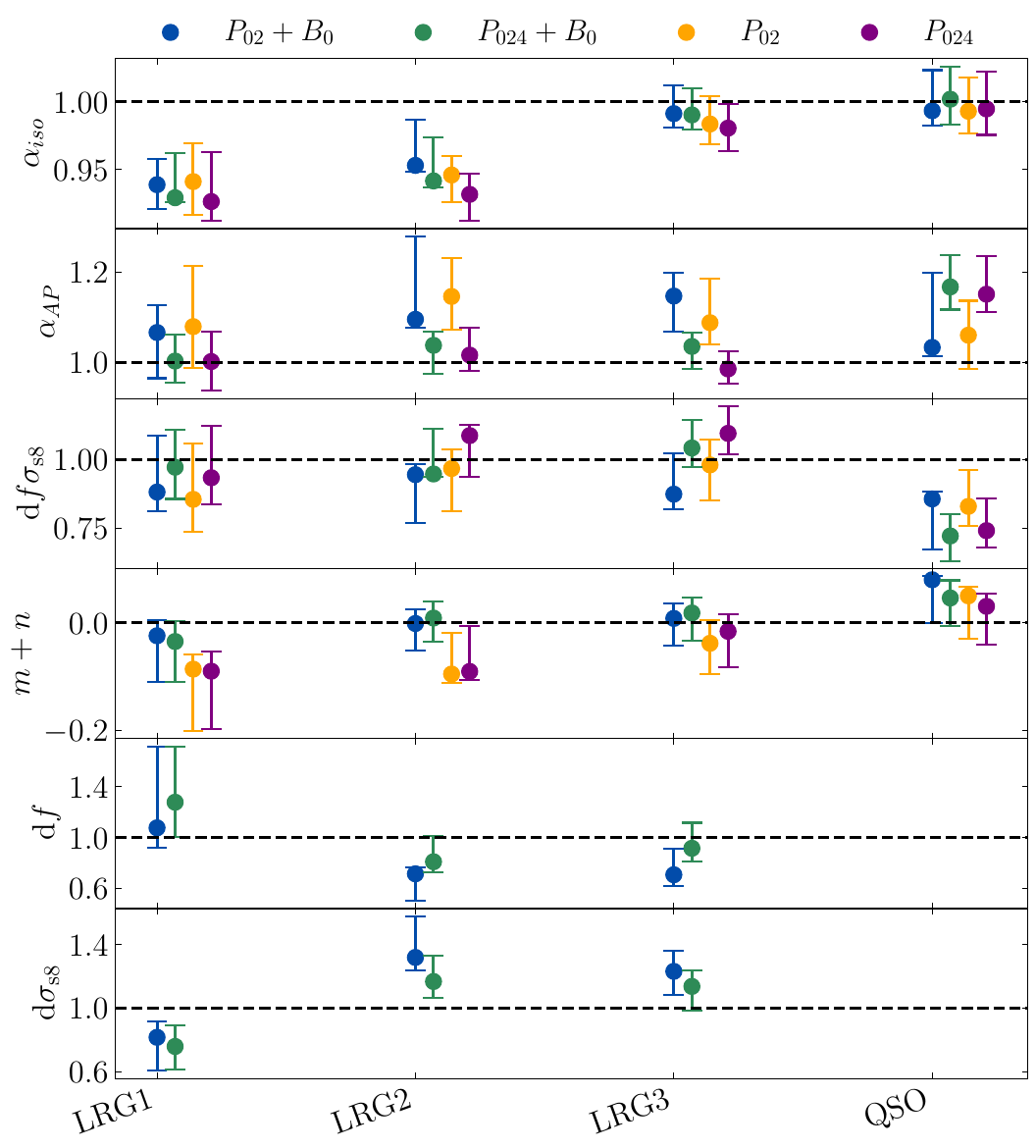}
\caption{Cosmological constraints obtained from the four blinded redshift bins (LRG1, LRG2, LRG3, QSO), by using each of the four combinations of summary statistics that we consider in this work: $P_{02},P_{024},P_{02}+B_0,P_{024}+B_0$. The parameters $f$ and $\sigma_\textrm{s8}$ are not shown separately, neither for the power spectrum, as usual, nor for the QSO redshift bin, where we saw that even in an analysis including the bispectrum the two parameters were markedly degenerate.}
\label{fig: blinded_results}
\end{figure}

 The blinded cosmological constraints for the parameters $\{\alpha_\textrm{iso},\alpha_\textrm{AP},f\sigma_\textrm{s8},m+n,f,\sigma_\textrm{s8}\}$ for the four combinations of summary statistics that we consider in this work ($P_{02},P_{024},P_{02}+B_0,P_{024}+B_0$) are shown together in Figure \ref{fig: blinded_results}.
 
These results for the blinded data determine the combination of summary statistics to use after unblinding. 
We aim at striking a 
balance between two aspects:
\begin{itemize}
    \item \textbf{Statistical errors:} In general, adding more multipoles or the bispectrum to  the data-vector 
    decreases the error bars
    \item \textbf{Systematic errors:} Some scales or summary statistics 
    are more prone to systematic errors than others.
    The systematic errors presented in Tables \ref{tab:P024B0syst}, \ref{tab:P02B0syst}, \ref{tab:P024syst}, \ref{tab:P02syst} allow us to quantify this.
\end{itemize}
We wish to select data-vectors which add significant information (and thus reduce the statistical error bars) while not increasing the systematic error budget. 

Figure \ref{fig: blinded_results} displays the constraints on the ShapeFit cosmological parameters for the different DESI redshift bins, and the different combinations of summary statistics considered (symbols in different colours). The addition of the power spectrum hexadecapole, $P_4$, significantly enhances the LRG constraints for the anisotropic parameter $\alpha_\textrm{AP}$, and, to a lesser extent, $f\sigma_\textrm{s8}.$
The comparison of Tables~\ref{tab:P024syst} vs~\ref{tab:P02syst} and Tables~\ref{tab:P024B0syst} vs~\ref{tab:P02B0syst} shows that  for LRGs the inclusion of $P_4$ does not increase the systematic error budget in any significant way.

Conversely, in the case of the QSO tracers, the inclusion of  $P_4$  does not reduce  appreciably the statistical errors but increases notably the systematic error budget
(specifically for the parameters $f\sigma_\textrm{s8}$ and $\alpha_\textrm{AP}).$ This is not unexpected, since the QSO tracers may not have sufficient number density for the signal-to-noise of the power spectrum hexadecapole to be big enough.

The inclusion of the bispectrum, aside from breaking the $f\sigma_\textrm{s8}$ degeneracy, reduces the error bars of the $\alpha_\textrm{iso}$ and $m+n$ parameters. Additionally, the $P_{024}+B_0$ multipoles combination features slightly less projection effects than the case without the power spectrum hexadecapole, particularly in the $\alpha_\textrm{AP},m+n,f\textrm{ and }\sigma_\textrm{s8}$ parameters.\footnote{The presence of projection effects can be inferred from the displacement of the maximum a posteriori (MAP) value from the centre of the error bars. When this shift is significant, it indicates that the data provide weak constraints on the parameter, since the posterior distribution is skewed.} 

Taking all these points into account, we define our baseline set of multipoles as $P_{024}+B_0$ for the three LRG redshift bins, and $P_{02}+B_0$ for the QSOs. Our baseline power spectrum-only analysis will likewise consist in $P_{024}$ for the LRGs and $P_{02}$ for QSOs.

\section{Nuisance Parameters}\label{app: nuisance}
In this section, we report the MAP values for the nuisance parameters obtained from the analysis described in section~\ref{sec: results}. Table~\ref{tab:nuisance} displays the results for the galaxy bias parameters, the shot noise amplitudes and the Fingers-of-God parameters for both types of analysis, with and without the bispectrum signal, as previously done in Table~\ref{tab:results} for the ShapeFit cosmological parameters.  The baseline choices of those results are the same to those presented before in section~\ref{sec: results}: for the LRG samples we vary $\sigma_{s8}$ and choose to report both $\{b_1,\,b_2$\}, as well as the combination of parameters best-measured, $\{b_1\sigma_{s8},\,b_2\sigma_{s8}^3$\}. On the other hand, for the QSO and for the LRG samples when only $P$ is employed, we keep $
\sigma_{s8}$ fixed, and re-interpret the best-fitting bias values as a product of $b_1$ and a power of $\sigma_{s8}$. We determine that the combination that for the power spectrum and bispectrum that keeps $b_x
\sigma_{s8}^n$ uncorrelated with $\sigma_{s8}$ is $n=1$ for $b_1$ (as expected from the Kaiser limit at large scales), and $n=3$ for $b_2$, as a resulting of combining the power spectrum and bispectrum. We then choose those combination of variables to be reported here. 

\begin{figure}
    \centering
    \includegraphics[width=1.\linewidth]{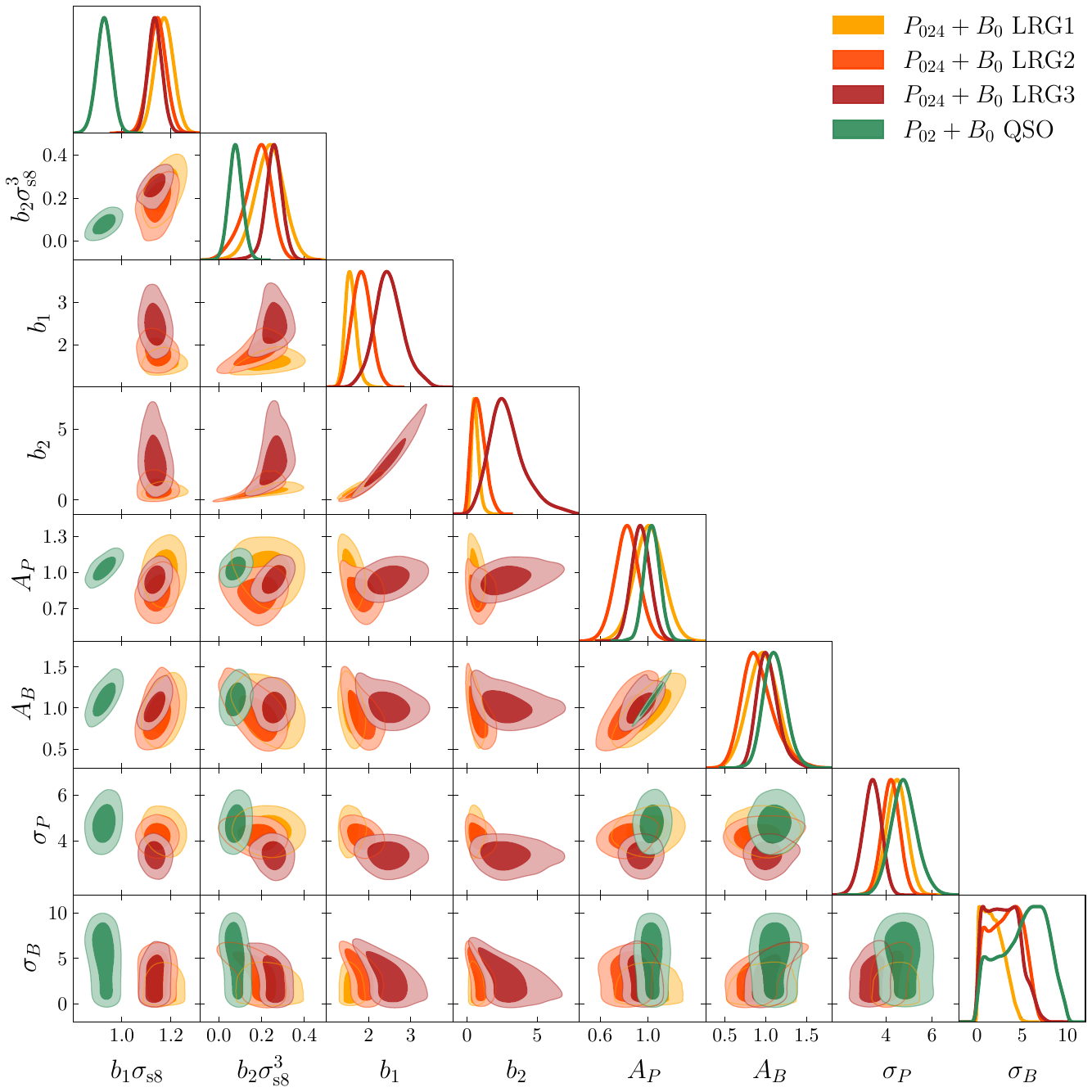}
    \caption{ Posteriors obtained for the ShapeFit analysis of $P+B$ for the nuisance parameters $b_1$, $b_2$, $A_P$, $A_B$, $\sigma_P$ and $\sigma_B$ (units of $\sigma_{P,B}$ are ${\rm Mpc}h^{-1}$ ), for each sample as labelled. The cosmological parameters are also varied in this fit (see Figure~\ref{fig: constraints_allbins}), but not shown for clarity. We also report the galaxy bias results in terms of $b_1\sigma_{s8}$ and $b_2\sigma_{s8}^3$, as these are the effective combinations better constrained for the $P+B$ case. The MAPs values of these contours are reported in Table~\ref{tab:nuisance}.}
    \label{fig:nuisance}
\end{figure}
Figure~\ref{fig:nuisance} displays these same results in a triangle-plot style.

\begin{table}[!ht]
\centering
        \small
    \resizebox{\columnwidth}{!}{
\begin{tabular}{|c|c|c|c|c|c|c|c|c|c|c|}
\hline
& Sample & $b_1\sigma_{s8}$ & $b_2\sigma_{s8}^3$& $b_1$ & $b_2$ & $A_P$ &   $A_B$ & $\sigma_P\,[{\rm Mpc}h^{-1}]$ & $\sigma_B\,[{\rm Mpc}h^{-1}]$ & $\chi^2/{\rm dof}$   \\ \hline\hline
\multirow{3}{*}{$P+B$} &  LRG1  &   $1.159_{-0.024}^{+0.084}$ & $0.213_{-0.044}^{+0.090}$ & $1.525_{-0.074}^{+0.180}$ & $0.484_{-0.129}^{+0.356}$  & $0.970_{-0.082}^{+0.161}$  &  $0.884_{-0.102}^{+0.265}$ &  $4.404_{-0.386}^{+0.497}$   & $0.521_{+0.010}^{+2.506}$  & 241/178\\ 
                      &  LRG2&    $1.140_{-0.031}^{+0.040}$ &$0.183_{-0.069}^{+0.058}$     & $1.882_{-0.274}^{+0.169}$ & $0.822_{-0.489}^{+0.487}$   &  $0.724_{+0.003}^{+0.203}$  & $0.787_{-0.066}^{+0.317}$  & $4.253_{-0.427}^{+0.335}$  & $3.672_{-2.598}^{+1.364}$ & 291/178 \\ 
                      & LRG3 &    $1.133_{-0.026}^{+0.030}$ &$0.233_{-0.014}^{+0.060}$   & $2.265_{-0.106}^{+0.551}$ & $1.863_{-0.197}^{+2.251}$   &$0.863_{+0.001}^{+0.155}$ & $0.998_{-0.102}^{+0.153}$& $3.339_{-0.355}^{+0.444}$  & $4.511_{-3.609}^{+0.163}$ & 253/178\\
                    &  QSO     &    $0.890_{+0.007}^{+0.069}$ &     $0.058_{-0.011}^{+0.052}$     & $-$ & $-$  & $0.981_{-0.013}^{+0.118}$ & $1.015_{-0.040}^{+0.231}$ &$4.563_{-0.340}^{+0.769}$ & $7.526_{-5.530}^{+0.190}$
                     & 193/166  \\ \hline
\multirow{3}{*}{$P$}   & LRG1       &  $1.136_{-0.078}^{-0.002}$  & $0.260_{-0.400}^{+0.500}$  & $-$ & $-$ &     $0.772_{+0.118}^{+0.427}$  & $-$& $4.057_{-1.192}^{-0.263}$ & $-$  & 33/31\\
                      & LRG2 &   $1.112_{-0.054}^{+0.023}$  &  $0.621_{-0.761}^{+0.139}$  & $-$ & $-$   &  $0.696_{+0.194}^{+0.503}$ &  $-$   & $4.083_{-1.218}^{-0.289}$  &  $-$  & 37/31\\ 
                      &  LRG3  &   $1.113_{-0.055}^{-0.021}$  &  $0.429_{-0.569}^{+0.331}$ & $-$ & $-$ &  $0.911_{-0.021}^{+0.288}$ &  $-$ & $3.258_{-0.392}^{+0.536}$ &  $-$ & 31/31\\
                      &  QSO     &    $0.904_{-0.043}^{+0.026}$ &  $0.047_{-0.213}^{+0.293}$  & $-$ & $-$ &     $0.992_{-0.031}^{+0.115}$  & $-$  &   $4.731_{-0.355}^{+0.973}$ & $-$   & 20/18\\\hline

\end{tabular}
}
\caption{Results for the nuisance parameters from our two main analyses, $P+B$ and $P$, for the LRG and QSO redshift bins, analogue to Table \ref{tab:results}. Each row displays the maximum a posteriori (MAP), together with the 1$\sigma$ region (which in this case only accounts for the statistical errors, as there is no systematic contribution calculated) centred in the MAP value. In the cases with significant projection effects, the MAP may fall outside of the $1\sigma$ region. We additionally report the corresponding values for $\chi^2$ over the degrees of freedom for each case.}
\label{tab:nuisance}
\end{table}

\section{Covariances and additional tables}
We will report here, upon acceptance, the covariance of the cosmological parameters for each tracer.

We also report here some numerical details not included in the main text. The cosmological parameter constraints for the non-baseline choices of data-vector (i.e. the ones involving $P_{024}$ in QSOs and $P_{02}$ in LRGs, for both power spectrum and joint power spectrum-bispectrum analyses), are shown in Table \ref{tab:non_baseline_results}.

\begin{table}[!ht]
\centering
        \small
    \resizebox{\columnwidth}{!}{
\begin{tabular}{|c|c|c|c|c|c|c|c|c|}
\hline
& Sample & $\alpha_\textrm{iso}$ & $\alpha_\textrm{AP}$& $df\sigma_\textrm{s8}$ &   $m+n$ & $df$ & $d\sigma_\textrm{s8}$ & $\chi^2/{\rm dof}$   \\ \hline\hline
\multirow{3}{*}{$P_{02}+B$} &  LRG1  &   $0.998_{-0.020}^{+0.018}$ & $0.961_{-0.036}^{+0.106}$   & $1.137_{-0.119}^{+0.131}$  &  $0.001_{-0.0795}^{+0.020}$ &  $1.014_{-0.217}^{+0.129}$   & $1.121_{-0.049}^{+0.197}$  & 231/165\\ 
                      &  LRG2&    $0.967_{-0.014}^{+0.029}$ &$1.064_{-0.093}^{+0.136}$     &     $0.993_{-0.164}^{+0.134}$  & $0.035_{-0.072}^{+0.023}$  & $0.745_{-0.173}^{+0.379}$  & $1.332_{-0.431}^{+0.200}$ & 271/165 \\ 
                      & LRG3 &    $1.002_{-0.014}^{+0.022}$ &$1.093_{-0.083}^{+0.085}$     &$0.896_{-0.111}^{+0.153}$ & $0.036_{-0.085}^{+0.009}$& $0.699_{-0.024}^{+0.418}$  & $1.281_{-0.404}^{-0.042}$ & 240/165\\\hline
                  $P_{024}+B$  &  QSO     &    $1.019_{-0.022}^{+0.026}$ &     $1.082_{-0.053}^{+0.029}$      &     $1.016_{-0.091}^{+0.093}$ & $0.067_{-0.046}^{+0.036}$     &   - & -  & 213/179  \\ \hline
\multirow{3}{*}{$P_{02}$}   & LRG1       &  $0.965_{-0.016}^{+0.026}$  & $1.042_{-0.075}^{+0.085}$  &     $0.935_{-0.047}^{+0.206}$  & $0.009_{-0.097}^{+0.010}$&- &-  & 24/18\\
                      & LRG2 &   $0.949_{-0.015}^{+0.022}$  &  $1.035_{-0.072}^{+0.078}$     &  $0.993_{-0.089}^{+0.164}$ &  $0.016_{-0.076}^{+0.032}$    & -&  -  & 20/18\\ 
                      &  LRG3  &   $0.994_{-0.019}^{+0.020}$  &  $1.056_{-0.070}^{+0.096}$  &  $0.951_{-0.127}^{+0.172}$ &  $-0.011_{-0.067}^{+0.037}$ & - &  - & 17/18\\\hline
                  $P_{024}$    &  QSO     &    $1.006_{-0.022}^{+0.031}$ &  $1.082_{-0.053}^{+0.029}$ &     $1.016_{-0.091}^{+0.093}$  & $0.067_{-0.046}^{+0.036}$  &   - & -   & 33/31\\\hline

\end{tabular}
}
\caption{Analogous to Table \ref{tab:results}, in this case showing the combinations of statistics that are not a part of our baseline analysis.
}
\label{tab:non_baseline_results}
\end{table}

The systematic error tables for the power spectrum analysis (not including the bispectrum), analogous to Tables \ref{tab:P024B0syst} and \ref{tab:P02B0syst}, are respectively Table \ref{tab:P024syst} and \ref{tab:P02syst}.

\begin{table}[!ht]
\centering
\begin{tabular}{|c|c|c|c|c|c|}
\hline
$[\%\sigma_\textrm{DR1}]$ & $P_{024}$ & $\sigma_{\alpha_\textrm{iso}}$ & $\sigma_{\alpha_\textrm{AP}}$& $\sigma_{f\sigma_\textrm{s8}}$  &   $\sigma_{m+n}$ \\ \hline\hline
\multirow{3}{*}{Modelling} &  LRG1      &   $<20$ &     $<20$   &     $<10$   &  $<20$        \\ 
                      &  LRG2     &    $\sim30$ &     $\sim20$   &     $\sim30$   &      $\sim20$      \\ 
                      & LRG3      &    $\sim40$ &     $<20$   &  $\sim20$   &      $\sim40$    \\
                    &  QSO     &    $\sim40$ &     $<10$      &     $\sim20$   &  $\sim20$      \\ \hline
\multirow{3}{*}{Observational}   & LRG1       &  $<10$  & $<10$  &     $<10$  &     $<10$      \\
                      & LRG2     &   $<10$  &      $<10$         &      $<10$  & $<10$     \\ 
                      &  LRG3     &   $<10$  &     $\sim60$     &     $\sim80$   & $<10$    \\
                      &  QSO     &    $\sim50$ &  $\sim50$ &     $\sim130$      &  $<20$      \\\hline
\multirow{3}{*}{HOD}   & LRG1       & $<10(11)$  & $<10(7)$     &   $<10(9)$  &   $<10(9)$         \\
                      & LRG2     & $<10(16)$ & $<10(9)$         &   $<10(14)$  &      $<20(13)$        \\ 
                      &  LRG3     &  $<10(16)$ & $<10(12)$       &   $<10(17)$  &     $<10(14)$   \\
                      &  QSO     &  $<10(4)$ & $<10(7)$       &   $<10(15)$   &  $<20(29)$   \\\hline
\multirow{3}{*}{Fiducial cosmology}   & LRG1  & $<10(17)$  & $<10(7)$     &   $<10(13)$  &   $<20(45)$         \\
                      & LRG2     & $<10(25)$ & $<10(9)$         &   $<20(19)$  &      $<20(61)$        \\ 
                      &  LRG3     &  $<10(25)$ & $<10(12)$       &   $<20(23)$  &     $<20(65)$   \\
                      &  QSO     &  $<10(24)$ & $<10(10)$       &   $<20(40)$   &  $<20(55)$   \\\hline\hline
\multirow{3}{*}{\textbf{Total}}&\textbf{LRG1}& $<20$&$<20$ &$\sim20$ &$<20$   \\
& \textbf{LRG2} &$\sim30$ & $\sim20$ & $\sim50$  & $\sim20$ \\ 
&  \textbf{LRG3}&$\sim40$   & $\sim60$&   $\sim90$  &   $\sim40$\\
&  \textbf{QSO}&$\sim64$   & $\sim50$&   $\sim132$  &  $\sim20$   \\\hline

\end{tabular}
\caption{Same as Table \ref{tab:P024B0syst}, with the only difference that the summary statistics considered here are the power spectrum monopole, quadrupole and hexadecapole ($P_{024}$), without the bispectrum.}
\label{tab:P024syst}
\end{table}

\begin{table}[!ht]
\centering
\begin{tabular}{|c|c|c|c|c|c|}
\hline
$[\%\sigma_\textrm{DR1}]$ & $P_{02}$ & $\sigma_{\alpha_\textrm{iso}}$ & $\sigma_{\alpha_\textrm{AP}}$& $\sigma_{f\sigma_\textrm{s8}}$  &   $\sigma_{m+n}$ \\ \hline\hline
\multirow{3}{*}{Modelling} &  LRG1      &   $\sim20$ &     $<20$   &     $<10$   &  $<10$        \\ 
                      &  LRG2     &    $\sim30$ &     $<10$          &     $<20$   &      $<20$      \\ 
                      & LRG3      &    $\sim40$ &     $<20$          &     $\sim40$   &      $\sim40$    \\
                    &  QSO     &    $\sim50$ &     $<20$      &     $<10$   &  $<20$      \\ \hline
\multirow{3}{*}{Observational}   & LRG1       &  $<10$  & $<10$  &     $<10$  &     $<10$      \\
                      & LRG2     &   $<10$  &      $<20$         &      $<20$  & $<10$     \\ 
                      &  LRG3     &   $<10$  &     $<10$     &     $<20$   & $<10$    \\
                      &  QSO     &    $\sim50$ &  $<10$ &     $\sim70$      &  $<20$      \\\hline
\multirow{3}{*}{HOD}   & LRG1       & $<10(10)$  & $<10(6)$     &   $<10(8)$  &   $<10(13)$         \\
                      & LRG2     & $<10(15)$ & $<10(9)$         &   $<10(12)$  &      $<10(18)$        \\ 
                      &  LRG3     &  $<10(15)$ & $<10(10)$       &   $<10(13)$  &     $<10(20)$   \\
                      &  QSO     &  $<10(4)$ & $<10(8)$       &   $<10(10)$   &  $<20(27)$   \\\hline
\multirow{3}{*}{Fiducial cosmology}   & LRG1  & $<10(18)$  & $<10(7)$     &   $<10(15)$  &   $<20(43)$         \\
                      & LRG2     & $<10(28)$ & $<10(10)$         &   $<20(22)$  &      $\sim20(59)$        \\ 
                      &  LRG3     &  $<10(26)$ & $<10(11)$       &   $<20(25)$  &     $\sim20(65)$   \\
                      &  QSO     &  $<20(28)$ & $<10(6)$       &   $<20(36)$   &  $<20(54)$   \\\hline\hline
\multirow{3}{*}{\textbf{Total}}&\textbf{LRG1}& $\sim20$&$<20$ &$\sim30$ &$<20$   \\
& \textbf{LRG2} &$\sim30$ & $<20$ & $\sim40$  & $\sim20$ \\ 
&  \textbf{LRG3}&$\sim40$   & $<20$&   $\sim57$  &   $\sim45$\\
&  \textbf{QSO}&$\sim71$   & $<20$&   $\sim70$  &  $<20$   \\\hline

\end{tabular}
\caption{Same as Table \ref{tab:P024B0syst}, with the only difference that the summary statistics considered here are the power spectrum monopole and quadrupole ($P_{02}$) only. 
}
\label{tab:P02syst}
\end{table}

\input{DESI-2025-0531_author_list33.affiliations}

\bibliographystyle{JHEP}
\bibliography{main}

\end{document}

%% file: DESI-2025-0531_author_list33.tex
\affiliation{Affiliations are in Appendix \ref{sec:affiliations}}

\author[1]{{S.~Novell-Masot}\orcidlink{0000-0002-7958-494},}
\author[2,3,1]{{H.~Gil-Mar\'in}\orcidlink{0000-0003-0265-6217},}
\author[4,1]{{L.~Verde}\orcidlink{0000-0003-2601-8770},}
\author[5]{{J.~Aguilar},}
\author[6]{{S.~Ahlen}\orcidlink{0000-0001-6098-7247},}
\author[5]{{S.~Bailey}\orcidlink{0000-0003-4162-6619},}
\author[7]{{S.~BenZvi}\orcidlink{0000-0001-5537-4710},}
\author[8,9]{{D.~Bianchi}\orcidlink{0000-0001-9712-0006},}
\author[10]{{D.~Brooks},}
\author[11,12]{{E.~Buckley-Geer},}
\author[13,14]{{A.~Carnero Rosell}\orcidlink{0000-0003-3044-5150},}
\author[5]{{E.~Chaussidon}\orcidlink{0000-0001-8996-4874},}
\author[5]{{T.~Claybaugh},}
\author[15]{{S.~Cole}\orcidlink{0000-0002-5954-7903},}
\author[5]{{A.~Cuceu}\orcidlink{0000-0002-2169-0595},}
\author[16]{{K.~S.~Dawson}\orcidlink{0000-0002-0553-3805},}
\author[17]{{A.~de la Macorra}\orcidlink{0000-0002-1769-1640},}
\author[7]{{R.~Demina},}
\author[18]{{A.~Dey}\orcidlink{0000-0002-4928-4003},}
\author[19,20]{{B.~Dey}\orcidlink{0000-0002-5665-7912},}
\author[10]{{P.~Doel},}
\author[5,21]{{S.~Ferraro}\orcidlink{0000-0003-4992-7854},}
\author[22]{{A.~Font-Ribera}\orcidlink{0000-0002-3033-7312},}
\author[23,24]{{J.~E.~Forero-Romero}\orcidlink{0000-0002-2890-3725},}
\author[3,25,26]{{E.~Gaztañaga},}
\author[5]{{S.~Gontcho A Gontcho}\orcidlink{0000-0003-3142-233X},}
\author[27]{{A.~X.~Gonzalez-Morales}\orcidlink{0000-0003-4089-6924},}
\author[12]{{G.~Gutierrez},}
\author[28,29]{{H.~K.~Herrera-Alcantar}\orcidlink{0000-0002-9136-9609},}
\author[30,31,32]{{K.~Honscheid}\orcidlink{0000-0002-6550-2023},}
\author[33]{{C.~Howlett}\orcidlink{0000-0002-1081-9410},}
\author[18]{{S.~Juneau}\orcidlink{0000-0002-0000-2394},}
\author[34]{{R.~Kehoe},}
\author[35]{{D.~Kirkby}\orcidlink{0000-0002-8828-5463},}
\author[5]{{T.~Kisner}\orcidlink{0000-0003-3510-7134},}
\author[5]{{A.~Kremin}\orcidlink{0000-0001-6356-7424},}
\author[36]{{C.~Lamman}\orcidlink{0000-0002-6731-9329},}
\author[5]{{M.~Landriau}\orcidlink{0000-0003-1838-8528},}
\author[37]{{L.~Le~Guillou}\orcidlink{0000-0001-7178-8868},}
\author[5]{{M.~E.~Levi}\orcidlink{0000-0003-1887-1018},}
\author[29]{{C.~Magneville},}
\author[38,22]{{M.~Manera}\orcidlink{0000-0003-4962-8934},}
\author[18]{{A.~Meisner}\orcidlink{0000-0002-1125-7384},}
\author[4,22]{{R.~Miquel},}
\author[39]{{J.~Moustakas}\orcidlink{0000-0002-2733-4559},}
\author[17]{{A.~Muñoz-Gutiérrez},}
\author[40]{{A.~D.~Myers},}
\author[25]{{S.~Nadathur}\orcidlink{0000-0001-9070-3102},}
\author[27,41]{{G.~Niz}\orcidlink{0000-0002-1544-8946},}
\author[42,17]{{H.~E.~Noriega}\orcidlink{0000-0002-3397-3998},}
\author[43,44,45]{{W.~J.~Percival}\orcidlink{0000-0002-0644-5727},}
\author[5,46,21]{{C.~Poppett},}
\author[47]{{F.~Prada}\orcidlink{0000-0001-7145-8674},}
\author[48]{{I.~P\'erez-R\`afols}\orcidlink{0000-0001-6979-0125},}
\author[30,49,32]{{A.~J.~Ross}\orcidlink{0000-0002-7522-9083},}
\author[50]{{G.~Rossi},}
\author[51,52,53]{{L.~Samushia}\orcidlink{0000-0002-1609-5687},}
\author[54]{{E.~Sanchez}\orcidlink{0000-0002-9646-8198},}
\author[5]{{D.~Schlegel},}
\author[55,56]{{M.~Schubnell},}
\author[57]{{H.~Seo}\orcidlink{0000-0002-6588-3508},}
\author[5]{{J.~Silber}\orcidlink{0000-0002-3461-0320},}
\author[18]{{D.~Sprayberry},}
\author[56]{{G.~Tarl\'{e}}\orcidlink{0000-0003-1704-0781},}
\author[17]{{M.~Vargas-Maga\~na}\orcidlink{0000-0003-3841-1836},}
\author[18]{{B.~A.~Weaver},}
\author[37]{{P.~Zarrouk}\orcidlink{0000-0002-7305-9578},}
\author[5]{{R.~Zhou}\orcidlink{0000-0001-5381-4372},}
\author[58]{{H.~Zou}\orcidlink{0000-0002-6684-3997}}

%% file: DESI-2025-0531_author_list33.affiliations.tex
% Author list file generated with: mkauthlist 0+unknown 
%% Affiliations file. Use \input to call it after \appendix

\section{Author Affiliations}
\label{sec:affiliations}

\noindent \hangindent=.5cm $^{1}${Institut de Ci\`encies del Cosmos (ICCUB), Universitat de Barcelona (UB), c. Mart\'i i Franqu\`es, 1, 08028 Barcelona, Spain.}

\noindent \hangindent=.5cm $^{2}${Departament de F\'{\i}sica Qu\`{a}ntica i Astrof\'{\i}sica, Universitat de Barcelona, Mart\'{\i} i Franqu\`{e}s 1, E08028 Barcelona, Spain}

\noindent \hangindent=.5cm $^{3}${Institut d'Estudis Espacials de Catalunya (IEEC), c/ Esteve Terradas 1, Edifici RDIT, Campus PMT-UPC, 08860 Castelldefels, Spain}

\noindent \hangindent=.5cm $^{4}${Instituci\'{o} Catalana de Recerca i Estudis Avan\c{c}ats, Passeig de Llu\'{\i}s Companys, 23, 08010 Barcelona, Spain}

\noindent \hangindent=.5cm $^{5}${Lawrence Berkeley National Laboratory, 1 Cyclotron Road, Berkeley, CA 94720, USA}

\noindent \hangindent=.5cm $^{6}${Physics Dept., Boston University, 590 Commonwealth Avenue, Boston, MA 02215, USA}

\noindent \hangindent=.5cm $^{7}${Department of Physics \& Astronomy, University of Rochester, 206 Bausch and Lomb Hall, P.O. Box 270171, Rochester, NY 14627-0171, USA}

\noindent \hangindent=.5cm $^{8}${Dipartimento di Fisica ``Aldo Pontremoli'', Universit\`a degli Studi di Milano, Via Celoria 16, I-20133 Milano, Italy}

\noindent \hangindent=.5cm $^{9}${INAF-Osservatorio Astronomico di Brera, Via Brera 28, 20122 Milano, Italy}

\noindent \hangindent=.5cm $^{10}${Department of Physics \& Astronomy, University College London, Gower Street, London, WC1E 6BT, UK}

\noindent \hangindent=.5cm $^{11}${Department of Astronomy and Astrophysics, University of Chicago, 5640 South Ellis Avenue, Chicago, IL 60637, USA}

\noindent \hangindent=.5cm $^{12}${Fermi National Accelerator Laboratory, PO Box 500, Batavia, IL 60510, USA}

\noindent \hangindent=.5cm $^{13}${Departamento de Astrof\'{\i}sica, Universidad de La Laguna (ULL), E-38206, La Laguna, Tenerife, Spain}

\noindent \hangindent=.5cm $^{14}${Instituto de Astrof\'{\i}sica de Canarias, C/ V\'{\i}a L\'{a}ctea, s/n, E-38205 La Laguna, Tenerife, Spain}

\noindent \hangindent=.5cm $^{15}${Institute for Computational Cosmology, Department of Physics, Durham University, South Road, Durham DH1 3LE, UK}

\noindent \hangindent=.5cm $^{16}${Department of Physics and Astronomy, The University of Utah, 115 South 1400 East, Salt Lake City, UT 84112, USA}

\noindent \hangindent=.5cm $^{17}${Instituto de F\'{\i}sica, Universidad Nacional Aut\'{o}noma de M\'{e}xico,  Circuito de la Investigaci\'{o}n Cient\'{\i}fica, Ciudad Universitaria, Cd. de M\'{e}xico  C.~P.~04510,  M\'{e}xico}

\noindent \hangindent=.5cm $^{18}${NSF NOIRLab, 950 N. Cherry Ave., Tucson, AZ 85719, USA}

\noindent \hangindent=.5cm $^{19}${Department of Astronomy \& Astrophysics, University of Toronto, Toronto, ON M5S 3H4, Canada}

\noindent \hangindent=.5cm $^{20}${Department of Physics \& Astronomy and Pittsburgh Particle Physics, Astrophysics, and Cosmology Center (PITT PACC), University of Pittsburgh, 3941 O'Hara Street, Pittsburgh, PA 15260, USA}

\noindent \hangindent=.5cm $^{21}${University of California, Berkeley, 110 Sproul Hall \#5800 Berkeley, CA 94720, USA}

\noindent \hangindent=.5cm $^{22}${Institut de F\'{i}sica d’Altes Energies (IFAE), The Barcelona Institute of Science and Technology, Edifici Cn, Campus UAB, 08193, Bellaterra (Barcelona), Spain}

\noindent \hangindent=.5cm $^{23}${Departamento de F\'isica, Universidad de los Andes, Cra. 1 No. 18A-10, Edificio Ip, CP 111711, Bogot\'a, Colombia}

\noindent \hangindent=.5cm $^{24}${Observatorio Astron\'omico, Universidad de los Andes, Cra. 1 No. 18A-10, Edificio H, CP 111711 Bogot\'a, Colombia}

\noindent \hangindent=.5cm $^{25}${Institute of Cosmology and Gravitation, University of Portsmouth, Dennis Sciama Building, Portsmouth, PO1 3FX, UK}

\noindent \hangindent=.5cm $^{26}${Institute of Space Sciences, ICE-CSIC, Campus UAB, Carrer de Can Magrans s/n, 08913 Bellaterra, Barcelona, Spain}

\noindent \hangindent=.5cm $^{27}${Departamento de F\'{\i}sica, DCI-Campus Le\'{o}n, Universidad de Guanajuato, Loma del Bosque 103, Le\'{o}n, Guanajuato C.~P.~37150, M\'{e}xico.}

\noindent \hangindent=.5cm $^{28}${Institut d'Astrophysique de Paris. 98 bis boulevard Arago. 75014 Paris, France}

\noindent \hangindent=.5cm $^{29}${IRFU, CEA, Universit\'{e} Paris-Saclay, F-91191 Gif-sur-Yvette, France}

\noindent \hangindent=.5cm $^{30}${Center for Cosmology and AstroParticle Physics, The Ohio State University, 191 West Woodruff Avenue, Columbus, OH 43210, USA}

\noindent \hangindent=.5cm $^{31}${Department of Physics, The Ohio State University, 191 West Woodruff Avenue, Columbus, OH 43210, USA}

\noindent \hangindent=.5cm $^{32}${The Ohio State University, Columbus, 43210 OH, USA}

\noindent \hangindent=.5cm $^{33}${School of Mathematics and Physics, University of Queensland, Brisbane, QLD 4072, Australia}

\noindent \hangindent=.5cm $^{34}${Department of Physics, Southern Methodist University, 3215 Daniel Avenue, Dallas, TX 75275, USA}

\noindent \hangindent=.5cm $^{35}${Department of Physics and Astronomy, University of California, Irvine, 92697, USA}

\noindent \hangindent=.5cm $^{36}${Center for Astrophysics $|$ Harvard \& Smithsonian, 60 Garden Street, Cambridge, MA 02138, USA}

\noindent \hangindent=.5cm $^{37}${Sorbonne Universit\'{e}, CNRS/IN2P3, Laboratoire de Physique Nucl\'{e}aire et de Hautes Energies (LPNHE), FR-75005 Paris, France}

\noindent \hangindent=.5cm $^{38}${Departament de F\'{i}sica, Serra H\'{u}nter, Universitat Aut\`{o}noma de Barcelona, 08193 Bellaterra (Barcelona), Spain}

\noindent \hangindent=.5cm $^{39}${Department of Physics and Astronomy, Siena College, 515 Loudon Road, Loudonville, NY 12211, USA}

\noindent \hangindent=.5cm $^{40}${Department of Physics \& Astronomy, University  of Wyoming, 1000 E. University, Dept.~3905, Laramie, WY 82071, USA}

\noindent \hangindent=.5cm $^{41}${Instituto Avanzado de Cosmolog\'{\i}a A.~C., San Marcos 11 - Atenas 202. Magdalena Contreras. Ciudad de M\'{e}xico C.~P.~10720, M\'{e}xico}

\noindent \hangindent=.5cm $^{42}${Instituto de Ciencias F\'{\i}sicas, Universidad Nacional Aut\'onoma de M\'exico, Av. Universidad s/n, Cuernavaca, Morelos, C.~P.~62210, M\'exico}

\noindent \hangindent=.5cm $^{43}${Department of Physics and Astronomy, University of Waterloo, 200 University Ave W, Waterloo, ON N2L 3G1, Canada}

\noindent \hangindent=.5cm $^{44}${Perimeter Institute for Theoretical Physics, 31 Caroline St. North, Waterloo, ON N2L 2Y5, Canada}

\noindent \hangindent=.5cm $^{45}${Waterloo Centre for Astrophysics, University of Waterloo, 200 University Ave W, Waterloo, ON N2L 3G1, Canada}

\noindent \hangindent=.5cm $^{46}${Space Sciences Laboratory, University of California, Berkeley, 7 Gauss Way, Berkeley, CA  94720, USA}

\noindent \hangindent=.5cm $^{47}${Instituto de Astrof\'{i}sica de Andaluc\'{i}a (CSIC), Glorieta de la Astronom\'{i}a, s/n, E-18008 Granada, Spain}

\noindent \hangindent=.5cm $^{48}${Departament de F\'isica, EEBE, Universitat Polit\`ecnica de Catalunya, c/Eduard Maristany 10, 08930 Barcelona, Spain}

\noindent \hangindent=.5cm $^{49}${Department of Astronomy, The Ohio State University, 4055 McPherson Laboratory, 140 W 18th Avenue, Columbus, OH 43210, USA}

\noindent \hangindent=.5cm $^{50}${Department of Physics and Astronomy, Sejong University, 209 Neungdong-ro, Gwangjin-gu, Seoul 05006, Republic of Korea}

\noindent \hangindent=.5cm $^{51}${Abastumani Astrophysical Observatory, Tbilisi, GE-0179, Georgia}

\noindent \hangindent=.5cm $^{52}${Department of Physics, Kansas State University, 116 Cardwell Hall, Manhattan, KS 66506, USA}

\noindent \hangindent=.5cm $^{53}${Faculty of Natural Sciences and Medicine, Ilia State University, 0194 Tbilisi, Georgia}

\noindent \hangindent=.5cm $^{54}${CIEMAT, Avenida Complutense 40, E-28040 Madrid, Spain}

\noindent \hangindent=.5cm $^{55}${Department of Physics, University of Michigan, 450 Church Street, Ann Arbor, MI 48109, USA}

\noindent \hangindent=.5cm $^{56}${University of Michigan, 500 S. State Street, Ann Arbor, MI 48109, USA}

\noindent \hangindent=.5cm $^{57}${Department of Physics \& Astronomy, Ohio University, 139 University Terrace, Athens, OH 45701, USA}

\noindent \hangindent=.5cm $^{58}${National Astronomical Observatories, Chinese Academy of Sciences, A20 Datun Rd., Chaoyang District, Beijing, 100012, P.R. China}